%% file: survey_on_SDN.tex
\pdfoutput=1
\documentclass[journal]{IEEEtran}
\usepackage{epsfig}
\usepackage{graphicx}
\usepackage[hyphens]{url}
\usepackage{xspace}
\usepackage{array}
\usepackage{tabularx,colortbl}
\usepackage{color}
\usepackage{makeidx}
\usepackage[colorlinks=true,citecolor=green,urlcolor=SteelBlue4,linkcolor=Firebrick4,hyperindex,breaklinks]{hyperref}
\usepackage[urlcolor=SteelBlue4,linkcolor=Firebrick4,hyperindex,breaklinks]{hyperref}
\usepackage[hyphenbreaks]{breakurl}
\usepackage[table,hyperref,x11names]{xcolor}
\usepackage{multirow}
\usepackage{comment}
\usepackage{array}
\usepackage{xcolor}

\newcommand{\coloredtext}[1]{#1\xspace}

\newcommand{\manapps}{network applications\xspace}

\newcommand{\ManaApps}{Network Applications\xspace}
\newcommand{\Manaaapps}{Network applications\xspace}

\makeatletter

\def\zapcolorreset{\let\reset@color\relax\ignorespaces}
\def\colorrows#1{\noalign{\aftergroup\zapcolorreset#1}\ignorespaces}

\makeatother

\begin{document}
%

%
\title{Software-Defined Networking: \\A Comprehensive Survey}

%
%
%

\author{Diego~Kreutz,~\IEEEmembership{Member,~IEEE,}
        Fernando~M.~V.~Ramos,~\IEEEmembership{Member,~IEEE,}
        Paulo~Verissimo,~\IEEEmembership{Fellow,~IEEE,}
        Christian~Esteve~Rothenberg,~\IEEEmembership{Member,~IEEE,}        
        Siamak~Azodolmolky,~\IEEEmembership{Senior Member,~IEEE,}        
        and~Steve~Uhlig,~\IEEEmembership{Member,~IEEE}
\thanks{D. Kreutz and F. Ramos are with the Department of Informatics of Faculty of Sciences, University of Lisbon, Lisbon,
	1749-016 Portugal e-mail: kreutz@lasige.di.fc.ul.pt, fvramos@fc.ul.pt.}
\thanks{P. Ver\'{i}ssimo is with the Interdisciplinary Centre for Security, Reliability and Trust (SnT), University of Luxembourg,
	4 rue Alphonse Weicker,  L-2721 Luxembourg. e-mail: paulo.verissimo@uni.lu.}    
\thanks{C. Esteve Rothenberg is with the School of Electrical and Computer Engineering (FEEC, University of Campinas, Brazil. e-mail: chesteve@dca.fee.unicamp.br.}
\thanks{S. Azodolmolky is with Gesellschaft f\"{u}r Wissenschaftliche Datenverarbeitung mbH G\"{o}ttingen (GWDG), Am Fa{\ss}berg 11, 37077 G\"{o}ttigen, Germany. e-mail: siamak.azodolmolky@gwdg.de.}
\thanks{S. Uhlig is with Queen Mary University of London. is with Queen Mary, University of London, Mile End Road, London E1 4NS, United Kingdom. e-mail steve@eecs.qmul.ac.uk.}
\thanks{Manuscript received May 31, 2014.}
}

%
%

\markboth{Version 2.01}%
{Kreutz \MakeLowercase{\textit{et al.}}: Software-Defined Networking: a comprehensive survey}
%



\maketitle


\newpage

\begin{abstract}
The Internet has led to the creation of a digital society, where
(almost) everything is connected and is accessible from anywhere.
However, despite their widespread adoption, traditional IP networks
are complex and very hard to manage.  It is both difficult to
configure the network according to pre-defined policies, and to
reconfigure it to respond to faults, load and changes.  To make
matters even more difficult, current networks are also vertically
integrated: the control and data planes are bundled together.
Software-Defined Networking (SDN) is an emerging paradigm that
promises to change this state of affairs, by breaking vertical
integration, separating the network's control logic from the
underlying routers and switches, promoting (logical) centralization of
network control, and introducing the ability to program the network.
The separation of concerns introduced between the definition of
network policies, their implementation in switching hardware, and the
forwarding of traffic, is key to the desired flexibility: by breaking
the network control problem into tractable pieces, SDN makes it easier
to create and introduce new abstractions in networking, simplifying
network management and facilitating network evolution.

In this paper we present a comprehensive survey on SDN.
We start by introducing the motivation for
SDN, explain its main concepts and how it differs from traditional
networking, its roots, and the standardization activities regarding this novel paradigm.
Next, we present the key building blocks of an SDN
infrastructure using a bottom-up, layered approach.  We provide an
in-depth analysis of the hardware infrastructure, southbound and
northbound APIs, network virtualization layers, network operating
systems (SDN controllers), network programming languages, and
network applications.  We also look at cross-layer problems such as
debugging and troubleshooting.  In an effort to anticipate the future
evolution of this new paradigm, we discuss the main ongoing research
efforts and challenges of SDN.  In particular, we address the design
of switches and control platforms -- with a focus on aspects such as
resiliency, scalability, performance, security and dependability -- as
well as new opportunities for carrier transport
networks and cloud providers. Last but not least, we analyze the
position of SDN as a key enabler of a software-defined environment.
\end{abstract}


\begin{IEEEkeywords}
Software-defined networking, OpenFlow, network virtualization, network operating systems, programmable networks, network hypervisor, programming languages, flow-based networking, scalability, dependability, carrier-grade networks, software-defined environments.
\end{IEEEkeywords}


%
\IEEEpeerreviewmaketitle

\input{text/0_introduction.tex}

\input{text/1_traditional_nets.tex}

\input{text/2_what_is_sdn.tex}
\input{text/3_history_of_sdn.tex}
\input{text/4_sdn_in_layers.tex}
\input{text/5_challenges.tex}
\input{text/8_conclusion.tex}
\section*{Acknowledgment}

The authors would like to thank the anonymous reviewers and a number of fellows that have contributed to this work. 
Jennifer Rexford for her feedback on an early version of this work and encouragement to get it finished. 
Srini Seetharaman for reviewing the draft and providing inputs to alternative SDN views. David Meyer for his thoughts on organizational challenges. 
\coloredtext{Thomas Nadeau for his inputs on OpenDaylight. 
Luis Miguel Contreras Murillo for his contributions to SDN standardization. 
In addition, we would like also to acknowledge the several contributions from the community, namely from Aurel A. Lazar, Carmelo Cascone, Gyanesh Patra, Haleplidis Evangelos, Javier Ancieta, Joe Stringer, Kostas Pentikousis, Luciano de Paula, Marco Canini, Philip Wette, Ramon Fontes, Raphael Rosa, Regivaldo Costa, Ricardo de Freitas Gesuatto, Wolfgang John.}

\ifCLASSOPTIONcaptionsoff
  \newpage
\fi



%
%
%


\bibliographystyle{IEEEtran}
\bibliography{IEEEabrv,refs_SDNs}

%

\begin{IEEEbiographynophoto}{Diego Kreutz}
received his Computer Science degree, MSc degree in Informatics, and MSc degree in Production Engineering from Federal University of Santa Maria.
Over the past 11 years he has worked as an Assistant Professor in the Lutheran University of Brazil and in the Federal University of Pampa, and as a researcher member of the Software/Hardware Integration Lab (LISHA) at Federal University of Santa Catarina.
Out of the academia, he has also experience as an independent technical consultant on network operations and management for small and medium enterprises and government institutions.
Currently, he is a PhD student at Faculty of Sciences of University of Lisbon, Portugal, involved in research projects related to intrusion tolerance, security, and future networks including the TRONE, and SecFuNet international projects. 
His main research interests are in network control platforms, software-defined networks, intrusion tolerance, system security and dependability, high performance computing, and cloud computing. 
\end{IEEEbiographynophoto}

\begin{IEEEbiographynophoto}{Fernando M. V. Ramos}
 is an Assistant Professor in the University of Lisbon. Previous academic positions include  those of Teaching Assistant (supervisor) in the University of Cambridge, in the ISEL and in the University of Aveiro. Over the past 12 years he has taught over a dozen courses: from physics (Electromagnetism) to EE (digital electronics, electric circuits, telecommunication systems and foundations) to CS (operating and distributed systems, computer networks, algorithms, programming languages).
Periods outside academia include working as a researcher in Portugal Telecom and in Telefonica Research.
He holds a PhD degree from the University of Cambridge where he worked on IPTV networks.
His current research interests are: software-defined networking, network virtualization, and cloud computing, with security and dependability as an orthogonal concern.
\end{IEEEbiographynophoto}


\begin{IEEEbiographynophoto}{Paulo Ver\'{i}ssimo}
is a Professor of the 	Interdisciplinary Centre for Security, Reliability and Trust (SnT), University of Luxembourg\footnote{This work was performed whilst this author was at the U. of Lisbon Faculty of Sciences}. He is currently Chair of the IFIP WG 10.4 on Dependable Computing and Fault-Tolerance and vice-Chair of the Steering Committee of the IEEE/IFIP DSN conference. PJV is Fellow of the IEEE and Fellow of the ACM. He is associate editor of the Elsevier Int'l Journal on Critical Infrastructure Protection. Ver\'{i}ssimo is currently interested in distributed architectures, middleware and algorithms for: adaptability and safety of real-time networked embedded systems; and resilience of secure and dependable large-scale systems. He is author of over 170 peer-refereed publications and co-author of 5 books. 
\end{IEEEbiographynophoto}

\begin{IEEEbiographynophoto}{Christian Esteve Rothenberg} is an Assistant Professor in the Faculty of Electrical and Computer Engineering at University of Campinas (UNICAMP), where he received his Ph.D. in 2010. 
From 2010 to 2013, he worked as Senior Research Scientist in the areas of IP systems and networking at CPqD Research and Development Center in Telecommunications (Campinas, Brazil), where he was technical lead of R\&D acitivities in the field of OpenFlow/SDN such as the RouteFlow project, the OpenFlow 1.3 Ericsson/CPqD softswitch, or the ONF Driver competition. 
Christian holds the Telecommunication Engineering degree from Universidad Polit\'{e}cnica de Madrid (ETSIT - UPM), Spain, and the M.Sc. (Dipl. Ing.) degree in Electrical Engineering and Information Technology from the Darmstadt University of Technology (TUD), Germany, 2006. Christian holds two international patents and has over 50 publications including scientific journals and top-tier networking conferences such as SIGCOMM and INFOCOM. Since April 2013, Christian is an ONF Research Associate.
\end{IEEEbiographynophoto}

\begin{IEEEbiographynophoto}{Siamak Azodolmolky}
received his Computer Engineering degree from Tehran University and his first MSc. degree in Computer Architecture from Azad University in 1994 and 1998 respectively. He was employed by Data Processing Iran Co. (IBM in Iran) as a Software Developer, Systems Engineer, and as a Senior R\& D Engineer during 1992-2001. He received his second MSc. degree with distinction from Carnegie Mellon University in 2006. He joined Athens Information Technology (AIT) as a Research Scientist and Software Developer in 2007, while pursuing his PhD degree. In August 2010, he joined the High Performance Networks research group of the School of Computer Science and Electronic Engineering (CSEE) of the University of Essex as a Senior Research Officer. He received his PhD from Universitat Polit\'{e}cnica de Catalunya (UPC) in 2011. He has been the technical investigator of various national and EU funded projects. Software Defined Networking (SDN) has been one of his research interests since 2010, in which he has been investigating the extension of OpenFlow towards its application in core transport (optical) networks. He has published more than 50 scientific papers in international conferences, journals, and books. Software Defined Networking with OpenFlow is one of his recent books. Currently, he is with Gesellschaft f\"{u}r Wissenschaftliche Datenverarbeitung mbH G\"{o}ttingen (GWDG) as a Senior Researcher and has lead SDN related activities since September 2012. He is a professional member of ACM and a senior member of IEEE.
\end{IEEEbiographynophoto}

\begin{IEEEbiographynophoto}{Steve Uhlig} obtained a Ph.D. degree in Applied Sciences from the University of Louvain, Belgium, in 2004. From 2004 to 2006, he was a Postdoctoral Fellow of the Belgian National Fund for Scientific Research (F.N.R.S.). His thesis won the annual IBM Belgium/F.N.R.S. Computer Science Prize 2005. Between 2004 and 2006, he was a visiting scientist at Intel Research Cambridge, UK, and at the Applied Mathematics Department of University of Adelaide, Australia. Between 2006 and 2008, he was with Delft University of Technology, the Netherlands. Prior to joining Queen Mary, he was a Senior Research Scientist with Technische Universit\"{a}t Berlin/Deutsche Telekom Laboratories, Berlin, Germany. Starting in January 2012, he is the Professor of Networks and Head of the Networks Research group at Queen Mary, University of London.
\end{IEEEbiographynophoto}




\end{document}

%% file: text/0_introduction.tex
\section{Introduction}



The distributed control and transport network protocols running inside the 
routers and switches are the key technologies that allow information, 
in the form of digital packets, to travel around the world. Despite 
their widespread adoption, traditional IP networks are \emph{complex and hard to manage}~\cite{benson2009}.
To express the desired high-level network policies, network operators
need to configure each individual network device separately using 
low-level and often vendor-specific commands. In addition to the 
configuration complexity, network environments have to endure the 
dynamics of faults and adapt to load changes. Automatic reconfiguration
and response mechanisms are virtually non-existent in current IP networks.
Enforcing the required policies in such a dynamic environment is therefore
highly challenging.

To make it even more complicated, current networks are also
\emph{vertically integrated}.  The control plane (that decides how to
handle network traffic) and the data plane (that forwards traffic
according to the decisions made by the control plane) are bundled inside
the networking devices, reducing flexibility and hindering innovation
and evolution of the networking infrastructure.  The transition from 
IPv4 to IPv6, started more than a decade ago and still largely incomplete, 
bears witness to this challenge, while in fact IPv6 represented \textit{merely} 
a protocol update. Due to the inertia of current IP networks, a new 
routing protocol can take 5 to 10 years to be fully designed, evaluated and deployed.
Likewise, a clean-slate approach to change the Internet architecture (e.g., 
replacing IP), is regarded as a \coloredtext{daunting task} -- simply not feasible in practice~\cite{raghavan2012},~\cite{ghodsi2011}. 
Ultimately, this situation has inflated the capital and operational expenses of 
running an IP network.

\begin{figure}[t!]
\centering
\includegraphics[width=0.95\columnwidth]{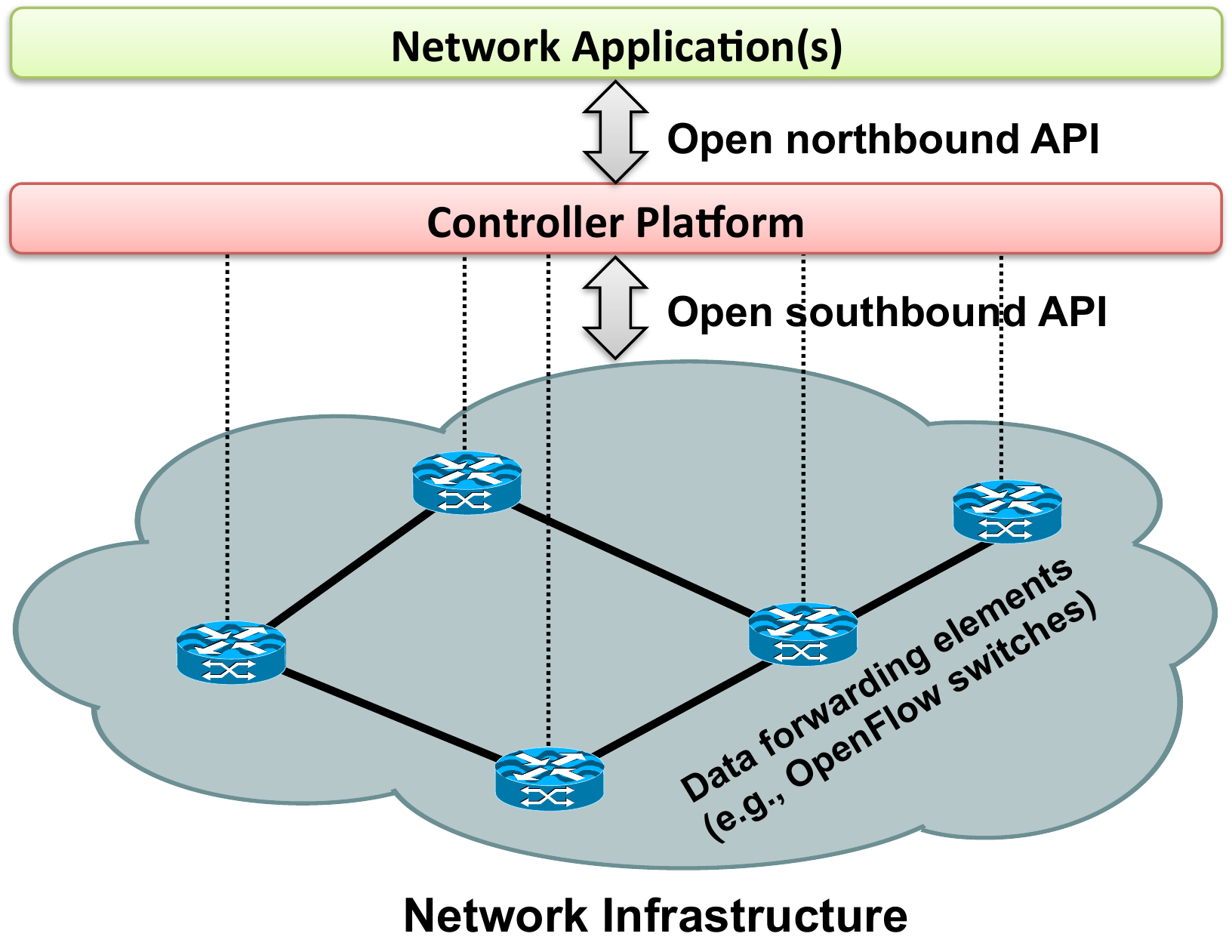}
\caption{Simplified view of an SDN architecture.}
\label{fig:sdn_simple}
\end{figure}

Software-Defined Networking (SDN)~\cite{mckeown2011,schenker2011} 
 is an emerging networking paradigm that gives hope to change the limitations of current
 network infrastructures. First, it breaks the vertical integration by separating the network's 
control logic (the control plane) from the underlying routers and switches that 
forward the traffic (the data plane). Second, with the separation of the control
and data planes, network switches become simple forwarding devices and
the control logic is implemented in a \textit{logically centralized} controller (or network operating system\footnote{We will use these two
terms interchangeably.}), simplifying policy enforcement and network 
(re)configuration and evolution~\cite{kim2013}. A simplified view of this 
architecture is shown in Figure~\ref{fig:sdn_simple}.  It is important
to emphasize that a logically centralized programmatic model does not
postulate a physically centralized system~\cite{koponen-1}.  In fact, the need to
guarantee adequate levels of performance, scalability, and reliability
would preclude such a solution.  Instead,
production-level SDN network designs resort to physically distributed
control planes~\cite{koponen-1,jain2013-1}.

The separation of the control plane and the data plane can be realized
by means of a well-defined programming interface between the switches
and the SDN controller.  The controller exercises direct control over
the state in the data-plane elements via this well-defined application programming interface (API), as
depicted in Figure~\ref{fig:sdn_simple}.  The most notable example of
such an API is OpenFlow~\cite{mckeown2008, onf2013-3}.  An OpenFlow
switch has one or more tables of packet-handling rules (flow table).  
Each rule matches a subset of the traffic and performs certain actions (dropping,
forwarding, modifying, etc.) on the traffic.
Depending on the rules installed by a controller application, an
OpenFlow switch can -- instructed by the controller -- behave like a
router, switch, firewall, or perform other roles (e.g., load balancer, traffic shaper, and in general those of a
middlebox).

An important consequence of the software-defined networking principles
is the \textit{separation of concerns} introduced between the
\textit{definition} of network policies, their \textit{implementation}
in switching hardware, and the \textit{forwarding} of traffic. This separation is key to the
desired flexibility, breaking the network control problem into
tractable pieces, and making it easier to create and introduce new
abstractions in networking, simplifying network management and
facilitating network evolution and innovation.

Although SDN and OpenFlow started as academic
experiments~\cite{mckeown2008}, they gained significant traction in the
industry over the past few years.  Most vendors of commercial switches
now include support of the OpenFlow API in their equipment. The SDN momentum 
was strong enough to make Google, Facebook, Yahoo, Microsoft, Verizon, and Deutsche Telekom fund Open Networking Foundation~(ONF)~\cite{onf2013-3} with the main
goal of promotion and adoption of SDN through open standards development. 
  As the initial concerns with SDN scalability were addressed~\cite{yeganeh2013}
-- in particular the myth that logical centralization implied a
physically centralized controller, an issue we will return to later on
-- SDN ideas have matured and evolved from an academic exercise to
a commercial success. Google, for example, has deployed a software-defined
network to interconnect its data centers across the globe. This production 
network has been in deployment for 3 years, helping the company to improve 
operational efficiency and significantly reduce costs~\cite{jain2013-1}.
VMware's network virtualization platform, NSX~\cite{vmware2013},
is another example.  NSX is a commercial solution that delivers a
fully functional network in software, provisioned independent of the
underlying networking devices, entirely based around SDN principles.  As
a final example, the world's largest IT companies (from carriers and
equipment manufacturers to cloud providers and financial-services
companies) have recently joined SDN consortia such as the ONF  and the OpenDaylight initiative~\cite{opendaylight2013}, 
another indication of the importance of SDN from an industrial perspective.

\coloredtext{A few recent papers have surveyed specific architectural aspects of SDN~\cite{lara2014,nunes2014,jarraya2014}.
An overview of OpenFlow and a short literature review can be found in~\cite{lara2014} and~\cite{nunes2014}.
These OpenFlow-oriented surveys present a relatively simplified three-layer stack composed of high-level network services, controllers, and the controller/switch interface.
In~\cite{jarraya2014}, the authors go a step further by proposing a taxonomy for SDN.
However, similarly to the previous works, the survey is limited in terms of scope and it does not provide an in-depth treatment of fundamental aspects of SDN. 
In essence, existing surveys lack a thorough discussion of the essential building blocks of an SDN such as the network operating systems, programming languages, and interfaces.
They also fall short on the analysis of cross-layer issues such as scalability, security, and dependability.
A more complete overview of ongoing research efforts, challenges, and related standardization activities  is also missing.}

\begin{figure*}[t]
\centering
\includegraphics[width=0.95\textwidth]{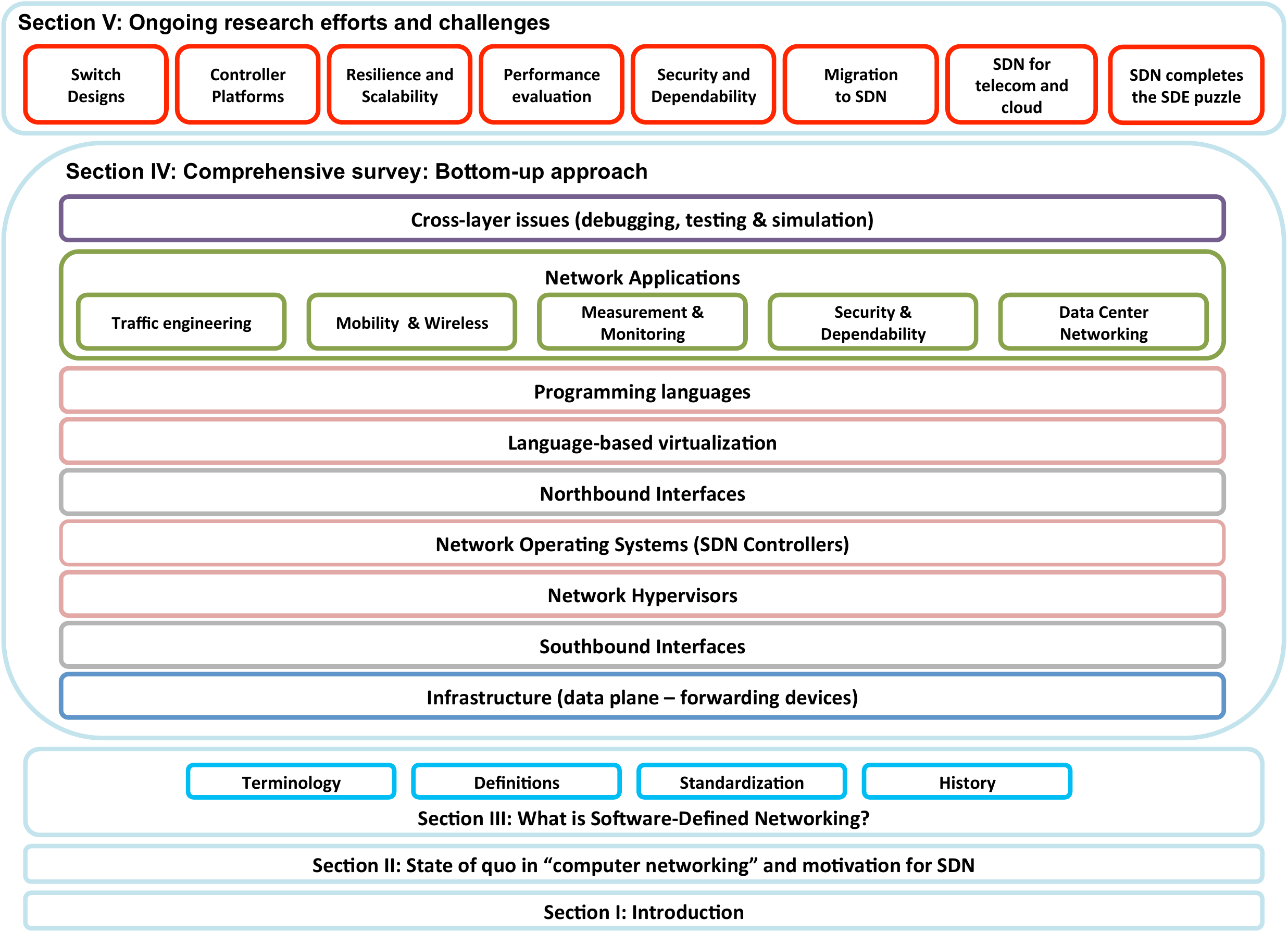}
\caption{Condensed overview of this survey on SDN.}
\label{fig:conclusion}
\end{figure*}

\coloredtext{In this paper, we present, to the best of our knowledge, the most comprehensive literature survey on SDN to date.
We organize this survey} as depicted in Figure~\ref{fig:conclusion}.  We start, in the next two sections, by explaining the context,
introducing the motivation for SDN and explaining the main concepts
of this new paradigm and how it differs from traditional networking.
Our aim in the early part of the survey is also to explain that SDN 
is not as novel as a technological advance.  Indeed, its existence 
is rooted at the intersection of a series of ``old'' ideas, 
technology drivers, and current and future needs. The concepts underlying SDN -- the 
separation of the control and data planes, the  flow abstraction upon
which forwarding decisions are made, the (logical) centralization of
network control, and the ability 
to program the network -- are not novel by themselves~\cite{feamster2013-2}.
However, the integration of already tested concepts with recent trends 
in networking -- namely the availability of merchant switch silicon and
the huge interest in feasible forms of network virtualization -- are
leading to this paradigm shift in networking. 
\coloredtext{As a result of the high industry interest and the potential to 
change the status quo of networking from multiple perspectives, a number of standardization efforts around SDN are ongoing, as we also discuss in Section~\ref{fig:sec-sdn-what}.}

\coloredtext{Section~\ref{sec:layeredapproach} is the core of this}
survey, presenting an extensive and comprehensive
analysis of the building blocks of an SDN infrastructure using a
bottom-up, layered approach.  The option for a layered approach is
grounded on the fact that SDN allows thinking of networking along two
fundamental concepts, which are common in other disciplines of 
computer science: a) separation of concerns (leveraging the concept of abstraction) and b) recursion.
Our layered, bottom-up approach divides the networking problem into eight parts: 1) hardware infrastructure, 2) southbound interfaces, 3) network virtualization (hypervisor layer between 
the forwarding devices and the network operating systems), 4) network operating 
systems (SDN controllers and control platforms), 5) northbound interfaces 
(to offer a common programming abstraction to the upper layers, mainly the network applications), 6) virtualization using slicing techniques provided by special purpose 
libraries or programming languages and compilers, 7) network 
programming languages, and finally 8) network applications. In addition, 
we also look at cross-layer problems such as debugging and troubleshooting 
mechanisms. The discussion in Section~\ref{sec:challenges} on ongoing research efforts, challenges, future work and opportunities concludes this paper.

%% file: text/1_traditional_nets.tex
\section{State of Quo in Networking}
\label{traditional_nets}

\begin{figure}[t!]
\centering
\includegraphics[width=0.75\columnwidth]{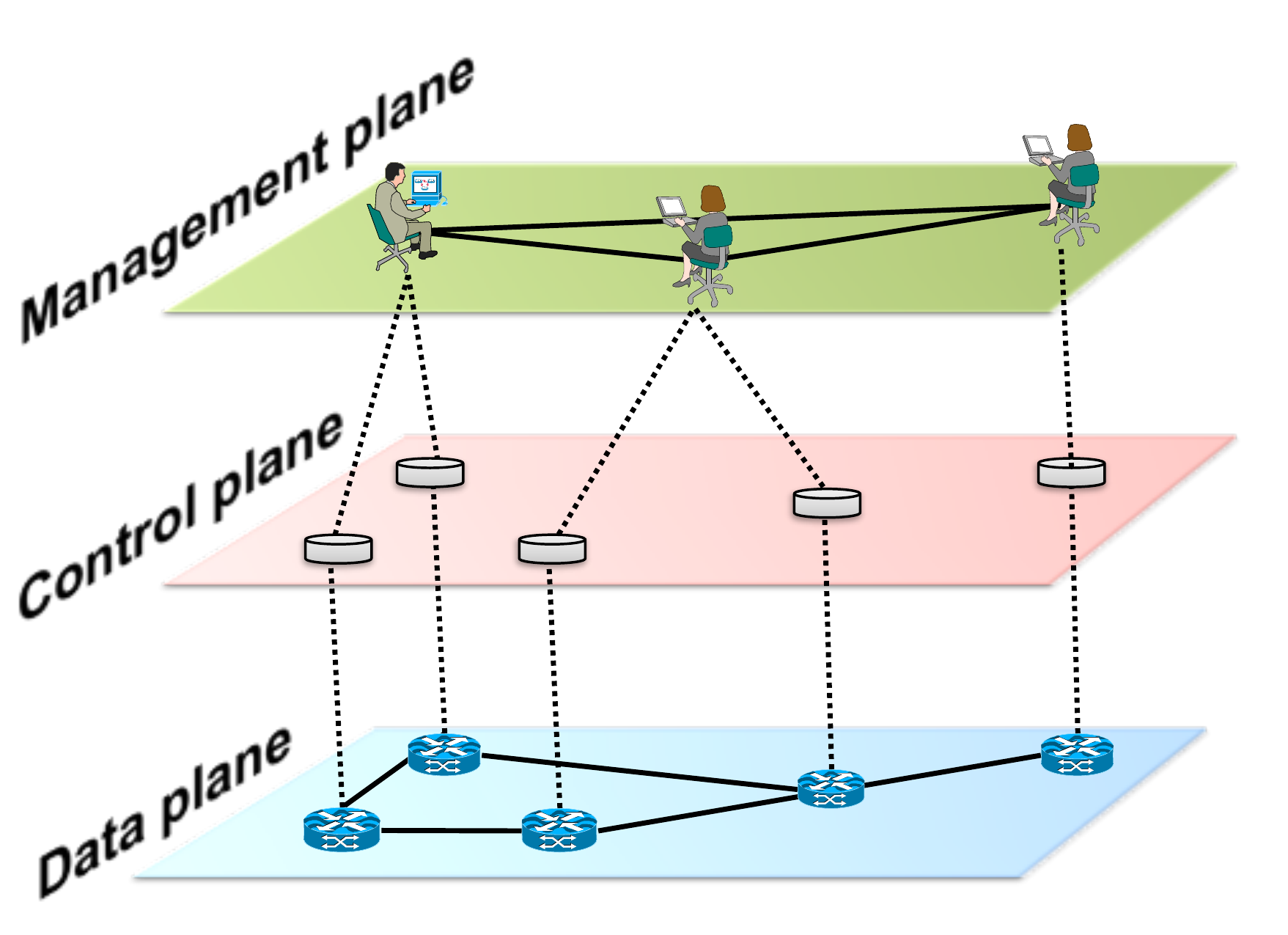}
\caption{Layered view of networking functionality.}
\label{fig:currentnetplanes}
\end{figure}

Computer networks can be divided in three planes of functionality: the
data, control and management planes (see Figure~\ref{fig:currentnetplanes}).  
The data plane corresponds to the networking devices, which are responsible 
for (efficiently) forwarding data.  The control plane represents the protocols 
used to populate the forwarding tables of the data plane elements. The management 
plane includes the software services, such as SNMP-based tools~\cite{presuhn2002}, 
used to remotely monitor and configure the control functionality.  Network policy 
is defined in the management plane, the control plane enforces the policy, and the 
data plane executes it by forwarding data accordingly.

In traditional IP networks, the control and data planes are tightly coupled, embedded in the same networking devices, and the whole structure is highly decentralized.
This was considered important for the design of the Internet in the early days: it seemed 
the best way to guarantee network resilience, which was a crucial design goal. In fact, 
this approach has been quite effective in terms of network performance, with a rapid 
increase of line rate and port densities.

\coloredtext{However, the outcome is a very complex and relatively static architecture, as has been often reported in the networking literature (e.g.,~\cite{benson2009,ghodsi2011,raghavan2012,kim2013,pan2011a})}. %
It is also the 
fundamental reason why traditional networks are rigid, and complex to manage and control.
These two characteristics are largely responsible for a vertically-integrated industry where 
innovation is difficult.

Network misconfigurations and related errors are extremely common in today's networks. 
For instance, more than 1000 configuration errors have been observed in BGP 
routers~\cite{feamster2005}. 
\coloredtext{From a single misconfigured device may result very undesired network behavior (including, among others, packet losses, forwarding loops, setting up of unintended paths, or service contract violations).} 
Indeed, while rare, a single misconfigured router is able to compromise the correct operation of the 
whole Internet for hours~\cite{barrett1997,butler2010}. 

To support network management, a small number of vendors offer proprietary solutions of 
specialized hardware, operating systems, and control programs (network applications). 
Network operators have to acquire and maintain different management solutions and the 
corresponding specialized teams.  The capital and operational cost of building and 
maintaining a networking infrastructure is significant, with long return on investment 
cycles, which hamper innovation and addition of new features and services (for instance 
access control, load balancing, energy efficiency, traffic engineering).  
To alleviate the lack of in-path functionalities within the network, a myriad of specialized 
components and middleboxes, such as firewalls, intrusion detection systems and deep packet inspection engines, proliferate 
in current networks. A recent survey of 57 enterprise networks shows that the number of 
middleboxes is already on par with the number of routers in current networks~\cite{sherry2012}.
Despite helping in-path functionalities, the net effect of middleboxes has been increased
complexity of network design and its operation.

%% file: text/2_what_is_sdn.tex
\section{What is Software-Defined Networking?}
\label{fig:sec-sdn-what}

The term SDN (Software-Defined Networking)  was originally coined to represent 
the ideas and work around OpenFlow at Stanford University~\cite{greene2009}. 
As originally defined, SDN refers to a network architecture where the forwarding 
state in the data plane is managed by a remote control plane decoupled from the former.
The networking industry has on many occasions shifted from this original view of SDN, 
by referring to anything that involves software as being SDN. 
We therefore attempt, in this section, to provide a much less ambiguous definition of software-defined 
networking.

We define an SDN as a network archi\-te\-cture with four pillars:\\
\begin{enumerate}
\item  The control and data planes are \textit{decoupled}. Control functionality is 
removed from network devices that will become simple (packet) forwarding elements.
\item  Forwarding decisions are flow-based, instead of des\-ti\-na\-tion-based. A flow is 
broadly defined by a set of packet field values acting as a match (filter) criterion 
and a set of actions (instructions).
\coloredtext{In the SDN/OpenFlow context, a flow is a sequence of packets between a source and a destination.
All packets of a flow receive identical service policies at the forwarding devices~\cite{newman1998IP,gude2008}.}
The flow abstraction allows unifying the behavior of 
different types of network devices, including routers, switches, firewalls, and \coloredtext{middleboxes~\cite{Jamjoom2014_4}}. 
Flow programming enables unprecedented flexibility, limited only to the capabilities of 
the implemented flow tables~\cite{mckeown2008}. 
\item Control logic is moved to an external entity, the so-called SDN controller or Network 
Operating System (NOS). The NOS is a software platform that runs on commodity server technology 
and provides the essential resources and abstractions to facilitate the programming of forwarding 
devices based on a logically centralized, abstract network view. Its purpose is therefore similar 
to that of a traditional operating system.
\item The network is \textit{programmable} through software applications running on top of the NOS that 
interacts with the underlying data plane devices. This is a fundamental characteristic of SDN, considered 
as its main value proposition.
\end{enumerate}

Note that the logical centralization of the control logic, in particular, offers several 
additional benefits.  First, it is simpler and less error-prone to modify network policies 
through high-level languages and software components, compared with low-level device specific
configurations. Second, a control program can automatically react to spurious changes of the 
network state and thus maintain the high-level policies intact. Third, the centralization of 
the control logic in a controller with global knowledge of the network state simplifies the
development of more sophisticated networking functions, services and applications.

Following the SDN concept introduced in~\cite{schenker2011}, an SDN can be defined by 
three fundamental abstractions: (\textit{i}) forwarding, (\textit{ii}) distribution, and 
(\textit{iii}) specification.
In fact, abstractions are essential tools of research in computer science and information technology, being already an ubiquitous feature of many computer architectures and systems~\cite{alkhatib2014}.

Ideally, the \textit{forwarding abstraction} should allow any forwarding behavior desired 
by the network application (the control program) while hiding details of the underlying hardware.
\coloredtext{OpenFlow is one realization of such abstraction}, which can be seen as the equivalent 
to a ``device driver'' in an operating system.

The \textit{distribution abstraction} should shield SDN applications from the vagaries 
of distributed state, making the distributed control problem a logically centralized one.  Its
realization requires a common distribution layer, which in SDN resides in the NOS. This layer has 
two essential functions. First, it is responsible for installing the control commands on the 
forwarding devices.  Second, it collects status information about the forwarding layer (network 
devices and links), to offer a global network view to network applications.

\begin{figure}[t!]
\centering
\includegraphics[width=0.95\columnwidth]{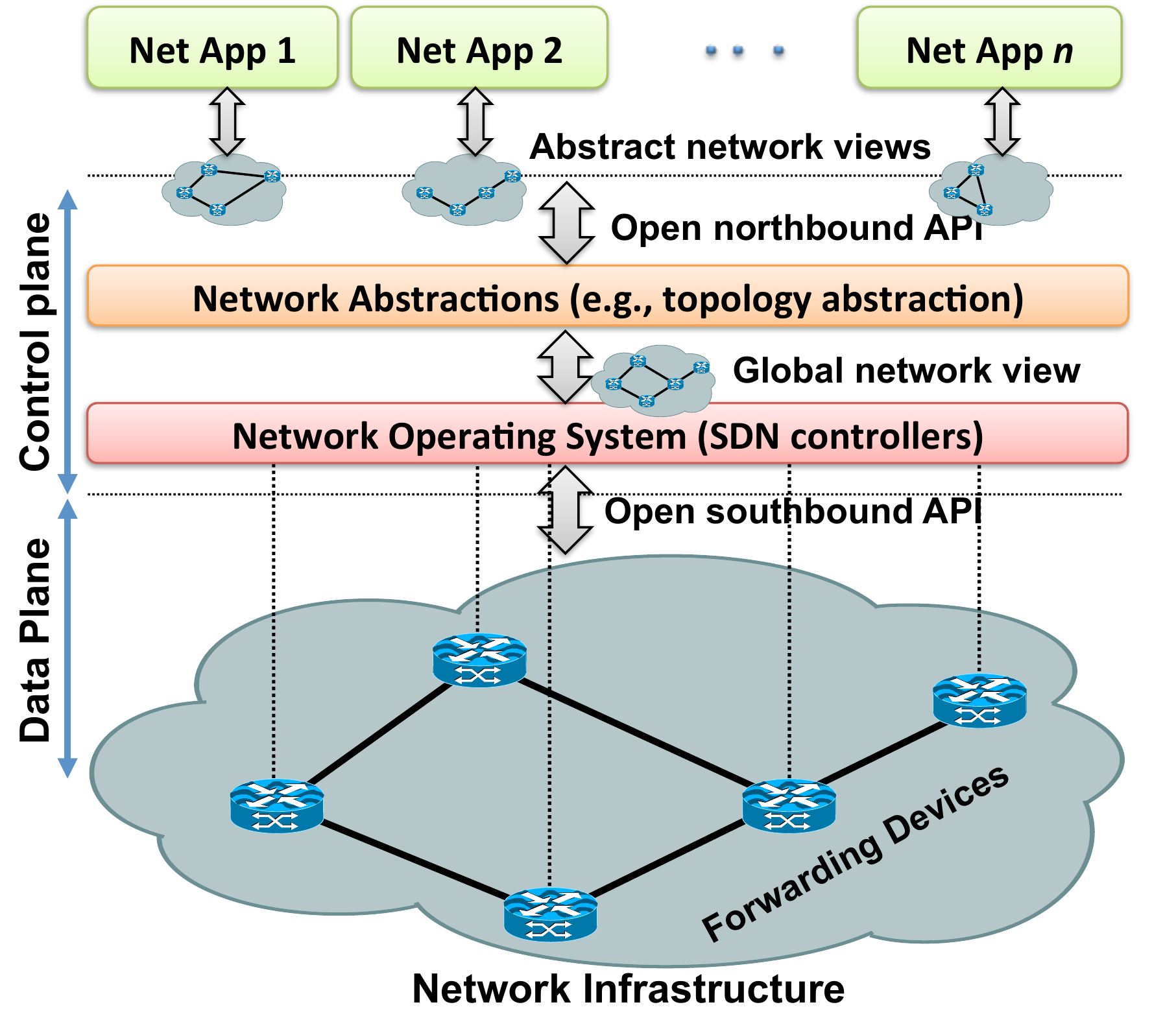}
\caption{SDN architecture and its fundamental abstractions.}
\label{fig:sdn_abstractions}
\end{figure}

The last abstraction is \textit{specification}, which should allow a network application to express 
the desired network behavior without being responsible for implementing that behavior itself.
This can be achieved through virtualization solutions, as well as network programming languages.
These approaches map the abstract configurations that the applications express based on a simplified, 
abstract model of the network, into a physical configuration for the global network view exposed by 
the SDN controller. Figure~\ref{fig:sdn_abstractions} depicts the SDN architecture, concepts and 
building blocks.

\begin{figure}[t!]
\centering
\includegraphics[width=0.95\columnwidth]{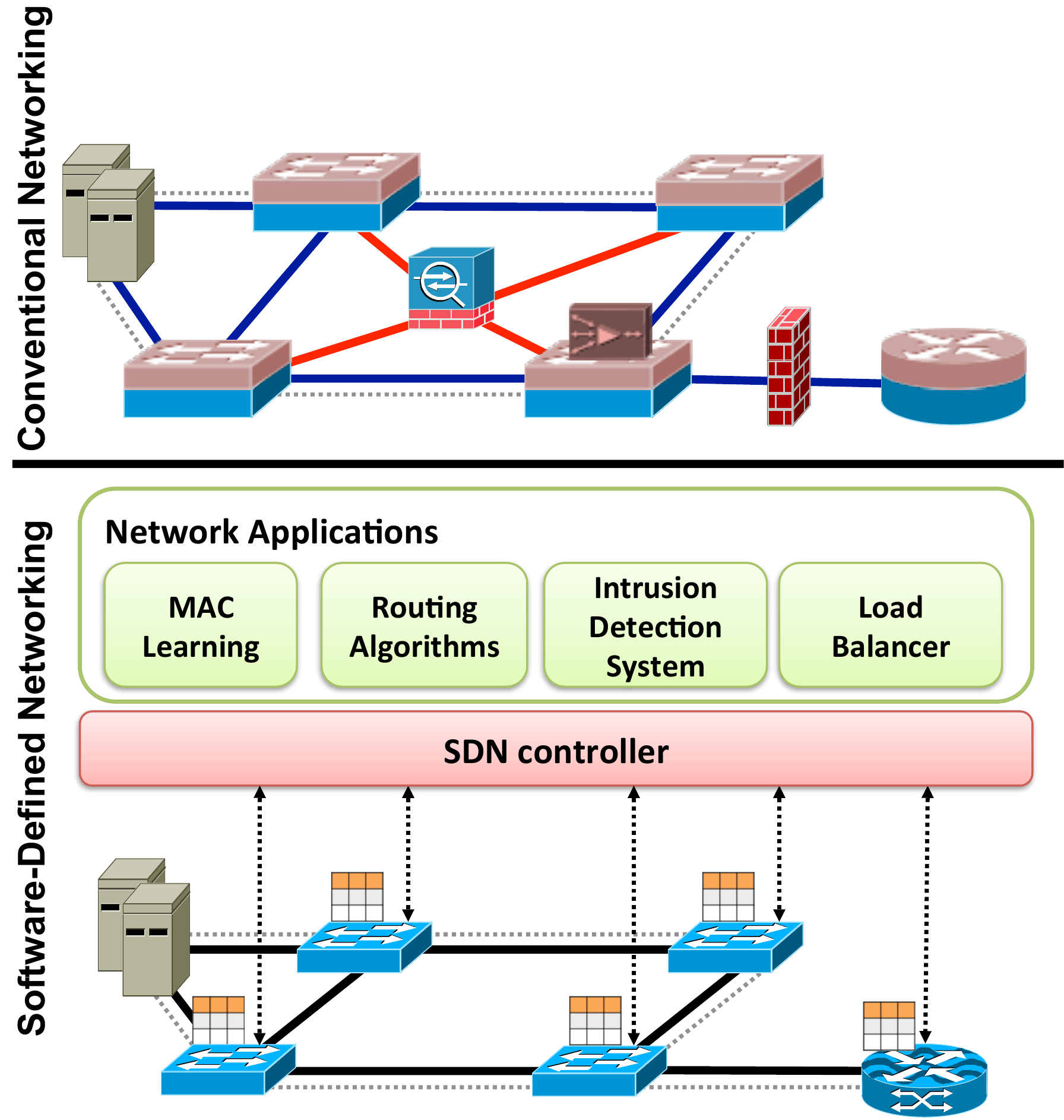}
\caption{Traditional networking versus Software-Defined Networking (SDN). With SDN, management becomes simpler and middleboxes services can be delivered as SDN controller applications.}
\label{fig:traditionalversusSDN}
\end{figure}

As previously mentioned, the strong coupling between control and data planes has made it difficult 
to add new functionality to traditional networks, \coloredtext{
a fact illustrated in Figure~\ref{fig:traditionalversusSDN}.
The coupling of the control and data planes (and its physical embedding in the network elements) makes the development and deployment of new networking features (e.g., routing algorithms) very hard since 
it would imply a modification of the control plane of all network devices -- through the installation of new firmware and, in some cases, hardware upgrades.
Hence, the new networking features are commonly introduced via expensive, specialized and hard-to-configure equipment (aka middleboxes) such as load balancers, intrusion detection systems (IDS), and firewalls, among others.  
 These middleboxes need to be placed strategically in the network, making it 
even harder to later change the network topology, configuration, and functionality.}

\coloredtext{In contrast, SDN decouples the control plane from the network devices and becomes an external entity: the network operating system or SDN controller.}
This approach has several advantages:
\begin{itemize}
\item It becomes easier to program these applications since the abstractions provided by the 
control platform and/or the network programming languages can be shared.
\item All applications can take advantage of the same network information (the global network view), 
leading (arguably) to more consistent and effective policy decisions while re-using control plane 
software modules.
\item These applications can take actions (i.e., reconfigure forwarding devices) from any part of 
the network. There is therefore no need to devise a precise strategy about the location of the new 
functionality.
\item The integration of different applications becomes more straightforward~\cite{Casado2014_4}. For instance, load 
balancing and routing applications can be combined sequentially, with load balancing decisions having 
precedence over routing policies.
\end{itemize}

\subsection{Terminology}

To identify the different elements of an SDN as unequivocally as possible, we now present the 
essential terminology used throughout this work.

\noindent \textit{Forwarding Devices (FD)}: Hardware- or software-based data plane devices that perform 
a set of elementary operations. The forwarding devices have well-defined instruction sets (e.g., flow 
rules) used to take actions on the incoming packets (e.g., forward to specific ports, drop, forward to 
the controller, rewrite some header). These instructions are defined by southbound interfaces (e.g., 
OpenFlow~\cite{mckeown2008}, ForCES~\cite{doria2010}, Protocol- Oblivious Forwarding (POF)~\cite{song2013}) and are installed in the forwarding devices by the SDN controllers implementing the southbound protocols.

\noindent \textit{Data Plane (DP)}: Forwarding devices are interconnected through wireless radio channels 
or wired cables. The network infrastructure comprises the interconnected forwarding devices, which 
represent the data plane.

\noindent \textit{Southbound Interface (SI)}: The instruction set of the forwarding devices is defined 
by the southbound API, which is part of the southbound interface. Furthermore, the SI also defines the 
communication protocol between forwarding devices and control plane elements. This protocol formalizes 
the way the control and data plane elements interact.

\noindent \textit{Control Plane (CP)}: Forwarding devices are programmed by control plane elements 
through well-defined SI embodiments. The control plane can therefore be seen as the ``network brain''.
All control logic rests in the applications and controllers, which form the control plane.

\noindent \textit{Northbound Interface (NI)}: The network operating system can offer an API to application 
developers. This API represents a northbound interface, i.e., a common interface for developing applications.
Typically, a northbound interface abstracts the low level instruction sets used by southbound interfaces 
to program forwarding devices.

\noindent \textit{Management Plane (MP)}: The management plane is the set of applications that leverage 
the functions offered by the NI to implement network control and operation logic.
This includes applications such as routing, firewalls, load balancers, monitoring, and so forth.
Essentially, a management application defines the policies, which are ultimately translated to 
southbound-specific instructions that program the behavior of the forwarding devices.

\subsection{Alternative and Broadening Definitions}
\label{sec:alt-SDN}

Since its inception in 2010~\cite{greene2009}, the original OpenFlow-centered SDN term 
has seen its scope broadened beyond architectures with a cleanly decoupled control plane 
interface. The definition of SDN will likely continue to broaden, driven by the industry 
business-oriented views on SDN -- irrespective of the decoupling of the control plane. In 
this survey, we focus on the original, ``canonical'' SDN definition based on the aforementioned
key pillars and the concept of layered abstractions. However, for the sake of completeness 
and clarity, we acknowledge alternative SDN definitions~\cite{nadeau2013}, including:

\noindent \textit{Control Plane / Broker SDN}: A networking approach that retains existing distributed 
control planes but offers new APIs that allow applications to interact (bidirectionally) with the network. 
An SDN controller --often called orchestration platform-- acts as a broker between the applications and 
the network elements. This approach effectively presents control plane data to the application and allows 
a certain degree of network programmability by means of ``plug-ins'' between the orchestrator function and 
network protocols. This API-driven approach corresponds to a hybrid model of SDN, since it enables the broker 
to manipulate and directly interact with the control planes of devices such as routers and switches. Examples 
of this view on SDN include recent standardization efforts at IETF \coloredtext{(see Section~\ref{sec:standardization})} and the design philosophy behind the OpenDaylight project~\cite{opendaylight2013} that goes beyond the OpenFlow split control mode.

\noindent \textit{Overlay SDN}: A networking approach where the (software- or hardware-based) network edge 
is dynamically programmed to manage tunnels between hypervisors and/or network switches, introducing an 
overlay network. In this hybrid networking approach, the distributed control plane providing the underlay 
remains untouched. The centralized control plane provides a logical overlay that utilizes the underlay 
as a transport network. This flavor of SDN follows a proactive model to install the overlay tunnels. The overlay
tunnels usually terminate inside virtual switches within hypervisors or in physical devices acting as gateways 
to the existing network. This approach is very popular in recent data center network virtualization \cite{chowdhury2010}, and are based on a variety of tunneling technologies \coloredtext{(e.g., STT~\cite{davie2014stt}, VXLAN~\cite{mahalingam2013}, NVGRE~\cite{sridharan2013}, LISP~\cite{maino2013lisp,hertoghs2014lisp}, GENEVE~\cite{Gross2014_4})}~\cite{jain2013}.

\coloredtext{Recently, other attempts to define SDN in a layered approach have appeared~\cite{haleplidis2014layers,jarraya2014}.
From a practical perspective and trying to keep backward compatibility with existing network management approaches, one  initiative at IRTF SDNRG~\cite{haleplidis2014layers} proposes a management plane at the same level of the control plane, i.e., it classifies solutions in two categories: control logic (with control plane southbound interfaces) and management logic (with management plane southbound interfaces).
In other words, the management plane can be seen as a control platform that accommodates traditional network management services and protocols, such as SNMP~\cite{presuhn2002}, BGP~\cite{rekhter2006}, PCEP~\cite{vasseur2009}, and NETCONF~\cite{enns2011-1}.
}

In addition the broadening definitions above, the term SDN is often used to define extensible network management planes (e.g., OpenStack~\cite{corradi2014}), \coloredtext{
whitebox / bare-metal switches with open operating systems (e.g., Cumulus Linux),} open-source dataplanes (e.g., Pica8 Xorplus~\cite{shang2014}, Quagga~\cite{jakma2014}), specialized programmable hardware devices (e.g., 
NetFPGA~\cite{netfpga2014}), virtualized software-based appliances \coloredtext{(e.g., Open Platform for Network Functions Virtualization - OPNFV~\cite{opnfv})}, in spite of 
lacking a decoupled control and data plane or common interface along its API. Hybrid SDN models are further discussed in Section~\ref{sec:hybrid}.

\coloredtext{

\subsection{Standardization Activities}
\label{sec:standardization}

The standardization landscape in SDN (and SDN-related issues) is already wide and is expected to keep evolving over time. 
While some of the activities are being carried out in Standard Development Organizations (SDOs), other related efforts are ongoing at industrial or community consortia (e.g., OpenDaylight, OpenStack, OPNFV), delivering results often considered candidates for \textit{de facto} standards.
These results often come in the form of open source implementations that have become the common strategy towards accelerating SDN and related cloud and networking technologies~\cite{floss-meets-sdn}.
The reason for this fragmentation is due to SDN concepts spanning different areas of IT and networking, both from a network segmentation point of view (from access to core) and from a technology perspective (from optical to wireless).

Table~\ref{tab:standardization} presents a summary of the main SDOs and organizations contributing to the standardization of SDN, as well as the main outcomes produced to date.
 
The Open Networking Foundation (ONF) was conceived as a member-driven organization to promote the adoption of SDN through the development of the OpenFlow protocol as an open standard to communicate control decisions to data plane devices. 
The ONF is structured in several working groups (WGs). Some WGs are focused on either defining extensions to the OpenFlow protocol in general, such as the Extensibility WG, or tailored to specific technological areas. 
Examples of the latter include the Optical Transport (OT) WG, the Wireless and Mobile (W\&M) WG, and the Northbound Interfaces (NBI) WG. Other WGs center their activity in providing new protocol capabilities to enhance the protocol itself, such as the Architecture WG or the Forwarding Abstractions (FA) WG.

Similar to how network programmability ideas have been considered by several Working Groups (WGs) of the Internet Engineering Task Force (IETF) in the past, the present SDN trend is also influencing a number of activities. 
A related body that focuses on research aspects for the evolution of the Internet, the Internet Research Task Force (IRTF), has created the Software Defined Networking Research Group (SDNRG).
This group investigates SDN from various perspectives with the goal of identifying the approaches that can be defined, deployed and used in the near term, as well as identifying future research challenges.

In the International Telecommunications Union's Telecommunication sector (ITU-T), some Study Groups (SGs) have already started to develop recommendations for SDN, 
and a Joint Coordination Activity on SDN (JCA-SDN) has been established to coordinate the SDN standardization work.

The Broadband Forum (BBF) is working on SDN topics through the Service Innovation \& Market Requirements (SIMR) WG. The objective of the BBF is to release recommendations for supporting SDN in multi-service broadband networks, including hybrid environments where only some of the network equipment is SDN-enabled.

The Metro Ethernet Forum (MEF) is approaching SDN with the aim of defining service orchestration with APIs for existing networks. 

At the Institute of Electrical and Electronics Engineers (IEEE), the 802 LAN/MAN Standards Committee has recently started some activities to standardize SDN capabilities on access networks based on IEEE 802 infrastructure through the P802.1CF project, for both wired and wireless technologies to embrace new control interfaces.

The Optical Internetworking Forum (OIF) Carrier WG released a set of requirements for Transport Software-Defined Networking. The initial activities have as main goal to describe the features and functionalities needed to support the deployment of SDN capabilities in carrier transport networks.

The Open Data Center Alliance (ODCA) is an organization working on unifying data center in the migration to cloud computing environments through interoperable solutions. Through the documentation of usage models, specifically one for SDN, the ODCA is defining new requirements for cloud deployment. 

The Alliance for Telecommunication Industry Solutions (ATIS) created a Focus Group for analyzing operational issues and opportunities associated with the programmable capabilities of network infrastructure. 

At the European Telecommunication Standards Institute (ETSI), efforts are being devoted to Network Function Virtualization (NFV) through a newly defined Industry Specification Group (ISG). 
NFV and SDN concepts are considered complementary, sharing the goal of accelerating innovation inside the network by allowing programmability, and altogether changing the network operational model through automation and a real shift to software-based platforms.

Finally, the mobile networking industry 3GPP consortium is studying the management of virtualized networks, an effort aligned with the ETSI NFV architecture and, as such, likely to leverage from SDN.

{\renewcommand{\arraystretch}{1.4}
\begin{table*}[!htp]
\caption{\coloredtext{OpenFlow standardization activities}}
\label{tab:standardization}
\begin{center}
\footnotesize
\begin{tabularx}{\linewidth}{p{0.8cm}p{3.8cm}Xp{4.6cm}}
\hline
\textbf{SDO} & \textbf{Working Group} & \textbf{Focus}  & \textbf{Outcomes} \\
\hline
\multirow{16}{*}{ONF} 
& Architecture \& Framework & SDN architecture, defining architectural components and interfaces & SDN Architecture~\cite{SDNARCH}  \\\cline{2-4}

& Northbound Interfaces & Definition of standard NBIs for SDN controllers	& \\\cline{2-4}

& Testing and Interoperability	& Specification of OpenFlow conformance test suites	& Conformance tests~\cite{CTSOSS} \\\cline{2-4}

& Extensibility	& Development of extensions to OpenFlow protocol, producing specifications of the OpenFlow switch (OF-WIRE) protocol &	OF-WIRE 1.4.0~\cite{OSS} \\\cline{2-4}

& Configuration \& Management	& OAM (operation, administration, and management) capabilities for OF protocol, producing specifications of the OF Configuration and Management (OF-CONFIG) protocol & OF-CONFIG 1.2~\cite{OFCONFIG} \par OpenFlow Notifications Framework~\cite{onf2013-2} \\\cline{2-4}

& Forwarding Abstractions & Development of hardware abstractions and simplification of behavioral descriptions mapping & OpenFlow Table Type Patterns~\cite{OTTP} \\\cline{2-4}

& Optical Transport	& Specification of SDN and control capabilities for optical transport networks by means of OpenFlow & Use cases~\cite{OTUC} \par Requirements~\cite{RATOSDN} \\\cline{2-4}

& Wireless \& Mobile & Specification of SDN and control capabilities for wireless and mobile networks by means of OpenFlow & \\\cline{2-4}

& Migration	& Methods to migrate from conventional networks to SDN-based networks based on OpenFlow & Use cases~\cite{MUCM} \\\cline{2-4}

& Market Education	& Dissemination of ONF initiatives in SDN and OpenFlow by releasing White Papers and Solution Briefs	& SDN White Paper~\cite{SDN_NNN} \\
\hline
\multirow{15}{*}{IETF} 
& Application-Layer Traffic Optimization (ALTO) &  Provides applications with network state information  &    Architectures for the coexistence of SDN and ALTO~\cite{xie2012} \\\cline{2-4}

& Forwarding and Control Element Separation (ForCES) & Protocol specifications for the communication between control and forwarding elements. & Protocol specification~\cite{doria2010} \\\cline{2-4}

& Interface to the Routing System (I2RS) & Real-time or event driven interaction with the routing system in an IP routed network	& Architecture~\cite{AIRS} \\\cline{2-4}

& Network Configuration (NETCONF) & Protocol specification for transferring configuration data to and from a device & NETCONF protocol~\cite{enns2004} \\\cline{2-4}

& Network Virtualization Overlays (NVO3) & Overlay networks for supporting multi-tenancy in the context of data center communications (i.e., VM communication) & Control plane requirements~\cite{NVENVA} \\\cline{2-4}

& Path Computation Element (PCE) & Path computation for traffic engineering and path selection based on constrains & ABNO framework~\cite{PCE} \par Cross stratum path computation~\cite{CSOPC} \\\cline{2-4}

& Source Packet Routing in Networking (SPRING) & Specification of a forwarding path at the source of traffic	 & OpenFlow interworking~\cite{khasnabish2014} \par SDN controlled use cases~\cite{kim2014} \\\cline{2-4}

& Abstraction and Control of Transport Networks (ACTN) BoF & Facilitate a centralized virtual network operation	 &   Virtual network controller framework~\cite{ceccarelli2014} \\
\hline
IRTF & Software-Defined Networking Research Group (SDNRG) & Prospection of SDN for the evolution of Internet & SDN operator perspective~\cite{boucadair2014} \par SDN Architecture~\cite{haleplidis2014} \par
 Service / Transport separation~\cite{contreras2014} \\
 \hline
\multirow{7}{*}{ITU-T} 
& SG 11 & Signalling requirements using SDN technologies in Broadband Access Networks & Q.Supplement-SDN~\cite{itutqsup2014} \par Q.SBAN~\cite{itutqsban2014} \\\cline{2-4}

& SG 13 & Functional requirements and architecture for SDN and networks of the future & Recommendation Y.3300~\cite{ituty33002014} \\\cline{2-4}

& SG 15 & Specification of a transport network control plane architecture to support SDN control of transport networks & \\\cline{2-4}

& SG 17 & Architectural aspects of security in SDN and security services using SDN & \\
\hline
BBF & Service Innovation and Market Requirements & Requirements and impacts of deploying SDN in broadband networks & SD-313~\cite{bforum2014}\\
\hline
MEF & The Third Network & Service orchestration in Network as a Service and NFV environments & \\
\hline
IEEE & 802 & Applicability of SDN to IEEE 802 infrastructure & \\
\hline
OIF & Carrier WG & Transport SDN networks & Requirements for SDN enabled transport networks~\cite{oif2013}\\
\hline
ODCA & SDN/Infrastructure  & Requirements for SDN in cloud environments & Usage model~\cite{odc2014}\\
\hline
ETSI & NFV ISG & Orchestration of network functions, including the combined control of computing, storage and networking resources & NFV Architecture~\cite{etsi2013}\\
\hline
ATIS & SDN Focus Group & Operational aspects of SDN and NFV & Operation of SDN~\cite{atis2014}\\
\hline
\end{tabularx}
\end{center}
\end{table*}
}

} 

%% file: text/3_history_of_sdn.tex
\subsection{History of Software-Defined Networking}

Albeit a fairly recent concept, SDN leverages on networking ideas with a longer history~\cite{feamster2013-2}.
In particular, it builds on work made on programmable networks, such as active networks~\cite{tennenhouse1997}\coloredtext{, programmable ATM networks~\cite{lazar1996,lazar1997} }, and on proposals for control and data plane separation, such as NCP~\cite{sheinbein1982} and RCP~\cite{caesar2005}.

In order to present an historical perspective, we summarize in Table~\ref{tab:history} different instances of SDN-related work prior to SDN, splitting it into five categories.
Along with the categories we defined, the second and third columns of the table mention past initiatives (pre-SDN, i.e., before the OpenFlow-based initiatives that sprung into the SDN concept), and recent developments that led to the definition of SDN.

\newcommand{\fcwidth}{3.2cm}
\newcommand{\scwidth}{9.2cm}
\newcommand{\tcwidth}{4.6cm}
{\renewcommand{\arraystretch}{1.4}
\begin{table*}[!htp]
\caption{Summarized overview of the history of programable networks}
\label{tab:history}
\begin{center}
\footnotesize
\begin{tabularx}{\linewidth}{p{\fcwidth}p{\scwidth}p{\tcwidth}}
\hline
\textbf{Category} & \textbf{Pre-SDN initiatives} & \textbf{More recent SDN developments} \\
\hline
\multirow{2}{*}{\begin{minipage}{\fcwidth}
		Data plane programmability
	\end{minipage}} 
& \multirow{2}{*}{\begin{minipage}{\scwidth}
		\coloredtext{xbind~\cite{lazar1996},}
		IEEE P1520~\cite{biswas1998},
		smart packets~\cite{schwartz1999},
		ANTS~\cite{wetherall1998},
		SwitchWare~\cite{alexander1998},
		Calvert~\cite{calvert1998},
		high performance router~\cite{wolf2000},
		NetScript~\cite{silva2001},
		Tennenhouse~\cite{tennenhouse2007}
	\end{minipage}} 
& \multirow{2}{*}{
	\begin{minipage}{\tcwidth}
		ForCES~\cite{doria2010},
		OpenFlow~\cite{mckeown2008},
		POF~\cite{song2013}
	\end{minipage}} \\
& & \\
\hline
\multirow{2}{*}{\begin{minipage}{\fcwidth}
		Control and data plane \\decoupling
	\end{minipage}} 
& \multirow{2}{*}{\begin{minipage}{\scwidth}
		NCP~\cite{sheinbein1982},
		\coloredtext{GSMP~\cite{rfc1987,rfc3294}},
		Tempest~\cite{merwe1998},
		ForCES~\cite{doria2010},
		RCP~\cite{caesar2005},
		SoftRouter~\cite{lakshman2004},
		PCE~\cite{vasseur2009},
		4D~\cite{greenberg2005},
		IRSCP~\cite{merwe2006}
	\end{minipage}} 
& \multirow{2}{*}{
	\begin{minipage}{\tcwidth}
		SANE~\cite{casado2006},
		Ethane~\cite{casado2007-1},
		OpenFlow~\cite{mckeown2008},
		NOX~\cite{gude2008},
		POF~\cite{song2013}
	\end{minipage}} \\
& & \\
\hline
\multirow{2}{*}{\begin{minipage}{\fcwidth}
		Network virtualization
	\end{minipage}} 
& \multirow{2}{*}{\begin{minipage}{\scwidth}
		Tempest~\cite{merwe1998},
		MBone~\cite{macedonia1994},
		6Bone~\cite{fink2004},
		RON~\cite{andersen2001},
		Planet Lab~\cite{chun2003},
		Impasse~\cite{anderson2005},
		GENI~\cite{peterson2006},
		VINI~\cite{bavier2006-1}
	\end{minipage}} 
& \multirow{2}{*}{
	\begin{minipage}{\tcwidth}
		Open vSwitch~\cite{pfaff2009},
		Mininet~\cite{lantz2010},
		FlowVisor~\cite{sherwood2010},
		NVP~\cite{koponen}
	\end{minipage}} \\
& & \\
\hline
\multirow{1}{*}{\begin{minipage}{\fcwidth}
		Network operating systems
	\end{minipage}} 
& \multirow{1}{*}{\begin{minipage}{\scwidth}
		Cisco IOS~\cite{bollapragada2000},
		JUNOS~\cite{junipernetworks2012}.
		ExtremeXOS~\cite{extremenetworks2014},
		SR OS~\cite{alcatellucent2014}
	\end{minipage}} 
& \multirow{1}{*}{
	\begin{minipage}{\tcwidth}
		NOX~\cite{gude2008},
		Onix~\cite{koponen-1},
		ONOS~\cite{krishnaswamy2013}
	\end{minipage}} \\
\hline
\multirow{1}{*}{\begin{minipage}{\fcwidth}
		Technology pull initiatives
	\end{minipage}} 
& \multirow{1}{*}{\begin{minipage}{\scwidth}
	Open Signaling~\cite{campbell1999}
	\end{minipage}} 
& \multirow{1}{*}{
	\begin{minipage}{\tcwidth}
	ONF~\cite{onf2013-3}
	\end{minipage}} \\
\hline
\end{tabularx}
\end{center}
\end{table*}
}

Data plane programmability has a long history. Active networks~\cite{tennenhouse1997} 
represent one of the early attempts on building new network architectures based on this concept.
The main idea behind active networks is for each node to have the capability to perform computations 
on, or modify the content of, packets. To this end, active networks propose two distinct approaches: 
programmable switches and capsules. The former does not imply changes in the existing packet or cell 
format. It assumes that switching devices support the downloading of programs with specific instructions 
on how to process packets. The second approach, on the other hand, suggests that packets should be 
replaced by tiny programs, which are encapsulated in transmission frames and executed at each node 
along their path.

ForCES~\cite{doria2010}, OpenFlow~\cite{mckeown2008} and POF~\cite{song2013}  represent recent approaches for designing and deploying programmable data plane devices.
In a manner different from active networks, these new proposals rely essentially on modifying forwarding devices to support flow tables, which can be dynamically configured by remote entities through simple operations such as adding, removing or updating flow rules, i.e., entries on the flow tables.

The earliest initiatives on separating data and control signalling date back to the 80s and 90s.
The network control point (NCP)~\cite{sheinbein1982} is probably the first attempt to separate 
control and data plane signalling. NCPs were introduced by AT\&T to improve the management and control 
of its telephone network. This change promoted a faster pace of innovation of the network and provided 
new means for improving its efficiency, by taking advantage of the global view of the network provided 
by NCPs. Similarly, other initiatives such as Tempest~\cite{merwe1998}, ForCES~\cite{doria2010},
RCP~\cite{caesar2005}, and PCE~\cite{vasseur2009} proposed the separation of the control 
and data planes for improved management in ATM, Ethernet, BGP, and MPLS networks, respectively.

More recently, initiatives such as SANE~\cite{casado2006}, Ethane~\cite{casado2007-1},
OpenFlow~\cite{mckeown2008}, NOX~\cite{gude2008} and POF~\cite{song2013} proposed the decoupling of the control and data planes 
for Ethernet networks. Interestingly, these recent solutions do not require significant modifications 
on the forwarding devices, making them attractive not only for the networking research community, but 
even more to the networking industry. OpenFlow-based devices~\cite{mckeown2008}, 
for instance, can easily co-exist with traditional Ethernet devices, enabling a progressive adoption 
(i.e., not requiring a disruptive change to existing networks).

Network virtualization has gained a new traction with the advent of SDN. Nevertheless, network virtualization 
also has its roots back in the 90s. The Tempest project~\cite{merwe1998} is one of the first initiatives to 
introduce network virtualization, by introducing the concept of switchlets in ATM networks. The core idea 
was to allow multiple switchlets on top of a single ATM switch, enabling multiple independent ATM networks 
to share the same physical resources. Similarly, MBone~\cite{macedonia1994} was one of the early initiatives that 
targeted the creation of virtual network topologies on top of legacy networks, or overlay networks. This work 
was followed by several other projects such as Planet Lab~\cite{chun2003}, GENI~\cite{peterson2006} 
and VINI~\cite{bavier2006-1}. It is also worth mentioning FlowVisor~\cite{sherwood} as one of the first recent initiatives to promote a 
hypervisor-like virtualization architecture for network infrastructures, resembling the hypervisor model 
common for compute and storage.
More recently, Koponen et al. proposed a Network Virtualization Platform (NVP~\cite{koponen}) for multi-tenant datacenters using SDN as a base technology.

The concept of a network operating system was reborn with the introduction of OpenFlow-based network 
operating systems, such as NOX~\cite{gude2008}, Onix~\cite{koponen-1} 
and ONOS~\cite{krishnaswamy2013}. Indeed, network operating systems have been in existence for decades.
One of the most widely known and deployed is the Cisco IOS~\cite{bollapragada2000}, which was originally 
conceived back in the early 90s. Other network operating systems worth mentioning are JUNOS~\cite{junipernetworks2012},
ExtremeXOS~\cite{extremenetworks2014} and SR OS~\cite{alcatellucent2014}. Despite being more specialized 
network operating systems, targeting network devices such as high-performance core routers, these NOSs abstract 
the underlying hardware to the network operator, making it easier to control the network infrastructure as well 
as simplifying the development and deployment of new protocols and management applications.

Finally, it is also worth recalling initiatives that can be seen as ``technology pull'' drivers.
Back in the 90s, a movement towards open signalling~\cite{campbell1999} started 
to happen. The main motivation was to promote the wider adoption of the ideas proposed by projects 
such as NCP~\cite{sheinbein1982} and Tempest~\cite{merwe1998}. The open signalling movement worked 
towards separating the control and data signalling, by proposing open and programmable interfaces.
Curiously, a rather similar movement can be observed with the recent advent of OpenFlow and SDN, with 
the lead of the Open Networking Foundation (ONF)~\cite{onf2013-3}. This type of movement is crucial 
to promote open technologies into the market, hopefully leading equipment manufacturers to support 
open standards and thus fostering interoperability, competition, and innovation.

For a more extensive intellectual history of programmable networks and SDN we forward the reader to 
the recent paper by Feamster et al.~\cite{feamster2013-2}.

%% file: text/4_sdn_in_layers.tex
\section{Software-Defined Networks: Bottom-up}
\label{sec:layeredapproach}

\begin{figure*}[ht!]
\centering
\includegraphics[width=0.85\textwidth]{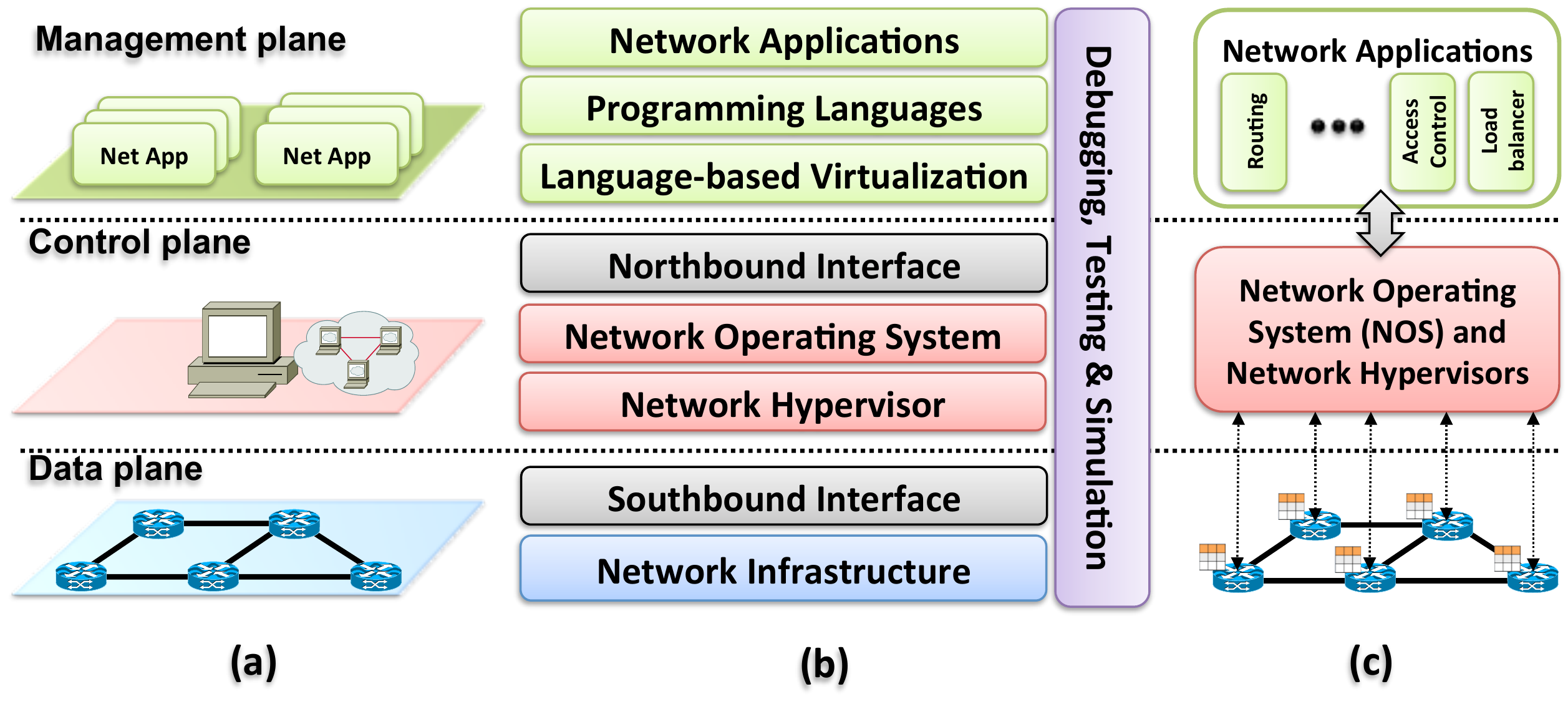}
\caption{Software-Defined Networks in (a) planes, (b) layers, and (c) system design architecture}
\label{fig:sdnlayers}
\end{figure*}

An SDN architecture can be depicted as a composition of different layers, 
as shown in Figure~\ref{fig:sdnlayers}~(b). Each layer has its own specific functions.
While some of them are always present in an SDN deployment, such as the southbound API, 
network operating systems, northbound API and \manapps, others may be 
present only in particular deployments, such as hypervisor- or language-based virtualization.

Figure~\ref{fig:sdnlayers} presents a tri-fold perspective of SDNs. The SDN layers are represented 
in the center (b) of the figure, as explained above. Figures~\ref{fig:sdnlayers} (a) and ~\ref{fig:sdnlayers} (c) depict
a plane-oriented view and a system design perspective, respectively.

The following sections introduce each layer, following a bottom-up approach. For each layer, the core properties and concepts are explained based on the different technologies and solutions.  Additionally, debugging and troubleshooting techniques and tools are discussed.

\subsection{Layer I: Infrastructure}
\label{sec:infrastructure}

An SDN infrastructure, similarly to a traditional network, is composed of a set of networking 
equipment (switches, routers and middlebox appliances). The main difference resides in 
the fact that those traditional physical devices are now simple forwarding elements 
without embedded control or software to take autonomous decisions. The network intelligence is 
removed from the data plane devices to a logically-centralized control system, i.e., the network 
operating system and applications, as shown in Figure ~\ref{fig:sdnlayers} (c).
More importantly, these new networks are built (conceptually) on top of open and standard 
interfaces (e.g., OpenFlow), a crucial approach for ensuring  configuration and communication 
compatibility and interoperability among different data and control plane devices. In other words, 
these open interfaces enable controller entities to dynamically program heterogeneous forwarding 
devices, something difficult in traditional networks, due to the large variety of 
proprietary and closed interfaces and the distributed nature of the control plane.

\begin{figure*}[ht!]
\centering
\includegraphics[width=0.85\textwidth]{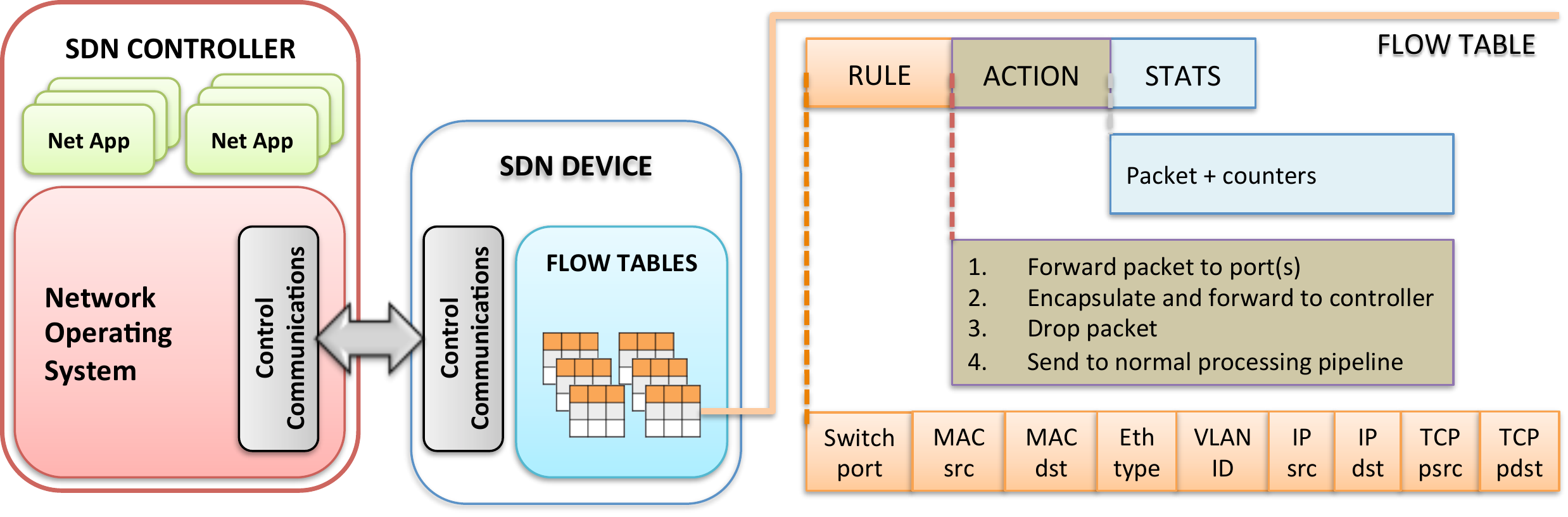}
\caption{OpenFlow-enabled SDN devices}
\label{fig:sdnopenflowswitch}
\end{figure*}

In an SDN/OpenFlow architecture, there are two main elements, the controllers and the forwarding 
devices, as shown in Figure~\ref{fig:sdnopenflowswitch}. A data plane device is a hardware or software 
element specialized in packet forwarding, while a controller is a software stack (the ``network brain'') 
running on a commodity hardware platform. An OpenFlow-enabled forwarding device is based on a pipeline of 
flow tables where each entry of a flow table has three parts: (1) a matching rule, (2) actions to be executed 
on matching packets, and (3) counters that keep statistics of matching packets. This high-level and simplified model derived from OpenFlow is currently  the most widespread design of SDN data plane devices. Nevertheless, other specifications of SDN-enabled forwarding devices are being pursued, including POF~\cite{song2013,song2013-1} and the  Negotiable Datapath Models (NDMs) from the ONF Forwarding Abstractions Working Group (FAWG)~\cite{onf2013}.

Inside an OpenFlow device, a path through a sequence of flow tables defines how packets should be handled. 
When a new packet arrives, the lookup process starts in the first table and ends either with a match in one of the tables of the pipeline 
or with a miss (when no rule is found for that packet). 
A flow rule can be defined by combining different matching fields, as illustrated in Figure~\ref{fig:sdnopenflowswitch}.
If there is no default 
rule, the packet will be discarded. However, the common case is to install a default rule which tells the switch to send the packet to the controller (or to the normal non-OpenFlow pipeline of the switch). The priority  of the rules follows the natural sequence number of the tables and the row order in a 
flow table. Possible actions include (1) forward the packet to outgoing port(s), (2) 
encapsulate it and forward it to the controller, (3) drop it, (4) send it to the 
normal processing pipeline, (5) send it to the next flow table or to special tables, such as group or metering tables introduced in the latest OpenFlow protocol.

As detailed in Table~\ref{tab:of-versions}, each version of the OpenFlow specification introduced new match fields including Ethernet, IPv4/v6, MPLS, TCP/UDP, etc. However, only a subset of those matching fields are mandatory to be compliant to a given protocol version. Similarly, many actions and port types are optional features. 
Flow match rules can be based on almost arbitrary combinations of bits of the different packet headers using bit masks for each field. Adding new matching fields has been eased with the extensibility capabilities introduced in OpenFlow version 1.2 through an OpenFlow Extensible Match (OXM) based on type-length-value (TLV) structures. To improve the overall protocol extensibility, with OpenFlow version 1.4 TLV structures have been also added to  ports, tables, and queues in replacement of the hard-coded counterparts of earlier protocol versions. 

{\renewcommand{\arraystretch}{1.4}
\begin{table*}[ht!]
\caption{Different match fields, statistics and capabilities have been added on each OpenFlow protocol revision. The number of required (Req) and optional (Opt) capabilities has grown considerably.}
\label{tab:of-versions}
\begin{center}
\footnotesize
\begin{tabular}{|c|l|l|c|c|c|c|c|c|c|c|}
\hline
\multirow{2}{*}{\textbf{OpenFlow Version}}        & \multirow{2}{*}{\textbf{Match fields}}           & \multirow{2}{*}{\textbf{Statistics}} & \multicolumn{2}{l|}{\textbf{\# Matches}}           & \multicolumn{2}{l|}{\textbf{\# Instructions}}    & \multicolumn{2}{l|}{\textbf{\# Actions}}          & \multicolumn{2}{l|}{\textbf{\# Ports}}           \\ \cline{4-11} 
                         &                                         &                             & Req                 & Opt                 & Req                & Opt                & Req                & Opt                 & Req                & Opt                \\ \hline  \hline
\multirow{4}{*}{v 1.0} & Ingress Port                            & Per table statistics        & \multirow{4}{*}{18} & \multirow{4}{*}{2}  & \multirow{4}{*}{1} & \multirow{4}{*}{0} & \multirow{4}{*}{2} & \multirow{4}{*}{11} & \multirow{4}{*}{6} & \multirow{4}{*}{2} \\ \cline{2-3}
                         & Ethernet: src, dst, type, VLAN                & Per flow statistics         &                     &                     &                    &                    &                    &                     &                    &                    \\ \cline{2-3}
                         & IPv4: src, dst, proto, ToS              & Per port statistics         &                     &                     &                    &                    &                    &                     &                    &                    \\ \cline{2-3}
                         & TCP/UDP: src port, dst port             & Per queue statistics        &                     &                     &                    &                    &                    &                     &                    &                    \\ \hline
\multirow{2}{*}{v 1.1} & Metadata, SCTP,  VLAN tagging       & Group statistics            & \multirow{2}{*}{23} & \multirow{2}{*}{2}  & \multirow{2}{*}{0} & \multirow{2}{*}{0} & \multirow{2}{*}{3} & \multirow{2}{*}{28} & \multirow{2}{*}{5} & \multirow{2}{*}{3} \\ \cline{2-3}
                         & MPLS: label, traffic class              & Action bucket statistics    &                     &                     &                    &                    &                    &                     &                    &                    \\ \hline
\multirow{2}{*}{v 1.2} & OpenFlow Extensible Match (OXM)         &                             & \multirow{2}{*}{14} & \multirow{2}{*}{18} & \multirow{2}{*}{2} & \multirow{2}{*}{3} & \multirow{2}{*}{2} & \multirow{2}{*}{49} & \multirow{2}{*}{5} & \multirow{2}{*}{3} \\ \cline{2-3}
                         & IPv6: src, dst, flow label, ICMPv6      &                             &                     &                     &                    &                    &                    &                     &                    &                    \\ \hline
\multirow{2}{*}{v 1.3} & \multirow{2}{*}{PBB, IPv6 Extension Headers} & Per-flow meter              & \multirow{2}{*}{14} & \multirow{2}{*}{26} & \multirow{2}{*}{2} & \multirow{2}{*}{4} & \multirow{2}{*}{2} & \multirow{2}{*}{56} & \multirow{2}{*}{5} & \multirow{2}{*}{3} \\ \cline{3-3}
                         &                                         & Per-flow meter band         &                     &                     &                    &                    &                    &                     &                    &                    \\ \hline
\multirow{2}{*}{v 1.4} & \multirow{2}{*}{---} & ---            & \multirow{2}{*}{\textbf{14}} & \multirow{2}{*}{\textbf{27}} & \multirow{2}{*}{\textbf{2}} & \multirow{2}{*}{\textbf{4}} & \multirow{2}{*}{\textbf{2}} & \multirow{2}{*}{\textbf{57}} & \multirow{2}{*}{\textbf{5}} & \multirow{2}{*}{\textbf{3}} \\ \cline{3-3}
                         &                                         & Optical port properties        &                     &                     &                    &                    &                    &                     &                    &                    \\ \hline
\end{tabular}
\end{center}
\end{table*}
}

%
%

\vspace{2mm}
\noindent \textit{Overview of available OpenFlow devices}

Several OpenFlow-enabled forwarding devices are available on 
the market, both as commercial and open source products (see Table~\ref{tab:openflowdevices}).
There are many off-the-shelf, ready to deploy, OpenFlow switches and routers, among other appliances.
Most of the switches available on the market have relatively small Ternary Content-Addressable Memory (TCAMs), with up to 8K entries. 
Nonetheless, this is changing at a fast pace. Some of the latest devices released in the market go 
far beyond that figure. Gigabit Ethernet (GbE) switches for common business purposes are already supporting up 
to 32K L2+L3 or 64K L2/L3 exact match flows~\cite{centecnetworks2013-1}. Enterprise class 10GbE switches 
are being delivered with more than 80K Layer 2 flow entries~\cite{nec2013-1}. Other switching 
devices using high performance chips (e.g., EZchip NP-4) provide optimized TCAM memory that 
 supports from 125K up to 1000K flow table entries~\cite{noviflow2013-1}. This is a clear sign that the size 
of the flow tables is growing at a pace aiming to meet the needs of future SDN deployments.

Networking hardware manufacturers have produced various kinds of OpenFlow-enabled devices, as is shown in Table~\ref{tab:openflowdevices}. 
These devices range from equipment for small businesses (e.g., GbE switches) to high-class data center equipment (e.g., high-density switch chassis with up to 100GbE connectivity for edge-to-core applications, with tens of Tbps of switching capacity).

{\renewcommand{\arraystretch}{1.4}
\begin{table*}[!htp]
\caption{OpenFlow enabled hardware and software devices}
\label{tab:openflowdevices}
\begin{center}
\footnotesize
\begin{tabularx}{\linewidth}{p{1.1cm}p{3.1cm}p{0.9cm}p{2.4cm}p{0.95cm}X}
\hline
\textbf{Group} & \textbf{Product} & \textbf{Type} & \textbf{Maker/Developer} & \textbf{Version} & \textbf{Short description} \\
\hline
\multirow{16}{*}{Hardware} 
& 8200zl and 5400zl~\cite{hp2013} & chassis & Hewlett-Packard & v1.0 & Data center class chassis (switch modules). \\\cline{2-6}
& Arista 7150 Series~\cite{aristanetworks2013} & switch & Arista Networks & v1.0 & Data centers hybrid Ethernet/OpenFlow switches. \\\cline{2-6}
& BlackDiamond X8~\cite{extremenetworks2013} & switch & Extreme Networks & v1.0 & Cloud-scale  hybrid Ethernet/OpenFlow switches. \\\cline{2-6}
& CX600 Series~\cite{co.2013} & router & Huawei & v1.0 & Carrier class MAN routers. \\\cline{2-6}
& EX9200 Ethernet~\cite{junipernetworks2013} & chassis & Juniper & v1.0 & Chassis based switches for cloud data centers. \\\cline{2-6}
& EZchip NP-4~\cite{yokneam2011} & chip & EZchip Technologies & v1.1 & High performance 100-Gigabit network processors. \\\cline{2-6}
& MLX Series~\cite{brocade2013} & router & Brocade & v1.0 & Service providers and enterprise class routers. \\\cline{2-6}
& NoviSwitch 1248~\cite{noviflow2013-1} & switch & NoviFlow & v1.3& High performance OpenFlow switch. \\\cline{2-6}
& NetFPGA~\cite{netfpga2014} & card & NetFPGA & v1.0 & 1G and 10G OpenFlow implementations. \\\cline{2-6}
& RackSwitch G8264~\cite{ibm2013} & switch & IBM & v1.0 & Data center switches supporting Virtual Fabric and OpenFlow. \\\cline{2-6}
& PF5240 and PF5820~\cite{nec2013-2} & switch & NEC & v1.0 & Enterprise class hybrid Ethernet/OpenFlow switches. \\\cline{2-6}
& Pica8 3920~\cite{pica8opennetworking2013} & switch & Pica8 & v1.0 & Hybrid Ethernet/OpenFlow switches. \\\cline{2-6}
& Plexxi Switch 1~\cite{plexxi2013} & switch & Plexxi & v1.0 & Optical multiplexing interconnect for data centers. \\\cline{2-6}
& V330 Series~\cite{centecnetworks2013} & switch & Centec Networks & v1.0 & Hybrid Ethernet/OpenFlow switches. \\\cline{2-6}
& Z-Series~\cite{cyan2013} & switch & Cyan & v1.0 & Family of packet-optical transport platforms.\\
\hline
\multirow{9}{*}{Software} 
& contrail-vrouter~\cite{networks2013} & vrouter &  Juniper Networks & v1.0 & Data-plane function to interface with a VRF.\\\cline{2-6}
& LINC~\cite{flowforwarding2013,rutka2013} & switch &  FlowForwarding & v1.4 & Erlang-based soft switch with OF-Config 1.1 support.\\\cline{2-6}
& ofsoftswitch13~\cite{cpqd2013} & switch & Ericsson, CPqD & v1.3 & OF 1.3 compatible user-space software switch implementation. \\\cline{2-6}
& Open vSwitch~\cite{listofcontributors2013,pfaff2009} & switch & Open Community & v1.0-1.3  & Switch platform designed for virtualized server environments.  \\\cline{2-6}
& OpenFlow Reference~\cite{openflowcommunity2009} & switch & Stanford & v1.0 & OF Switching capability to a Linux PC with multiple NICs. \\\cline{2-6}
& OpenFlowClick~\cite{mundada2009} & vrouter & Yogesh Mundada & v1.0 & OpenFlow switching element for Click software routers.\\\cline{2-6}
& Switch Light~\cite{bigswitchnetworks2013} & switch & Big Switch & v1.0  & Thin switching software platform for physical/virtual switches. \\\cline{2-6}
& Pantou/OpenWRT~\cite{yiakoumis2011} & switch & Stanford  & v1.0 & Turns a wireless router into an OF-enabled switch. \\\cline{2-6}
& XorPlus~\cite{shang2014} & switch &  Pica8 & v1.0 & Switching software for high performance ASICs.\\
\hline
\end{tabularx}
\end{center}
\end{table*}
}

Software switches are emerging as one of the most promising 
solutions for data centers and virtualized network infrastructures~\cite{weissberger2013,schenker2013,casado2013}.
Examples of software-based OpenFlow switch implementations include Switch Light~\cite{bigswitchnetworks2013}, ofsoftswitch13~\cite{cpqd2013}, Open vSwitch~\cite{listofcontributors2013}, OpenFlow Reference~\cite{openflowcommunity2009}, Pica8~\cite{pica8opennetworking2013-1}, Pantou~\cite{yiakoumis2011}, and XorPlus~\cite{shang2014}.
Recent reports show that the number of virtual access ports is already larger than physical access ports on data centers~\cite{casado2013}. 
Network virtualization has been one of the drivers behind this trend.
Software switches such as Open vSwitch have been used for moving network functions to the edge (with the core performing traditional IP forwarding), thus enabling network virtualization~\cite{koponen}.

An interesting observation is the number of small, start-up enterprises devoted to SDN, such as 
Big Switch, Pica8, Cyan, Plexxi, and NoviFlow. This seems to imply that SDN is springing a more competitive and open 
networking market, one of its original goals.
Other effects of this openness triggered by SDN include the emergence of so-called ``bare metal switches'' or ``whitebox switches'', where the software and hardware are sold separately and the end-user is free to load an operating system of its choice~\cite{onie2013}. 


\subsection{Layer II: Southbound Interfaces}
\label{sec:southboundAPIs}

Southbound interfaces (or southbound APIs) are the connecting bridges between control and forwarding 
elements, thus being the crucial instrument for clearly separating control and data plane functionality. However, 
these APIs are still tightly tied to the forwarding elements of the underlying physical or virtual 
infrastructure.

Typically, a new switch can take two years to be ready for commercialization if built from scratch, 
with upgrade cycles that can take up to nine months. The software development for a new product can 
take from six months to one year~\cite{kato2013}. The initial investment 
is high and risky. 
As a central component of its design the southbound APIs represent one of the major barriers for the introduction 
and acceptance of any new networking technology. 
In this light, the emergence of SDN southbound API proposals such as OpenFlow~\cite{mckeown2008} is seen as welcome by many in the industry.
These standards promote interoperability, allowing the deployment of vendor-agnostic network devices. 
This has already been demonstrated by the interoperability between OpenFlow-enabled equipments from different vendors.

As of this writing, OpenFlow is the most widely accepted and deployed open southbound standard 
for SDN. It provides a common specification to implement OpenFlow-enabled forwarding devices, and for the
communication channel between data and control plane devices (e.g., switches and controllers).
The OpenFlow protocol provides three information sources for network operating systems.
First, event-based messages are sent by forwarding devices to the controller when a link or port 
change is triggered. Second, flow statistics are generated by the forwarding devices and collected 
by the controller. Third, packet-in messages are sent by forwarding devices to the controller when 
they do not known what to do with a new incoming flow or because there is an explicit ``send to 
controller'' action in the matched entry of the flow table. These information channels are the essential 
means to provide flow-level information to the network operating system.

Albeit the most visible, OpenFlow is not the only available southbound interface for SDN.
There are other API proposals such as ForCES~\cite{doria2010}, OVSDB~\cite{pfaff2013-1}, POF~\cite{song2013,song2013-1}, OpFlex~\cite{smith2014}, \coloredtext{OpenState~\cite{bianchi2014},  Revised OpenFlow Library (ROFL)~\cite{Sune2014},  Hardware Abstraction Layer (HAL)~\cite{Parniewicz2014,Belter2014}, and Programmable Abstraction of
Datapath (PAD)~\cite{Belter2014-1}.}
ForCES proposes a more flexible approach to traditional network management without changing the current architecture of the network, i.e., without the need of a logically-centralized external controller. 
The control and data planes are separated but can potentially be kept in the same network element.
However, the control part of the network element can be upgraded on-the-fly with third-party firmware.


OVSDB~\cite{pfaff2013-1} is another type of southbound API, designed to provide 
advanced management capabilities for Open vSwitches. Beyond OpenFlow's capabilities to configure 
the behavior of flows in a forwarding device, an Open vSwitch offers other networking functions.
For instance, it allows the control elements to create multiple virtual switch instances, set QoS 
policies on interfaces, attach interfaces to the switches, configure tunnel interfaces on OpenFlow 
data paths, manage queues, and collect statistics.
Therefore, the OVSDB is a complementary protocol to OpenFlow for Open vSwitch.

%

One of the first direct competitors of OpenFlow is POF~\cite{song2013,song2013-1}.
One of the main goals of POF is to enhance the current SDN forwarding plane.
With OpenFlow, switches have to understand the protocol headers to extract the required bits to be matched with the flow tables entries.
This parsing represents a significant burden for data plane devices, in particular if we consider that OpenFlow version 1.3 already contains more  than 40 header fields. 
Besides this inherent complexity, backward compatibility issues may arise every time new header fields are included in or removed from the protocol.
To achieve its goal, POF proposes a generic flow  instruction set (FIS) that makes the forwarding plane protocol-oblivious.
A forwarding element does not need to know, by itself, anything about the packet format in advance.
Forwarding devices are seen as white boxes with only processing and forwarding  capabilities. 
In POF, packet parsing is a controller task that results in a sequence of generic keys and table lookup instructions that are installed in the forwarding elements.
The behavior of data plane devices is therefore completely under the control of the SDN controller.
Similar to a CPU in a computer system, a POF switch is application- and protocol-agnostic. 

A recent southbound interface proposal is OpFlex~\cite{smith2014}.
Contrary to OpenFlow (and similar to ForCES), one of the ideas behind OpFlex is to distribute part of the 
complexity of managing the network back to the forwarding devices, with the aim of improving scalability.
Similar to OpenFlow, policies are logically centralized and abstracted from the underlying implementation.
The differences between OpenFlow and OpFlex are a clear illustration of one of the important questions to be answered when devising a southbound interface: where to place each piece of the overall functionality. 

\coloredtext{In contrast to OpFlex and POF, OpenState~\cite{bianchi2014} and ROFL~\cite{Sune2014} do not propose a new set of instructions for programming data plane devices.
OpenState proposes extended finite machines (stateful programming abstractions) as an extension (super-set) of the OpenFlow match/action abstraction. 
Finite state machines allow the implementation of several stateful tasks inside forwarding devices, i.e., without augmenting the complexity or overhead of the control plane.
For instance, all tasks involving only local state, such as MAC learning operations, port knocking or stateful edge firewalls can be performed directly on the forwarding devices without any extra control plane communication and processing delay.
ROFL, on the other hand, proposes an abstraction layer that hides the details of the different OpenFlow versions, thus providing a clean API for software developers, simplifying application development.}

\coloredtext{
HAL~\cite{Parniewicz2014,Belter2014} is not exactly a southbound API, but is closely related.
Differently from the aforementioned approaches, HAL is rather a translator that enables a southbound API such as OpenFlow to control heterogeneous hardware devices.
It thus sits between the southbound API and the hardware device.
Recent research experiments with HAL have demonstrated the viability of SDN control in access networks such as Gigabit Ethernet passive optical networks (GEPONs)~\cite{Clegg2014} and cable networks (DOCSIS)~\cite{Fuentes2014}.
A similar effort to HAL is the Programmable Abstraction of Datapath (PAD)~\cite{Belter2014-1}, a proposal that goes a bit further by also working as a southbound API by itself.
More importantly, PAD allows a more generic programming of forwarding devices by enabling the control of datapath behavior using generic byte operations, defining protocol headers and providing function definitions.}


\subsection{Layer III: Network Hypervisors}
\label{sec:virtualizationhypervisor}

Virtualization is already a consolidated technology in modern computers. The fast 
developments of the past decade have made virtualization of computing platforms mainstream.
Based on recent reports, the number of virtual servers has already exceeded the number of physical 
servers~\cite{bittman2013,koponen}.

Hypervisors enable distinct virtual machines to share the same hardware resources. In a 
cloud infrastructure-as-a-service (IaaS), each user can have its own virtual resources, from computing to 
storage.
This enabled new revenue and business models where users allocate resources on-demand, from a shared 
physical infrastructure, at a relatively low cost.
At the same time, providers make better use of the capacity of their installed physical 
infrastructures, creating new revenue streams without significantly increasing their CAPEX and OPEX 
costs. One of the interesting features of virtualization technologies today is the fact that virtual 
machines can be easily migrated from one physical server to another and can 
be created and/or destroyed on-demand, enabling the provisioning of elastic services with flexible 
and easy management.
Unfortunately, virtualization has been only partially realized in practice. Despite the great advances 
in virtualizing computing and storage elements, the network is still mostly statically configured in a box-by-box manner~\cite{chowdhury2010}.

The main network requirements can be captured along two dimensions: network topology and address space.
Different workloads require different network topologies and services, such as flat L2 or L3 services, or even more complex L4-L7 
services for advanced functionality.
Currently, it is very difficult for a single physical topology to support the diverse demands 
of applications and services. Similarly, address space is hard to change in current networks. Nowadays, 
virtualized workloads have to operate in the same address of the physical infrastructure. 
Therefore, it is hard to keep the original network configuration for a tenant, virtual machines can not migrate to arbitrary locations, and the addressing scheme is fixed and hard to change. 
For example, IPv6 cannot be used by the VMs of a tenant if the underlying physical forwarding devices support only IPv4.


To provide complete virtualization the network should provide similar properties to the computing layer~\cite{chowdhury2010}.
The network infrastructure should be able to support arbitrary network topologies and addressing schemes.
Each tenant should have the ability to configure both the computing nodes and the network simultaneously.
Host migration should automatically trigger the migration of the corresponding virtual network ports.
One might think that long standing virtualization primitives such as VLANs (virtualized L2 domain), NAT (Virtualized IP address space), and 
MPLS (virtualized path) are enough to provide full and automated network virtualization.
However, these  technologies are anchored on a box-by-box basis configuration, i.e., there is no single unifying 
abstraction that can be leveraged to configure (or reconfigure) the network in a global manner.
As a consequence, current network provisioning can take months, while computing provisioning takes only 
minutes~\cite{koponen,cearley2013,peng2012,zhang2014}.

There is hope that this situation will change with SDN and the availability of new tunneling techniques (e.g., VXLAN~\cite{mahalingam2013}, NVGRE~\cite{sridharan2013}).
For instance, solutions such as FlowVisor~\cite{sherwood2009,sherwood2010,azodolmolky2012},  FlowN~\cite{drutskoy2012}, NVP~\cite{koponen}, OpenVirteX~\cite{al-shabibi2014,Al_Shabibi2014_4}, IBM SDN VE~\cite{racherla2014,li2014}, \coloredtext{RadioVisor~\cite{Gudipati2014_4}, AutoVFlow~\cite{Yamanaka2014}, eXtensible Datapath Daemon (xDPd)~\cite{berlin2014,doriguzzicorin2014}, optical transport network virtualization~\cite{Szyrkowiec2014}, and version-agnostic OpenFlow slicing mechanisms~\cite{Depaoli2014}, } have been recently proposed, evaluated and deployed in real scenarios for on-demand provisioning of virtual networks.

\vspace{2mm}
\noindent \textit{Slicing the network}

FlowVisor is one of the early technologies to virtualize a SDN.
Its basic idea is to allow multiple logical networks share the same OpenFlow networking infrastructure.
For this purpose, it provides an abstraction layer that makes it easier to slice a data plane based on off-the-shelf OpenFlow-enabled switches, allowing multiple and diverse networks to co-exist. 

Five slicing dimensions are considered in FlowVisor: bandwidth, topology, traffic, device CPU and forwarding 
tables. Moreover, each network slice supports a controller, i.e., multiple controllers can co-exist on top 
of the same physical network infrastructure. Each controller is allowed to act only on its own network slice.
In general terms, a slice is defined as a particular set of flows on the data plane. 
From a system design perspective, FlowVisor is a transparent proxy that intercepts OpenFlow messages between 
switches and controllers. It partitions the link bandwidth and flow tables of each switch. Each slice receives 
a minimum data rate and each guest controller gets its own virtual flow table in the switches.

Similarly to FlowVisor, OpenVirteX~\cite{al-shabibi2014,Al_Shabibi2014_4} acts as a proxy between the network operating system and the forwarding 
devices. However, its main goal is to provide virtual SDNs through both topology, address, and control function
virtualization. 
All these properties are necessary in multi-tenant environments where virtual networks 
need to be managed and migrated according to the computing and storage virtual resources. Virtual network 
topologies have to be mapped onto the underlying forwarding devices, with virtual addresses allowing tenants to completely manage their address space without depending on the underlying network elements addressing schemes.

\coloredtext{AutoSlice~\cite{bozakov2012} is another SDN-based virtualization proposal. 
Differently from FlowVisor, it focuses on the automation of the deployment and operation of vSDN (virtual SDN) topologies with minimal mediation or arbitration by the substrate network operator.
Additionally, AutoSlice targets also scalability aspects of network hypervisors by optimizing resource utilization and by mitigating the flow-table limitations through a precise monitoring of the flow traffic statistics. 
Similarly to AutoSlice, AutoVFlow~\cite{Yamanaka2014} also enables multi-domain network virtualization.
However, instead of having a single third party to control the mapping of vSDN topologies, as is the case of AutoSlice, AutoVFlow uses a multi-proxy architecture that allows network owners to implement flow space virtualization in an autonomous way by exchanging information among the different domains.
}

FlowN~\cite{drutskoy2012,drutskoy2013} is based on a slightly different concept. Whereas FlowVisor can be compared to a full virtualization technology, FlowN 
is analogous to a container-based virtualization, i.e., a lightweight virtualization approach. FlowN 
was also primarily conceived to address multi-tenancy in the context of cloud platforms. It is designed to 
be scalable and allows a unique shared controller platform to be used for managing multiple domains 
in a cloud environment. Each tenant has full control over its virtual networks and is free to deploy 
any network abstraction and application on top of the controller platform.

\coloredtext{The compositional SDN hypervisor~\cite{Jin2014_4} was designed with a different set of goals. 
Its main objective is to allow the cooperative (sequential or parallel) execution of applications developed with different programming languages or conceived for diverse control platforms.
It thus offers interoperability and portability in addition to the typical functions of network hypervisors.
}

\vspace{2mm}
\noindent \textit{Commercial multi-tenant network hypervisors}

None of the aforementioned approaches is designed to address all challenges of multi-tenant 
data centers. For instance, tenants want to be able to migrate their enterprise solutions to cloud 
providers without the need to modify the network configuration of their home network. Existing 
networking technologies and migration strategies have mostly failed to meet both the tenant and the 
service provider requirements.
A multi-tenant environment should be anchored in a network hypervisor capable of abstracting the 
underlaying forwarding devices and physical network topology from the tenants. Moreover, each tenant 
should have access to control abstractions and manage its own virtual networks independently and 
isolated from other tenants.

With the market demand for network virtualization and the recent research on SDN showing promise as an enabling technology, different commercial virtualization platforms based on SDN concepts have started to appear.
VMWare has proposed a network virtualization platform (NVP)~\cite{koponen} that provides the necessary abstractions to allow the creation of independent virtual networks for large-scale multi-tenant environments.
NVP is a complete network virtualization solution that allows the creation of virtual networks, each with independent service model, topologies, and addressing architectures over the same physical network.
With NVP, tenants do not need to know anything about the underlying network topology, configuration or other specific aspects of the forwarding devices. 
NVP's network hypervisor translates the tenants configurations and requirements into low level instruction sets to be installed on the forwarding devices. 
For this purpose, the platform uses a cluster of SDN controllers to manipulate the forwarding tables of the Open vSwitches in the host's hypervisor. 
Forwarding decisions are therefore made exclusively on the network edge.
After the decision is made, the packet is tunneled over the physical network to the receiving host hypervisor (the physical network sees nothing but ordinary IP packets).

IBM has also recently proposed SDN VE~\cite{racherla2014,li2014}, another commercial and enterprise-class network virtualization platform.
SDN VE uses OpenDaylight as one \coloredtext{ of the building blocks of the so-called Software-Defined Environments (SDEs), a trend further discussed in Section~\ref{sec:challenges}.}
This solution also offers a complete implementation framework for network virtualization. 
Like NVP, it uses a host-based overlay approach, achieving advanced network abstraction that enables application-level network services in large-scale multi-tenant environments. 
Interestingly, SDN VE 1.0 is capable of supporting in one single instantiation up to 16,000 virtual networks and 128,000 virtual machines~\cite{racherla2014,li2014}.

To summarize, currently there are already a few network hypervisor proposals leveraging the advances of SDN.
\coloredtext{There are, however, still several issues to be addressed.
These include, among others, the improvement of virtual-to-physical mapping techniques~\cite{Ghorbani2014_4}, the definition of the level of detail that should be exposed at the logical level, and the support for nested virtualization~\cite{Casado2014_4}.}
We anticipate, however, this ecosystem to expand in the near future since network virtualization will most likely play a key role in future virtualized environments, similarly to the expansion we have been witnessing in virtualized computing.



\subsection{Layer IV: Network Operating Systems / Controllers}
\label{sec:controllers}

Traditional operating systems provide abstractions (e.g., high-level programming APIs) for accessing 
lower-level devices, manage the concurrent access to the underlying resources (e.g., hard drive, network 
adapter, CPU, memory), and provide security protection mechanisms. These functionalities and resources 
are key enablers for increased productivity, making the life of system and application developers easier. Their 
widespread use has significantly contributed to the evolution of various ecosystems (e.g., programming 
languages) and the development of a myriad of applications.

In contrast, networks have so far been managed and configured using lower level, device-specific instruction 
sets and mostly closed proprietary network operating systems (e.g., Cisco IOS and Juniper JunOS). 
Moreover, the idea of operating systems abstracting device-specific characteristics 
and providing, in a transparent way, common functionalities is still mostly absent in networks.
For instance, nowadays designers of routing protocols need to deal with complicated distributed algorithms when solving networking problems.
Network practitioners have therefore been solving the same problems over and over again.

SDN is promised to facilitate network management and ease the burden of solving networking problems by means of the logically-centralized control offered by a network operating system (NOS)~\cite{gude2008}.
As with traditional operating systems, the crucial value of a NOS is to provide abstractions, essential services, 
and common application programming interfaces (APIs) to developers. 
Generic functionality as network state and network topology information, device discovery, and distribution of network configuration can be provided as services of the NOS.
With NOSs, to define network policies a developer no longer needs to care about the low-level details of data distribution among routing elements, for instance.
Such systems can arguably create a new environment capable of fostering innovation at a faster pace by reducing the inherent complexity of creating new network protocols and \manapps.

A NOS (or controller) is a critical element in an SDN architecture as it is the key supporting piece 
for the control logic (applications) to generate the network configuration based on the policies defined by the network operator. 
Similar to a traditional operating system, the control platform abstracts the lower-level details of connecting and interacting with forwarding devices (i.e., of materializing the network policies).

\vspace{2mm}
\noindent \textit{Architecture and design axes}
\vspace{2mm}

There is a very diverse set of controllers and control platforms with different design and architectural choices~\cite{koponen-1,opendaylight2013,junipernetworks2013-1,hp2013-1,phemius2013,erickson2013}.
Existing controllers can be categorized based on many aspects. 
From an architectural point of view, one of the most relevant is if they are centralized or distributed.
This is one of the key design axes of SDN control platforms, so we start by discussing this aspect next. 

\vspace{2mm}
\noindent \textit{Centralized vs. Distributed}

A centralized controller  is a single entity that manages all forwarding devices of the network.
Naturally, it represents a single point of failure and may have scaling limitations.
A single controller may not be enough to manage a network with a large number of data plane elements. 
Centralized controllers such as NOX-MT ~\cite{tootoonchian2012},
Maestro~\cite{cai2011}, Beacon~\cite{erickson2013}, and Floodlight~\cite{openflowhub.org2012}    
have been designed as highly concurrent systems, to achieve the throughput required by enterprise class 
networks and data centers. 
These controllers are based on multi-threaded designs to explore the parallelism of multi-core computer architectures.
As an example, Beacon can deal with more than 12 
million flows per second by using large size computing nodes of cloud providers such as Amazon~\cite{erickson2013}. Other centralized controllers such as Trema~\cite{takamiya2012}, 
Ryu NOS~\cite{nippontelegraphandtelephonecorporation2012}, Meridian~\cite{banikazemi2013}, and 
ProgrammableFlow~\cite{nec2013,nec2013-2} target specific environments such as data centers, cloud infrastructures, and carrier grade networks.
\coloredtext{Furthermore, controllers such as Rosemary~\cite{shin2014} offer specific functionality and guarantees, namely security and isolation of applications.
By using a container-based architecture called micro-NOS, it achieves its primary goal of isolating applications and preventing the propagation of failures throughout the SDN stack.
}

Contrary to a centralized design, a distributed network operating system can be scaled up to meet the requirements 
of potentially any environment, from small to large-scale networks.
A distributed controller can be 
a centralized cluster of nodes or a physically distributed set of elements. While the first alternative 
can offer high throughput for very dense data centers, the latter can be more resilient to different 
kinds of logical and physical failures.
A cloud provider that spans multiple data centers interconnected by a wide area network may require a hybrid approach, with clusters of controllers inside each data center and 
distributed controller nodes in the different sites~\cite{jain2013-1}.

Onix~\cite{koponen-1}, HyperFlow~\cite{tootoonchian2010},
HP VAN SDN~\cite{hp2013-1}, ONOS~\cite{krishnaswamy2013}, DISCO~\cite{phemius2013},
 \textit{yanc}~\cite{monaco2013}, \coloredtext{PANE~\cite{ferguson2013}, SMaRtLight~\cite{Botelho2014}, and Fleet~\cite{Matsumoto2014_4}} are examples of distributed controllers. 
Most distributed controllers offer weak consistency semantics, which means that data updates on distinct nodes will \emph{eventually} be updated on 
all controller nodes.
This implies that there is a period of time in which distinct nodes may read different 
values (old value or new value) for a same property.
Strong consistency, on the other hand, ensures that all 
controller nodes will read the most updated property value after a write operation.
Despite its impact on system performance, strong consistency offers a simpler interface to application developers.
To date, only \coloredtext{Onix, ONOS, and SMaRtLight provide this data consistency model.} 


Another common property of distributed controllers is fault tolerance.
When one node fails, another neighbor node should take over the duties and devices of the failed node. 
So far, despite some controllers tolerating crash failures, they do not tolerate
arbitrary failures, which means that any node with an abnormal behavior will not be replaced by a potentially 
well behaved one.

A single controller may be enough to manage a small network, however it represents a single point 
of failure.
Similarly, independent controllers can be spread across the network, each of them managing 
a network segment, reducing the impact of a single controller failure. Yet, if the control plane
availability is critical, a cluster of controllers can be used to achieve a higher degree of availability 
and/or for supporting more devices.
Ultimately, a distributed controller can improve the control plane 
resilience, scalability and reduce the impact of problems caused by network partition, for instance.
SDN resiliency as a whole is an open challenge that will be further discussed in Section \ref{sec:resiliency}.

\vspace{2mm}
\noindent \textit{Dissecting SDN Controller Platforms}
\vspace{2mm}

To provide a better architectural overview and understanding the design a network operating system, Table~\ref{tab:controllerdesign} summarizes some of the most relevant architectural and design properties of 
SDN controllers and control platforms. 
We have focused on the elements, services and interfaces of a selection of production-level, well-documented controllers and control platforms.
Each line in the table represent a component we consider important in a modular and scalable control platform.
We observe a highly diversified environment, with different properties and components being used by distinct control platforms.
This is not surprising, given an environment with many competitors willing to be at the forefront of SDN development.
Note also that not all components are available on all platforms.
For instance, east/westbound APIs are not required in centralized controllers such as Beacon. 
In fact, some platforms have very specific niche markets, such as telecom companies and cloud providers, so the requirements will be different.

{\renewcommand{\arraystretch}{1.4}
\begin{table*}[!htp]
\caption{Architecture and design elements of control platforms}
\label{tab:controllerdesign}
\begin{center}
\footnotesize
\begin{tabularx}{\textwidth}{|p{1.9cm}|X|X|X|X|X|}
\hline
\textbf{Component} & \textbf{OpenDaylight} & \textbf{OpenContrail} & \textbf{HP VAN SDN} & \textbf{Onix} & \textbf{Beacon} \\\hline

\textbf{Base network \hfill {\color{white}x} services}     & Topology/Stats/Switch Manager, Host \hfill {\color{white}x} Tracker, Shortest \hfill {\color{white}x} Path Forwarding & Routing, Tenant \hfill {\color{white}x} Isolation & Audit Log, Alerts, \hfill{\color{white}x} Topology, Discovery & Discovery, Multi-\hfill{\color{white}x} consistency Storage, {\color{white}x} Read State, Register \hfill{\color{white}x} for updates  & Topology, device \hfill {\color{white}x} manager, and routing \\\hline

\textbf{East/Westbound APIs}       & \textit{---} & Control Node (XMPP-\hfill{\color{white}x} like control channel) & Sync API & Distribution I/O module & \textit{Not present} \\\hline

\textbf{Integration Plug-ins}  & OpenStack Neutron & CloudStack, OpenStack & OpenStack & \textit{---} & \textit{---} \\\hline

\textbf{Management Interfaces} & GUI/CLI, REST API & GUI/CLI & REST API Shell / GUI Shell & \textit{---} & Web \\\hline

\textbf{Northbound APIs}           & REST, RESTCONF~\cite{bierman2014restconf}, Java APIs & REST APIs (configuration, operational, and analytic) & REST API, GUI Shell & Onix API (general \hfill {\color{white}x} purpose) & API (based on \hfill {\color{white}x} OpenFlow events) \\\hline

\textbf{Service abstraction layers} & Service Abstraction\hfill {\color{white}x} Layer (SAL) & \textit{---} & Device Abstraction API & Network Information \hfill{\color{white}x} Base (NIB) Graph \hfill{\color{white}x} with Import/Export\hfill {\color{white}x} Functions & \textit{---} \\\hline

\textbf{Southbound APIs or \hfill {\color{white}x} connectors} & OpenFlow, OVSDB,\hfill{\color{white}x}  SNMP, PCEP, BGP, \hfill{\color{white}x} NETCONF & \textit{---} & OpenFlow, L3 Agent, \hfill {\color{white}x} L2 Agent & OpenFlow, OVSDB & OpenFlow \\\hline

\end{tabularx}
\end{center}
\end{table*}
}

\begin{figure*}[ht]
\centering
\includegraphics[width=0.95\textwidth]{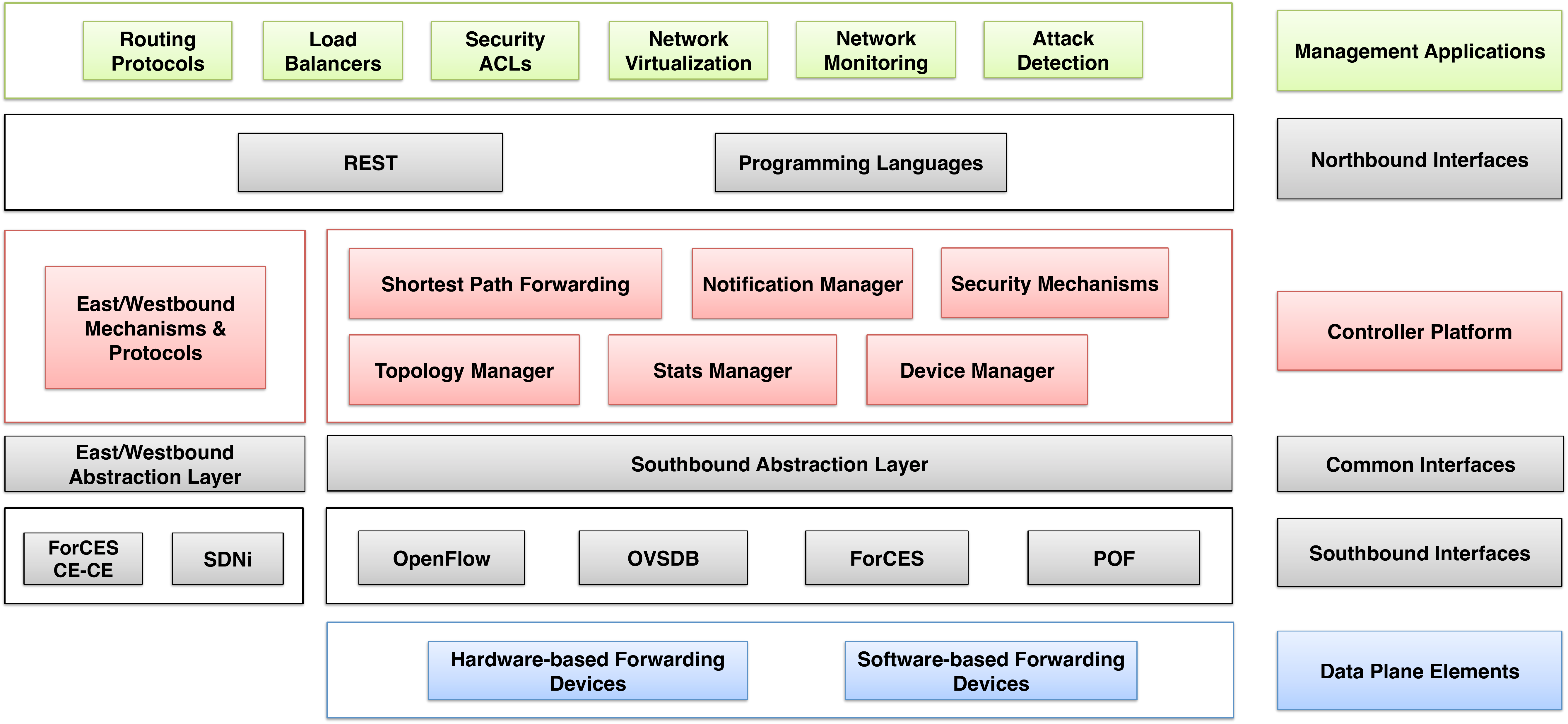}
\caption{SDN control platforms: elements, services and interfaces}
\label{fig:controlplatformarch}
\end{figure*}

Based on the analysis of the different SDN controllers proposed to date (both those presented in Table~\ref{tab:controllerdesign} and others, such as NOX~\cite{gude2008}, Meridian~\cite{banikazemi2013}, ForCES~\cite{doria2010}, and  FortNOX~\cite{porras2012}), we extract several common elements and provide a first attempt to clearly and systematically dissect an SDN control platform in Figure~\ref{fig:controlplatformarch}.

There are at least three relatively well-defined layers in most of the existing control platforms: 
(\textit{i}) the application, orchestration and services; (\textit{ii})  the core controller functions, and 
(iii) the elements for southbound communications. The connection at the upper-level layers is based on northbound interfaces such as REST APIs~\cite{richardson2008restful}  and programming languages such as FML~\cite{hinrichs2009}, Frenetic~\cite{foster2011} and NetCore~\cite{monsanto2012}. 
On the lower-level part of a control platform, southbound APIs and protocol plugins interface the forwarding elements. 
The core of a controller platform can be characterized as a combination of its base network service functions and the various interfaces.

\vspace{2mm}
\noindent \textit{Core controller functions}

The base network service functions are what we consider the essential functionality all controllers should provide.
As an analogy, these functions are like base services of operating systems, 
such as program execution, I/O operations control, communications, protection, and so on. 
These services are used by other operating 
system level services and user applications. 
In a similar way, functions such as topology, statistics, notifications and device management, together with shortest path forwarding and security mechanisms are essential network control functionalities that network applications may use in building its logic.
For instance, the notification manager should be able to receive, process, and forward events (e.g., alarm notifications, security alarms, state changes)~\cite{onf2013-2}.
Security mechanisms are another example, as they are critical components to provide basic isolation and security enforcement between services and applications.
For instance, rules generated by high priority services should not be overwritten with rules created by applications with a lower priority.


\vspace{2mm}
\noindent \textit{Southbound}

On the lower-level of control platforms, the southbound APIs can be seen as a layer of device drivers.
They provide a common interface for the upper layers, while allowing a control platform to use different 
southbound APIs (e.g., OpenFlow, OVSDB, ForCES) and protocol plugins to manage existing or new physical or 
virtual devices (e.g., SNMP, BGP, NetConf). This is essential both for backward compatibility and heterogeneity, 
i.e., to allow multiple protocols and device management connectors. Therefore, on the data plane, a mix of physical devices, virtual devices (e.g., Open vSwitch~\cite{listofcontributors2013,pfaff2009}, vRouter~\cite{singla2013}) and a variety of device interfaces (e.g., OpenFlow, OVSDB, of-config~\cite{onf2013-1}, NetConf, and SNMP) can co-exist.

Most controllers support only OpenFlow as a southbound API. Still, a few of them, such as OpenDaylight, 
Onix and HP VAN SDN Controller, offer a wider range of southbound APIs and/or protocol plugins. Onix 
supports both the OpenFlow and OVSDB protocols. The HP VAN SDN Controller has other southbound connectors 
such as L2 and L3 agents. OpenDaylight goes a step beyond by providing a Service Layer Abstraction (SLA) 
that allows several southbound APIs and protocols to co-exist in the control platform. For instance, its 
original architecture was designed to support at least seven different protocols and plugins: OpenFlow, OVSDB~\cite{pfaff2013-1}, NETCONF~\cite{enns2011-1}, PCEP~\cite{vasseur2009}, SNMP~\cite{harrington2002}, BGP~\cite{rekhter2006} and LISP Flow Mapping~\cite{opendaylight2013}. Hence, OpenDaylight is one of 
the few control platforms being conceived to support a broader integration of technologies in a single 
control platform.

\vspace{2mm}
\noindent \textit{Eastbound and Westbound}

\begin{figure}[ht]
\centering
\includegraphics[width=0.99\columnwidth]{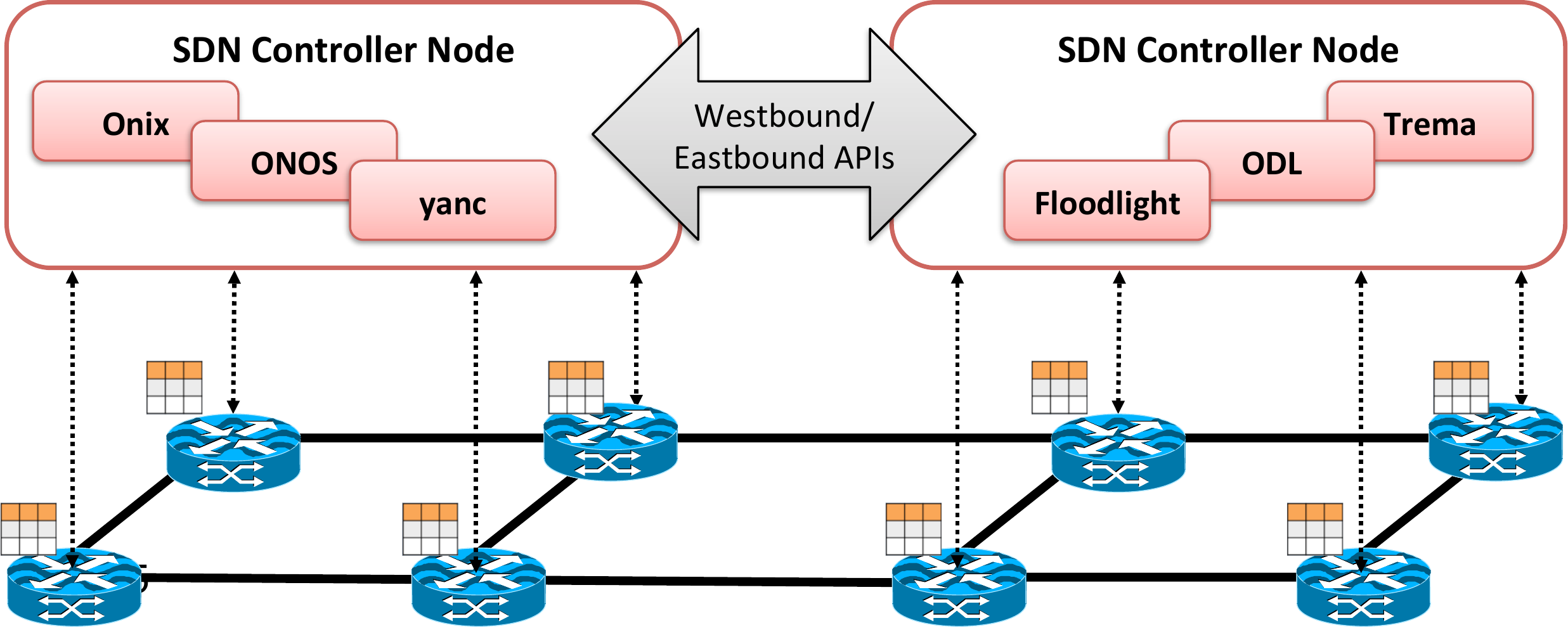}
\caption{Distributed controllers: east/westbound APIs.}
\label{fig:eastwestbounds}
\end{figure}

East/westbound APIs, as illustrated in Figure~\ref{fig:eastwestbounds}, are a special case of interfaces 
required by distributed controllers.
Currently, each controller implements its own east/westbound API. The functions of these interfaces include
import/export data between controllers, algorithms for data consistency models, and monitoring/notification 
capabilities (e.g., check if a controller is up or notify a take over on a set of forwarding devices).

Similarly to southbound and northbound interfaces, east/westbound APIs are essential components of 
distributed controllers. 
To identify and provide common compatibility and interoperability between 
different controllers, it is necessary to have standard east/westbound interfaces. 
For instance, SDNi~\cite{yin2012} defines 
common requirements to coordinate flow setup and exchange reachability information across multiple 
domains. 
In essence, such protocols can be used in an orchestrated and interoperable 
way to create more scalable and dependable distributed control platforms. Interoperability can be 
leveraged to increase the diversity of the control platform element. Indeed, diversity increases the 
system robustness by reducing the probability of common faults, such as software faults~\cite{garcia2013}.

Other proposals that try to define interfaces between controllers include Onix data import/export functions~\cite{koponen-1}, ForCES CE-CE interface~\cite{doria2010,wang2011-1}, 
ForCES Intra-NE cold-standby mechanisms for high availability~\cite{ogawa2013}, and distributed data stores~\cite{botelho2013}. 
An east/westbound API requires advanced data distribution mechanisms such as the Advanced Message Queuing Protocol  (AMQP)~\cite{vinoski2006} used by
DISCO~\cite{phemius2013}, \coloredtext{techniques for distributed concurrent and consistent policy composition~\cite{canini2013}}, transactional databases and DHTs~\cite{ghodsi2006} (as used in Onix~\cite{koponen-1}), or advanced algorithms for strong consistency and fault 
tolerance~\cite{botelho2013,Botelho2014}.

In a multi-domain setup, east/westbound APIs may require also more specific communication protocols 
between SDN domain controllers~\cite{stallings2013}. Some of the essential functions of such 
protocols are to \textit{coordinate flow setup} originated by applications, \textit{exchange reachability 
information} to facilitate inter-SDN routing, \textit{reachability update} to keep the network state 
consistent, among others.

\coloredtext{Another important issue regarding east/westbound interfaces is heterogeneity. 
For instance, besides communicating with peer SDN controllers, controllers may also need to communicate with subordinate controllers (in a hierarchy of controllers) and non-SDN controllers~\cite{ONF2014SDNarch}, as is the case of ClosedFlow~\cite{Hand2014_4}.
To be interoperable, east/westbound interfaces thus need to accommodate different controller interfaces, with their specific set of services, and the diverse characteristics of the underlying infrastructure, including the diversity of technology, the geographic span and scale of the network, and the distinction between WAN and LAN -- potentially across administrative boundaries. 
In those cases, different information has to be exchanged between controllers, including adjacency and capability discovery, topology information (to the extent of the agreed contracts between administrative domains), billing information, among many others~\cite{ONF2014SDNarch}. 

Lastly, a ``SDN compass'' methodology~\cite{sdn-compass} suggests a finer distinction between eastbound and westbound horizontal interfaces, referring to westbound interfaces as SDN-to-SDN protocols and controller APIs while eastbound interfaces would be used to refer to standard protocols used to communicate with legacy network control planes (e.g., PCEP~\cite{vasseur2009}, GMPLS~\cite{mannie2004gmpls}).
}

\vspace{2mm}
\noindent \textit{Northbound}

Current controllers offer a quite broad variety of northbound APIs, such as ad-hoc APIs, RESTful APIs~\cite{richardson2008restful}, 
multi-level programming interfaces, file systems, among other more specialized APIs such as NVP NBAPI~\cite{koponen-1,koponen} and SDMN API~\cite{pentikousis2013}. 
 Section~\ref{sec:layer-north} is devoted to a more detailed discussion on the evolving layer of northbound APIs.  
 A second kind of northbound interfaces are those stemming out of SDN programming languages such as  
Frenetic~\cite{foster2011}, Nettle~\cite{voellmy2011-1},
NetCore~\cite{monsanto2012}, Procera~\cite{voellmy2012}, 
 Pyretic~\cite{monsanto2013}, \coloredtext{NetKAT~\cite{Anderson2014_4} and other query-based languages~\cite{Narayana2014_4}.}  
 Section~\ref{sec:programminglanguages} gives a more detailed 
overview of the several existing programming languages for SDN.


{\renewcommand{\arraystretch}{1.4}
\begin{table*}[!htp]
\caption{Controllers classification}
\label{tab:controllers}
\begin{center}
\footnotesize
\begin{tabularx}{\linewidth}{p{2.94cm}Xp{2.29cm}p{1.47cm}p{0.75cm}p{1.35cm}p{2.1cm}p{0.9cm}}
\hline
\textbf{Name} & \textbf{Architecture} & \textbf{Northbound API} & \textbf{Consistency} & \textbf{Faults} & \textbf{License} & \textbf{Prog. language} & \textbf{Version}\\
\hline
Beacon~\cite{erickson2013}  & centralized multi-threaded  & ad-hoc API & no & no & GPLv2 & Java & v1.0 \\
\hline
DISCO~\cite{phemius2013} & distributed & REST & --- & yes & --- & Java & v1.1\\
\hline
 Fleet~\cite{Matsumoto2014_4} & distributed & ad-hoc & no & no & --- & --- & v1.0 \\
\hline
Floodlight~\cite{openflowhub.org2012} & centralized multi-threaded  & RESTful API & no & no & Apache & Java & v1.1 \\\hline
HP VAN SDN~\cite{hp2013-1} & distributed & RESTful API & weak & yes & --- & Java & v1.0 \\
\hline
HyperFlow~\cite{tootoonchian2010}  & distributed               & --- & weak & yes & --- & C++ & v1.0 \\
\hline
Kandoo~\cite{yeganeh2012}     & hierarchically distributed & --- & no & no & --- & C, C++, Python & v1.0 \\
\hline
Onix~\cite{koponen-1}       & distributed               & NVP NBAPI   & weak, strong & yes & commercial & Python, C & v1.0 \\
\hline
Maestro~\cite{cai2011}    & centralized multi-threaded  & ad-hoc API & no & no & LGPLv2.1 & Java & v1.0 \\
\hline
Meridian~\cite{banikazemi2013} & centralized multi-threaded & extensible API layer & no & no & --- & Java & v1.0\\
\hline
MobileFlow~\cite{pentikousis2013} & --- & SDMN API & --- & --- & --- & --- & v1.2 \\
\hline
MuL~\cite{Saikia:2013:MuL}        & centralized multi-threaded  & multi-level interface & no & no & GPLv2 & C & v1.0 \\
\hline
NOX~\cite{gude2008}       & centralized               & ad-hoc API  & no & no & GPLv3 & C++ & v1.0 \\
\hline
NOX-MT~\cite{tootoonchian2012}     & centralized multi-threaded  & ad-hoc API & no & no & GPLv3 & C++ & v1.0 \\
\hline
NVP Controller~\cite{koponen} & distributed & --- & --- & --- & commercial & --- & --- \\
\hline
OpenContrail~\cite{junipernetworks2013-1} & --- & REST API & no & no & Apache 2.0 & Python, C++, Java & v1.0\\
\hline
OpenDaylight~\cite{opendaylight2013} & distributed & REST, RESTCONF & weak & no & EPL v1.0 & Java & v1.\{0,3\} \\
\hline
ONOS~\cite{krishnaswamy2013} & distributed & RESTful API & weak, strong & yes & --- & Java & v1.0 \\
\hline

PANE~\cite{ferguson2013} & {distributed} & {PANE API} & {yes} & {---} & {---} & {---} & {---} \\
\hline

POX~\cite{mccauley2012}       & centralized               & ad-hoc API  & no & no & GPLv3 & Python & v1.0 \\\hline
ProgrammableFlow \cite{shimonishi2010} & centralized & --- & --- & --- & --- & C & v1.3 \\
\hline

{Rosemary~\cite{shin2014}} & {centralized} & {ad-hoc} & {---} & {---} & {---} & {---} & {v1.0} \\
\hline

Ryu NOS~\cite{nippontelegraphandtelephonecorporation2012} & centralized multi-threaded & ad-hoc API & no & no & Apache 2.0 & Python & v1.\{0,2,3\}\\
\hline

{SMaRtLight~\cite{Botelho2014}} & {distributed} & {RESTful API} & {no} & {no} & {Apache} & {Java} & {v1.0} \\
\hline

SNAC~\cite{appenzeller2011} & centralized & ad-hoc API & no & no & GPL & C++ & v1.0 \\
\hline

Trema~\cite{takamiya2012}      & centralized multi-threaded  & ad-hoc API & no & no & GPLv2 & C, Ruby & v1.0 \\
\hline
\textit{Unified Controller}~\cite{racherla2014} & --- & REST API & --- & --- & commercial & --- & v1.0 \\
\hline
\textit{yanc}~\cite{monaco2013} & distributed & file system & --- & --- & --- & --- & --- \\
\hline

\end{tabularx}
\end{center}
\end{table*}
}

\vspace{2mm}
\noindent \textit{Wrapping up remarks and platforms comparison}

Table~\ref{tab:controllers} shows a summary of some of the existing controllers with their respective 
architectures and characteristics.
As can be observed, most controllers are centralized and multi-threaded.
Curiously, the northbound API is very diverse. In particular, five controllers (Onix, Floodlight, MuL, Meridian and SDN Unified Controller) pay a bit more attention to this interface, as a statement
of its importance. Consistency models and fault tolerance are only present in Onix, HyperFlow, HP VAN SDN,  ONOS and SMaRtLight. Lastly, when it comes to the OpenFlow standard as southbound API, only Ryu supports its three 
major versions (v1.0, v1.2 and v1.3).

To conclude, it is important to emphasize that the control platform is one of the critical points for 
the success of SDN~\cite{casemore2012-1}.
One of the main issues that needs to be addressed in this respect is interoperability. 
This is rather interesting, as it was the very first problem that southbound APIs (such as OpenFlow) tried to solve.
For instance, while WiFi and LTE (Long-Term Evolution) networks~\cite{Kwan2010survey} need specialized control platforms such as MobileFlow~\cite{pentikousis2013} or SoftRAN~\cite{gudipati2013}, 
data center networks have different requirements that can be met with platforms such as Onix~\cite{koponen-1} or OpenDaylight~\cite{opendaylight2013}. 
For this reason, in environments where diversity of networking infrastructures is a reality, coordination and cooperation between different controllers is crucial. 
Standardized APIs for multi-controller and multi-domain deployments are therefore seen as an important step to achieve this goal.

\subsection{Layer V: Northbound Interfaces}
\label{sec:layer-north}

The North- and Southbound interfaces are two key abstractions of the SDN ecosystem. 
The southbound interface has already a widely accepted proposal (OpenFlow), but a common northbound 
interface is still an open issue.
At this moment it may still be a bit too early to define a standard northbound interface, as use-cases are still being worked out~\cite{dix2013}.
Anyway, it is to be expected a common (or a \emph{de facto}) northbound interface to arise as SDN evolves.
An abstraction that would allow network applications not to depend on specific implementations is important to explore the full potential of SDN 

The northbound interface is mostly a software ecosystem, not a hardware one as is the case of the southbound APIs.
In these ecosystems, the implementation is commonly the forefront driver, while standards emerge later and 
are essentially driven by wide adoption~\cite{guis2012}. Nevertheless, an initial and 
minimal standard for northbound interfaces can still play an important role for the future of SDN. 
Discussions about this issue have already begun ~\cite{dix2013,guis2012,salisbury2012-1,ferro2012,casemore2012,pepelnjak2012,johnson2012,little2013-1}, and a common consensus is that northbound APIs are indeed important but that it is indeed too early to define a single standard right now.
The experience from the development of different controllers will certainly be the basis for coming up with a common application level interface.

Open and standard northbound interfaces are crucial to promote application portability and interoperability among the different the control platforms.
A northbound API can be compared to the POSIX standard~\cite{posixieee} in operating systems, representing an abstraction that guarantees programming language and controller independence.
NOSIX~\cite{wundsam2012} is one of the first examples of an effort in this direction. It tries to 
define portable low-level (e.g., flow model) application interfaces, making southbound APIs such as OpenFlow 
look like ``device drivers''. However, NOSIX is not exactly a general purpose northbound interface, but 
rather a higher-level abstraction for southbound interfaces. Indeed, it could be part of the common 
abstraction layer in a control platform as the one described in Section~\ref{sec:controllers}.

Existing controllers such as Floodlight, Trema, NOX, Onix, and OpenDaylight propose and define their own northbound 
APIs~\cite{salisbury2012-1,chua2012}.
However, each of them has its own specific definitions.
Programming languages such as Frenetic~\cite{foster2011}, Nettle~\cite{voellmy2011-1}, NetCore~\cite{monsanto2012}, Procera~\cite{voellmy2012}, Pyretic~\cite{reich2013} and \coloredtext{NetKAT~\cite{Anderson2014_4}} also abstract the inner details of the controller functions and data plane behavior from the application developers.
Moreover, as we explain in 
Section~\ref{sec:programminglanguages}, programming languages can provide a wide range of powerful 
abstractions and mechanisms such as application composition, transparent data plane fault tolerance, 
and a variety of basic building blocks to ease software module and application development.

SFNet~\cite{yap2010} is another example of a northbound interface. It is a high-level API that 
translates application requirements into lower level service requests. However, SFNet has a limited scope, 
targeting queries to request the congestion state of the network and services such as bandwidth reservation 
and multicast.

Other proposals use different approaches to allow applications to 
interact with controllers. The \textit{yanc} control platform~\cite{monaco2013} explores 
this idea by proposing a general control platform based on Linux and abstractions such as the virtual file 
system (VFS). This approach simplifies the development of SDN applications as programmers are able to use
a traditional concept (files) to communicate with lower level devices and sub-systems.
 
Eventually, it is unlikely that a single northbound interface emerges as the winner, as the requirements for 
different network applications are quite different. APIs for security applications are likely to be different 
from those for routing or financial applications. One possible path of evolution for northbound APIs are 
vertically-oriented proposals, before any type of standardization occurs, a challenge the ONF has started 
to undertake \coloredtext{in the NBI WG in parallel to open-source SDN developments~\cite{floss-meets-sdn}. 
The ONF architectural work~\cite{ONF2014SDNarch} includes the possibility of northbound APIs providing resources to enable dynamic and granular control of the network resources from customer applications, eventually across different business and organizational boundaries.} 

\coloredtext{There are also other kind of APIs, such as those provided by the PANE controller~\cite{ferguson2013}.
Designed to be suitable for the concept of participatory networking, PANE allows network administrators to define module-specific quotas and access control policies on network resources. 
The controller provides an API that allows end-host applications to dynamically and autonomously request network resources. 
For example, audio (e.g., VoIP) and video applications can easily be modified to use the PANE API to reserve bandwidth for certain quality guarantees during the communication session. 
PANE includes a compiler and verification engine to ensure that bandwidth requests do not exceed the limits set by the administrator and to avoid starvation, i.e., other applications shall not be impaired by new resource requests.
}


\subsection{Layer VI: Language-based Virtualization}
\label{sec:virtualizationslicing}

Two essential characteristics of virtualization solutions are the capability of expressing modularity 
and of allowing different levels of abstractions while still guaranteeing desired properties such as protection.
For instance, virtualization techniques can allow different views of a single physical infrastructure.
As an example, one virtual ``big switch'' could represent a combination of several underlying forwarding devices.
This intrinsically simplifies the task of application developers as they do not need to think about the sequence of switches where forwarding rules have to be installed, but rather see the network as a 
simple ``big switch''.
Such kind of abstraction significantly simplify the development 
and deployment of complex network applications, such as advanced security related services.

Pyretic~\cite{reich2013} is an interesting example of a programming language that offers this type of high-level abstraction of network topology.
It incorporates this concept of abstraction by introducing network objects.
These objects consist of an abstract network topology and the sets of policies applied to it.
Network objects simultaneously hide information and offer the required services.


Another form of language-based virtualization is static slicing. 
This a scheme where the network is sliced by a compiler, based on application layer definitions.
The output of the compiler is a monolithic control program that has already slicing definitions and configuration commands for the network.
In such a case, there is no need for a hypervisor to dynamically manage the network slices.
Static slicing can be valuable for deployments with specific requirements, in particular those where higher performance and simple isolation guarantees are preferrable to
dynamic slicing.

One example of static slicing approach it the Splendid isolation~\cite{gutz2012}.
In this solution the network slices are made of 3 components:
(a) \textit{topology}, consisting of switches, ports, and links;
(b) \textit{mapping} of slice-level switches, ports and links on the network infrastructure;
(c) \textit{predicates on packets}, where each port of the slice's edge switches has an associated predicate.
The topology is a simple graph of the sliced nodes, ports and links.
Mapping will translate the abstract topology elements into the corresponding physical ones.
The predicates are used to indicate whether a packet is permitted or not to enter a specific slice. 
Different applications can be associated to each slice.
The compiler takes the combination of slices (topology, mapping, and predicates) and respective programs to generate a global configuration for the entire network. 
It also ensures that properties such as isolation are enforced among slices, i.e., no packets of a slice A can traverse to a slice B unless 
explicitly allowed.

Other solutions, such as libNetVirt~\cite{turull2012}, try to integrate heterogeneous technologies for creating 
static network slices. libNetVirt is a library designed to provide a flexible way to create and manage virtual 
networks in different computing environments. Its main idea is similar to the OpenStack Quantum 
project~\cite{quantumcommunicty2012}. While Quantum is designed for OpenStack (cloud environments), 
libNetVirt is a more general purpose library which can be used in different environments. Additionally, 
it goes one step beyond OpenStack Quantum by enabling QoS capabilities in virtual networks~\cite{turull2012}.
The libNetVirt library has two layers: (1) a generic network interface; and (2) technology specific device drivers 
(e.g., VPN, MPLS, OpenFlow). On top of the layers are the \manapps and virtual network descriptions.
The OpenFlow driver uses a NOX controller to manage the underlying infrastructure, using OpenFlow rule-based flow 
tables to create isolated virtual networks. By supporting different technologies, it can be used as a bridging 
component in heterogeneous networks.

Table~\ref{tab:virtualizationsolutions} summarizes the hypervisor and non-hypervisor based virtualization 
technologies. 
As can be observed, only libNetVirt supports heterogeneous technologies, not restricting its 
application to OpenFlow-enabled networks. FlowVisor, AutoSlice and OpenVirteX allow multiple controllers, 
one per network slice. FlowN provides a container-based approach where multiple applications from different 
users can co-exist on a single controller. FlowVisor allows QoS provisioning guarantees by using VLAN PCP bits for priority queues.
SDN VE and NVP also provide their own provisioning methods for guaranteeing QoS.


{\renewcommand{\arraystretch}{1.4}
\begin{table*}[!htp]
\caption{Virtualization solutions}
\label{tab:virtualizationsolutions}
\begin{center}
\footnotesize
\begin{tabularx}{\textwidth}{p{3.8cm}p{3.1cm}p{3.1cm}p{2.6cm}X}
\hline
\textbf{Solution} & \textbf{Multiple controllers} & \textbf{Slicing} & \textbf{QoS ``guarantees''} & \textbf{Multi-technology} \\\hline

{AutoVFlow~\cite{Yamanaka2014}} & {yes, one per tenant} & {flow space virtualization} & {no}  & {no, OF only} \\
\hline

AutoSlice~\cite{bozakov2012} & yes, one per slice & VLAN tags & no & no, OF only\\\hline

{Compositional Hypervisor~\cite{Jin2014_4}} & {---} & {---} & {no}  & {no, OF only} \\
\hline

FlowVisor~\cite{sherwood2009,azodolmolky2012} & yes, one per slice & virtual flow tables per slice &  yes (VLAN PCP bits) & no, OF only \\\hline
FlowN~\cite{drutskoy2012,drutskoy2013} & no (contained applications) & VLAN tags  & no & no, OF only \\\hline
IBM SDN VE~\cite{racherla2014} & yes, a cluster of controllers & logical datapaths & yes (priority-based) & yes (VXLAN, OVS, OF) \\\hline
libNetVirt~\cite{turull2012} & no, one single controller & VLAN tags & no  & yes (e.g., VPN, MPLS, OF)\\\hline
NVP's Hypervisor~\cite{koponen} & yes, a cluster of controller & logical datapaths & yes & no, OVS only \\\hline
OpenVirteX~\cite{Al_Shabibi2014_4} & yes, one per slice & virtual flow tables per slice & \textit{unknown}  & no, OF only \\\hline 
Pyretic~\cite{reich2013} & no, one single controller & compiler time OF rules & no & no, OF only \\\hline
Splendid Isolation~\cite{gutz2012} & no, one single controller & compiler time VLANs & no  & no, OF only \\\hline

{RadioVisor~\cite{Gudipati2014_4}} & {yes, one per slice} & {3D resource grid slices} & {---}  & {Femto API~\cite{smallcellforum2013}} \\
\hline

{xDPd virtualization~\cite{doriguzzicorin2014}} & {yes, one per slice} & {flow space virtualization} & {no}  & {no, OF only} \\
\hline

\end{tabularx}
\end{center}
\end{table*}
}

\subsection{Layer VII: Programming languages}
\label{sec:programminglanguages}

Programming languages have been proliferating for decades. Both academia and industry have evolved 
from low-level hardware-specific machine languages, such as assembly for x86 architectures, to high-level 
and powerful programming languages such as Java and Python. The advancements towards more portable and 
reusable code has driven a significant shift on the computer industry~\cite{guzdial2008,farooq2014}.

Similarly, programmability in networks is starting to move from low level machine languages such as OpenFlow (``assembly'') to high-level programming languages~\cite{foster2011,hinrichs2009,voellmy2011-1,monsanto2012,voellmy2012,monsanto2013,koponen}.
Assembly-like machine languages, such as OpenFlow~\cite{mckeown2008} and POF~\cite{song2013,song2013-1}, essentially mimic the behavior of forwarding devices, forcing developers to spend too much time on low-level details
rather than on the problem solve.
Raw OpenFlow programs have to deal with hardware behavior details such as overlapping rules, the priority ordering of rules, and data-plane inconsistencies that arise from in-flight packets whose flow rules are under installation~\cite{foster2011,monsanto2012,ferguson2012}.
The use of these low-level languages makes it difficult to reuse software, to create modular and extensive code, and leads to a more error-prone development process~\cite{monsanto2013,nelson2014,katta2012}.

Abstractions provided by high level programming languages can significantly help address many 
of the challenges of these lower-level instruction sets~\cite{foster2011,hinrichs2009,voellmy2011-1,monsanto2012,voellmy2012,monsanto2013}.
In SDNs, high-level programming languages can be designed and used to:
\begin{enumerate}
\item create higher level abstractions for simplifying the task of programming forwarding devices;
\item enable more productive and problem-focused environments for network software programmers, speeding up development and innovation;
\item promote software modularization and code reusability in the network control plane;
\item foster the development of network virtualization.
\end{enumerate}

Several challenges can be better addressed by programming languages in SDNs.
For instance, in pure OpenFlow-based SDNs, it is hard to ensure that multiple tasks of a single application 
(e.g., routing, monitoring, access control) do not interfere with each other.
For example, rules generated for one task should not override the functionality of another task~\cite{foster2011,ferguson2012}.
Another example is when multiple applications run on a single controller~\cite{monsanto2013,ferguson2012,porras2012,shin2013-1,son2013}.
Typically, each application generates rules based on its own needs and policies without further knowledge about the rules generated by other applications. As a consequence, conflicting rules can be generated and installed in forwarding devices, which can create problems for network operation.
Programming languages and runtime systems can help to solve these problems that would be otherwise hard to prevent.

Important software design techniques such as code modularity and reusability are very hard to achieve using low-level programming models~\cite{monsanto2013}.
Applications thus built are monolithic and consist of building blocks that can not be reused in other applications.
The end result is a very time consuming and error prone development process.

Another interesting feature that programming language abstractions provide is the capability of 
creating and writing programs for virtual network topologies~\cite{reich2013,gutz2012}.
This concept is similar to object-oriented programming, where objects abstract both data and specific functions 
for application developers, making it easier to focus on solving a particular problem without worrying about 
data structures and their management. For instance, in an SDN context, instead of generating and installing 
rules in each forwarding device, one can think of creating simplified virtual network topologies that represent 
the entire network, or a subset of it. For example, the application developer should be able to abstract the 
network as an atomic big switch, rather than a combination of several underlying physical devices.
The programming languages or runtime systems should be responsible for generating and installing the lower-level 
instructions required at each forwarding device to enforce the user policy across the network. With such kind of 
abstractions, developing a routing application becomes a straightforward process. Similarly, a single physical 
switch could be represented as a set of virtual switches, each of them belonging to a different virtual network.
These two examples of abstract network topologies would be much harder to implement with low-level 
instruction sets. In contrast, a programming language or runtime system can more easily provide abstractions 
for virtual network topologies, as has already been demonstrated by languages such as Pyretic~\cite{reich2013}.

\vspace{2mm}
\noindent \textit{High-level SDN programming languages}

\coloredtext{High-level  programming languages can be powerful tools as a mean for implementing and providing abstractions for different important properties and functions of SDN such as network-wide structures, distributed updates, modular composition, virtualization, and formal verification~\cite{Casado2014_4}.
}

Low-level instruction sets suffer from several problems.
To address some of these challenges, higher-level programming languages have been proposed, with 
diverse goals, such as:
\begin{itemize}
\item Avoiding low-level and device-specific configurations and dependencies spread across the network, as happens
in traditional network configuration approaches; 
\item Providing abstractions that allow different management tasks to be accomplished through easy to understand and maintain network policies;
\item Decoupling of multiple tasks (e.g., routing, access control, traffic engineering);
\item Implementing higher-level programming interfaces to avoid low-level instruction sets;
\item Solving forwarding rules problems, e.g., conflicting or incomplete rules that can prevent a switch event to be triggered, in an automated way;
\item Addressing different race condition issues which are inherent to distributed systems;
\item Enhancing conflict-resolution techniques on environments with distributed decision makers;
\item Provide native fault-tolerance capabilities on data plane path setup;
\item Reducing the latency in the processing of new flows;
\item Easing the creation of stateful applications (e.g., stateful firewall).
\end{itemize}

Programming languages can also provide specialized abstractions to cope with other management requirements, 
such as monitoring\coloredtext{~\cite{voellmy2012,foster2011,tootoonchian2010-1,Narayana2014_4}}.
For instance, the runtime system of a programming language can do all the ``laundry work'' of installing rules, polling 
the counters, receiving the responses, combining the results as needed, and composing monitoring queries in 
conjunction with other policies. Consequently, application developers can take advantage of the simplicity and 
power of higher level query instructions to easily implement monitoring modules or applications.

Another aspect of paramount importance is the portability of the programming language, necessary so that developers do not 
need to re-implement applications for different control platforms.
The portability of a programming language can be considered as a significant added value to the control plane ecosystem. 
Mechanisms such as decoupled back-ends could be key architectural ingredients to enable platform portability.
Similarly to the Java virtual machine, a portable northbound interface will easily allow applications to run on different controllers without requiring any modification.
As an example, the Pyretic language requires only a standard socket interface and a simple OpenFlow client on the target controller 
platform~\cite{monsanto2013}.

Several programming languages have been proposed for SDNs, as summarized in Table~\ref{tab:programminglanguages}.
The great majority propose abstractions for OpenFlow-enabled networks.
The predominant programming paradigm is the declarative one, with a single exception, Pyretic, which is an imperative language. 
Most declarative languages are functional, while but there are instances of the logic and reactive types.
The purpose -- i.e., the specific problems they intend to solve -- and the expressiveness power vary from language to 
language, while the end goal is almost always the same: to provide higher-level abstractions to facilitate the 
development of network control logic.

{\renewcommand{\arraystretch}{1.4}
\begin{table*}[!htp]
\caption{Programming languages}
\label{tab:programminglanguages}
\begin{center}
\footnotesize
\begin{tabularx}{\linewidth}{p{2cm}p{4cm}X}
\hline
\textbf{Name} & \textbf{Programming paradigm} & \textbf{Short description/purpose} \\\hline
FatTire~\cite{reitblatt2013}  & declarative (functional) & Uses regular expressions to allow programmers to describe network paths and respective fault-tolerance requirements. \\\hline
Flog~\cite{katta2012}  & declarative (logic), event-driven & Combines ideas of FML and Frenetic, providing an event-driven and forward-chaining logic programming language. \\\hline
FlowLog~\cite{nelson2014} & declarative (functional) & Provides a finite-state language to allow different analysis, such as model-checking.\\\hline
FML~\cite{hinrichs2009} & declarative (dataflow, reactive) & High level policy description language (e.g., access control).  \\\hline
Frenetic~\cite{foster2011} & declarative (functional) & Language designed to avoid race conditions through well defined high level programming abstractions.  \\\hline
HFT~\cite{ferguson2012}  & declarative (logic, functional) & Enables hierarchical policies description with conflict-resolution operators, well suited for decentralized decision makers. \\\hline
Maple~\cite{voellmy2013} & declarative (functional) & Provides a highly-efficient multi-core scheduler that can scale efficiently to controllers with 40+ cores. \\\hline
Merlin~\cite{soule2013} & declarative (logic) & Provides mechanisms for delegating management of sub-policies to tenants without violating global constraints. \\\hline
nlog~\cite{koponen} & declarative (functional) &  Provides mechanisms for data log queries over a number of tables. Produces immutable tuples for reliable detection and propagation of updates. \\\hline

NetCore~\cite{monsanto2012}  & declarative (functional) & High level programming language that provides means for expressing packet-forwarding policies in a high level.  \\\hline

{NetKAT~\cite{Anderson2014_4}}  & {declarative (functional)} & {It is based on Kleene algebra for reasoning about network structure and supported by solid foundation on equational theory.} \\\hline

Nettle~\cite{voellmy2011-1}   & declarative (functional, reactive) & Based on functional reactive programming  principles in order to allow programmers to deal with streams instead of events.   \\\hline

Procera~\cite{voellmy2012}  & declarative (functional, reactive) & Incorporates a set of high level abstractions to make it easier to describe reactive and temporal behaviors.  \\\hline
Pyretic~\cite{monsanto2013}  & imperative & Specifies network policies at a high level of abstraction, offering transparent composition and topology mapping. \\\hline
\end{tabularx}
\end{center}
\end{table*}
}

Programming languages such as FML~\cite{hinrichs2009}, 
Nettle~\cite{voellmy2011-1}, and Procera~\cite{voellmy2012}
are functional and reactive.
Policies and applications written in these languages are based on reactive actions triggered by events (e.g., a new host connected to the network, or the current network load).
Such languages allow users to declaratively express different network configuration rules such as access control lists (ACLs), virtual LANs (VLANs), and many others.
Rules are essentially expressed as allow-or-deny policies, which 
are applied to the forwarding elements to ensure the desired network behavior.

Other SDN programming languages such as Frenetic~\cite{foster2011},
Hierarchical Flow Tables (HFT)~\cite{ferguson2012}, NetCore~\cite{monsanto2012}, and
Pyretic~\cite{monsanto2013}, were designed with the simultaneous goal of efficiently expressing packet-forwarding policies and dealing with overlapping rules of different applications, offering advanced operators for parallel and sequential composition of software modules.
To avoid overlapping conflicts, Frenetic disambiguates rules 
with overlapping patterns by assigning different integer priorities, while HFT uses hierarchical policies with 
enhanced conflict-resolution operators.

\textit{See-every-packet} abstractions and race-free semantics also represent interesting features provided 
by programming languages (such as Frenetic~\cite{foster2011}). 
The former ensures that \emph{all} control packets will be available for analysis, sooner or later, while the latter provides the mechanisms for
suppressing unimportant packets. As an example, packets that arise from a network race condition, such as due to a concurrent flow 
rule installation on switches, can be simply discarded by the runtime system.

Advanced operators for parallel and sequential composition help bind (through internal workflow operators) 
the key characteristics of programming languages such as Pyretic~\cite{monsanto2013}. Parallel 
composition makes it possible to operate multiple policies on the same set of packets, while sequential 
composition facilitates the definition of a sequential workflow of policies to be processed on a set of 
packets. Sequential policy processing allows multiple modules (e.g., access control and 
routing) to operate in a cooperative way.
By using sequential composition complex applications can be built out of a combination of different modules (in a similar way as pipes can be used to build sophisticated Unix applications).

Further advanced features are provided by other SDN programming languages.
FatTire~\cite{reitblatt2013} is an example of a declarative language that heavily relies on regular 
expressions to allow programmers to describe network paths with fault-tolerance requirements. For instance, 
each flow can have its own alternative paths for dealing with failure of the primary paths. 
Interestingly, this feature is provided in a very programmer-friendly way, with the application programmer having only to use regular 
expressions with special characters, such as an asterisk. In the particular case of FatTire, an asterisk will produce the same behavior 
as a traditional regular expression, but translated into alternative traversing paths.

Programming languages such as FlowLog~\cite{nelson2014} and Flog~\cite{katta2012} bring 
different features, such as model checking, dynamic verification and stateful middleboxes. For instance, using a programming language such as Flog, it is possible to build a 
stateful firewall application with only five lines of code~\cite{katta2012}.

Merlin~\cite{soule2013} is one of the first examples of unified framework for controlling different network components, such as forwarding devices, middleboxes, and end-hosts.
An important advantage is backward-compatibility with existing systems.
To achieve this goal, Merlin generates specific code for each type of component.
Taking a policy definition as input, Merlin's compiler determines forwarding 
paths, transformation placement, and bandwidth allocation.
The compiled outputs are sets of component-specific 
low-level instructions to be installed in the devices.
Merlin's policy language also allows operators to delegate the control of a sub-network to tenants, while ensuring isolation.
This delegated control is expressed by means of policies that can be further refined 
by each tenant owner, allowing them to customize policies for their particular needs. 

Other recent initiatives (e.g., systems programming languages~\cite{casey2013}) 
target problems such as detecting anomalies to improve the security of network protocols (e.g., OpenFlow), 
and optimizing horizontal scalability for achieving high throughput in applications running on multicore 
architectures~\cite{voellmy2013}.
\coloredtext{Nevertheless, there is still scope for further investigation and development on programming languages. 
For instance, one recent research has revealed that current policy compilers generate unnecessary (redundant) rule updates, most of which modify only the priority field~\cite{Wen2014_4}.
}

Most of the value of SDN will come from the network managements applications built on top of the infrastructure.
Advances in high-level programming languages are a fundamental component to the success of a prolific SDN 
application development ecosystem. To this end, efforts are undergoing to shape forthcoming standard 
interfaces (cf.~\cite{kuzniar2013}) and towards the realization of integrated development 
environments (e.g., NetIDE~\cite{facca2013}) with the goal of fostering the development of a myriad of SDN applications.
We discuss these next.

\subsection{Layer VIII: \ManaApps}

\Manaaapps can be seen as the ``network brains''. 
They implement the control-logic that will be translated into commands to be installed in the data plane, dictating the behavior of the forwarding devices.
Take a simple application as routing as an example. 
The logic of this application is to define the path through which packets will flow from 
a point A to a point B.
To achieve this goal a routing application has to, based on the topology input, decide on the path to use and instruct the controller to install the 
respective forwarding rules in all forwarding devices on the chosen path, from A to B.

Software-defined networks can be deployed on any traditional network environment, from home and enterprise networks to data centers and Internet exchange points.
Such variety of environments has led to a wide array of \manapps.
Existing \manapps perform traditional functionality such as routing, load balancing, and security policy enforcement, but also explore novel approaches, such as reducing power consumption. 
Other examples include fail-over and reliability functionalities to the data plane, end-to-end QoS enforcement, network virtualization, mobility management in wireless networks, among many others.
\coloredtext{The variety of \manapps, combined with real use case deployments, is expected to be one of the major forces on fostering a broad adoption of SDN~\cite{Reinecke2014}.}

%

Despite the wide variety of use cases, most SDN applications can be grouped in one of five categories: traffic engineering, mobility and wireless, measurement and monitoring, security and dependability and data center networking.
Tables~\ref{tab:managementapplications1} and~\ref{tab:managementapplications2} summarize several applications categorized as such, stating their main purpose, controller where it was implemented/evaluated, and southbound API used.

{\renewcommand{\arraystretch}{1.4}
\begin{table*}[!htp]
\caption{\ManaApps}
\label{tab:managementapplications1}
\begin{center}
\footnotesize
\begin{tabularx}{\linewidth}{p{2cm}p{3.2cm}p{5.1cm}Xp{3.3cm}}
\hline
\textbf{Group} & \textbf{Solution/Application} & \textbf{Main purpose} & \textbf{Controller} & \textbf{Southbound API} \\
\hline
\multirow{20}{*}{\begin{minipage}{2cm}Traffic \\engineering\end{minipage}} 
& ALTO VPN~\cite{scharf2013} & on-demand VPNs & NMS~\cite{stiemerling2014ALTODeployment,alimi2013} & SNMP \\
& Aster*x~\cite{handigol2009}       & load balancing             &  NOX & OpenFlow      \\
& ElasticTree~\cite{heller2010}   & energy aware routing       &  NOX & OpenFlow      \\

& {FlowQoS~\cite{Seddiki2014_4}} & {QoS for broadband access networks} & {POX} & {OpenFlow}\\

& Hedera~\cite{al-fares2010}        & scheduling / optimization  &  --- & OpenFlow  \\
& In-packet Bloom filter~\cite{macapuna2010}        & load balancing  &  NOX & OpenFlow  \\

& {MicroTE~\cite{benson2011mTE}} & {traffic engineering with minimal overhead} & {NOX} & {OpenFlow}\\

& {Middlepipes~\cite{Jamjoom2014_4}} & {Middleboxes as a PaaS} & {middlepipe controller} & {---}\\

& OpenQoS~\cite{egilmez2012} & dynamic QoS routing for multimedia apps & Floodlight  & OpenFlow \\

& {OSP~\cite{sgambelluri2013opt}} & {fast recovery through fast-failover groups} & {NOX} & {OpenFlow}\\

& {PolicyCop~\cite{bari2013pc}} & {QoS policy management framework} & {Floodlight} & {OpenFlow}\\

& {ProCel~\cite{Nagaraj2014_4}} & {Efficient traffic handling for software EPC} & {ProCel controller} & {---}\\

& {Pronto~\cite{Xiong2014pronto,Xiong2014pronto2}} & {Efficient queries on distributed data stores} & {Beacon} & {OpenFlow}\\

& Plug-n-Serve~\cite{handigol2009-1}  & load balancing             &  NOX & OpenFlow      \\
& QNOX~\cite{jeong2012}          & QoS enforcement                &  NOX & Generalized OpenFlow \\

& {QoS for SDN~\cite{Sharma2014}} & {QoS over heterogeneous networks} & {Floodlight} & {OpenFlow}\\

& QoS framework~\cite{kim2010} & QoS enforcement                &  NOX & OF with QoS extensions \\
& QoSFlow~\cite{ishimori2013} & multiple packet schedulers to improve QoS & --- & OpenFlow\\

& {QueuePusher~\cite{Palma2014}} & {Queue management for QoS enforcement} & {Floodlight} & {OpenFlow}\\

& SIMPLE~\cite{qazi2013-1}  & middlebox-specific ``traffic steering'' & Extended POX & OpenFlow  \\
& ViAggre SDN~\cite{skoldstrom2013-1} & divide and spread forwarding tables & NOX & OpenFlow\\

\hline
\multirow{10}{*}{\begin{minipage}{2cm}Mobility \\\& \\Wireless\end{minipage}} 
& AeroFlux~\cite{schulz-zander2014-ons,schulz-zander2014-hotsdn} & scalable hierarchical WLAN control plane & Floodlight & OpenFlow, Odin \\
& CROWD~\cite{ali-ahmad2013} & overlapping of LTE and WLAN cells & --- & OpenFlow \\
& CloudMAC~\cite{vestin2013} & outsourced processing of WLAN MACs & --- & OpenFlow \\

& {C-RAN~\cite{Dawson2014}} & {RAN~\cite{gudipati2013} virtualization for mobile nets} & {---} & {---} \\

& FAMS~\cite{yamasaki2011} & flexible VLAN system based on OpenFlow & ProgrammableFlow & OpenFlow \\
& MobileFlow~\cite{pentikousis2013} & flow-based model for mobile networks & MobileFlow & SDMN API\\
& Odin~\cite{schulz-zander2014-atc} & programmable virtualized WLANs & Floodlight & OpenFlow, Odin \\
& OpenRAN~\cite{yang2013} & vertical programmability and virtualization & --- & --- \\
& OpenRoads~\cite{yap2010-1} & control of the data path using OpenFlow & FlowVisor & OpenFlow \\
& SoftRAN~\cite{gudipati2013} & load balancing and interference management & --- & Femto API~\cite{smallcellforum2013,Chandrasekhar2008} \\

\hline
\multirow{10}{*}{\begin{minipage}{2cm}Measurement \\\& \\Monitoring\end{minipage}} 
& BISmark~\cite{kim2013} & active and passive measurements & Procera framework & OpenFlow \\

& {DCM~\cite{Yu2014_4}} & {distributed and coactive traffic monitoring} & {DCM controller} & {OpenFlow} \\

& FleXam~\cite{shirali-shahreza2013} & flexible sampling extension for OpenFlow & --- & --- \\
& FlowSense~\cite{yu2013} & measure link utilization in OF networks & --- & OpenFlow \\
& measurement model~\cite{jose2011} & model for OF switch measurement tasks & --- & OpenFlow \\

& OpenNetMon~\cite{niels2014open} & monitoring of QoS parameters to improve TE & POX & OpenFlow \\

& {OpenSample~\cite{suh2014os}} & {low-latency sampling-based measurements} & {Floodlight} & {modified sFlow~\cite{sflow.orgforum2012}} \\

& OpenSketch\,\cite{yu2013-1} & separated measurement data plane & OpenSketch & ``OpenSketch sketches'' \\
& OpenTM~\cite{tootoonchian2010-1} & traffic matrix estimation tool & NOX & OpenFlow \\
& PaFloMon\,\cite{argyropoulos2012} & passive monitoring tools defined by users & FlowVisor & OpenFlow \\

& {PayLess~\cite{chowdhury2014payless}} & {query-based real-time monitoring framework} & {Floodlight} & {OpenFlow} \\

\hline
\multirow{7}{*}{\begin{minipage}{2cm}Data Center \\Networking\end{minipage}} 
& Big Data Apps ~\cite{wang2012} & optimize network utilization & --- & OpenFlow   \\
& CloudNaaS~\cite{benson2011} & networking primitives for cloud applications & NOX &  OpenFlow  \\
& FlowComb~\cite{das2013} & predicts application workloads & NOX & OpenFlow \\
& FlowDiff~\cite{arefin2013} & detects operational problems & FlowVisor & OpenFlow \\
& LIME~\cite{keller2012} & live network migration & Floodlight & OpenFlow      \\
& NetGraph ~\cite{raghavendra2012} & graph queries for network management & --- & OpenFlow, SNMP   \\
& OpenTCP~\cite{ghobadi2013} & dynamic and programmable TCP adaptation & --- & --- \\

\hline

\end{tabularx}
\end{center}
\end{table*}
}

{\renewcommand{\arraystretch}{1.4}
\begin{table*}[!htp]
\caption{\ManaApps}
\label{tab:managementapplications2}
\begin{center}
\footnotesize
\begin{tabularx}{\linewidth}{p{2cm}p{3.2cm}p{5.1cm}Xp{3.3cm}}
\hline
\textbf{Group} & \textbf{Solution/Application} & \textbf{Main purpose} & \textbf{Controller} & \textbf{Southbound API} \\
\hline
\multirow{20}{*}{\begin{minipage}{2cm}Security \\\& \\Dependability\end{minipage}} 
& Active security~\cite{hand2013} & integrated security using feedback control & Floodlight & OpenFlow \\
& AVANT-GUARD~\cite{shin2013-3} & DoS security specific extensions to OF & POX & OpenFlow \\
& CloudWatcher~\cite{shin2012}  & framework for monitoring clouds & NOX & OpenFlow      \\

& Cognition~\cite{tantar2014evolve} & cognitive functions to enhanced security mechanisms in network applications & {---} & {---} \\

& DDoS detection~\cite{braga2010-1}     & attacks detection and mitigation    &  NOX & OpenFlow      \\
& Elastic IP \& Security~\cite{stabler2012}  & SDN-based elastic IP and security groups & NOX & OpenFlow   \\
& Ethane~\cite{casado2007-1}   & flow-rule enforcement (match/action)&  Ethane controller & first instance of OpenFlow  \\

& {FlowNAC~\cite{Matias2014}} & {flow-based network access control} & {NOX} & {OpenFlow} \\

& FortNOX ~\cite{porras2012}    & security flow rules prioritization &  NOX & OpenFlow      \\
& FRESCO ~\cite{shin2013-1}    & framework for security services composition & NOX & OpenFlow      \\
& LiveSec~\cite{wang2012-1}       & security policy enforcement          &  NOX & OpenFlow      \\

& {MAPPER~\cite{Sapio2014}} & {fine-grained access control} & {---} & {---}\\

& NetFuse~\cite{wang2013} & protection against OF traffic overload & --- & OpenFlow \\
& OF-RHM~\cite{jafarian2012}    & random host mutation (defense) &  NOX & OpenFlow      \\
& OpenSAFE~\cite{ballard2010} & direct spanned net traffic in arbitrary ways & NOX & OpenFlow \\

& OrchSec~\cite{zaalouk2014} & architecture for developing security apps & \textit{any} & Flow-RT~\cite{sflowrt2014}, OpenFlow\\

& Reliable multicasting~\cite{kotani2012} & reduce packet loss when failures occur & Trema & OpenFlow \\
& SANE~\cite{casado2006}       & security policy enforcement         &  SANE controller & SANE header (pre-OF) \\

& {SDN RTBH~\cite{Giotis2014}} & {DoS attack mitigation} & {POX} & {OpenFlow}\\

& VAVE~\cite{yao2011} & source address validation with a global view & NOX & OpenFlow \\

\hline
\end{tabularx}
\end{center}
\end{table*}
}

\vspace{2mm}
\noindent \textit{Traffic engineering}

Several traffic engineering applications have been proposed, including 
ElasticTree~\cite{heller2010}, Hedera~\cite{al-fares2010},
OpenFlow-based server load balancing~\cite{wang2011},
Plug-n-Serve~\cite{handigol2009-1} and Aster*x~\cite{handigol2009},
In-packet Bloom filter~\cite{macapuna2010},
SIMPLE~\cite{qazi2013-1}, QNOX~\cite{jeong2012}, QoS framework~\cite{kim2010}, 
\coloredtext{QoS for SDN~\cite{Sharma2014}}, 
ALTO~\cite{scharf2013}, ViAggre SDN~\cite{skoldstrom2013-1},
\coloredtext{ProCel~\cite{Nagaraj2014_4}, FlowQoS~\cite{Seddiki2014_4}, and Middlepipes~\cite{Jamjoom2014_4}.
In addition to these, recent proposals include optimization of rules placement~\cite{Nguyen2014_4}, the use of MAC as an universal label for efficient routing in data centers~\cite{Schwabe2014_4}, among other techniques for flow management, fault tolerance, topology update, and traffic characterization~\cite{Akyildiz2014traffic}}.
The main goal of most applications is to engineer traffic with the aim of minimizing power consumption, maximizing aggregate network utilization, providing optimized load balancing, and other generic traffic optimization techniques.

Load balancing was one of the first applications envisioned for SDN/OpenFlow.
Different algorithms and techniques have been proposed for this purpose~\cite{wang2011,handigol2009,handigol2009-1}. 
One particular concern is the scalability of these solutions.
A technique to allow this type of applications to scale is to use wildcard-based rules to perform proactive load balancing~\cite{wang2011}.
Wildcards can be utilized for aggregating clients requests based on the ranges of IP prefixes, for instance, allowing the distribution and directing of large groups of client requests without requiring controller intervention for every new flow.
In tandem, operation in reactive mode may still be used when traffic bursts are detected.
The controller application needs to monitor the network traffic and use some sort of 
threshold in the flow counters to redistribute clients among the servers when bottlenecks are likely to happen.

SDN load-balancing also simplifies the placement of network services in the network~\cite{handigol2009-1}.
Every time a new server is installed, the load-balancing service can take the appropriate actions to seamlessly distribute the traffic among the 
available servers, taking into consideration both the network load and the available computing capacity of the respective servers.
This simplifies network management and provides more flexibility to network operators.

Existing southbound interfaces can be used for actively monitoring the data plane load.
This information can be leveraged to optimize the energy consumption of the network~\cite{heller2010}. 
By using specialized optimization algorithms and diversified configuration options, it is possible to meet the 
infrastructure goals of latency, performance, and fault tolerance, for instance, while reducing power consumption.
With the use of simple techniques, such as shutting down links and devices intelligently in response to traffic load dynamics, data center operators can save up to 50\% of the network energy in normal traffic 
conditions~\cite{heller2010}. 

One of the important goals of data center networks is to avoid or mitigate the effect of network bottlenecks on the operation of the computing services offered.
Linear bisection bandwidth is a technique that can be adopted for traffic patterns that stress the network by exploring path diversity in a data center 
topology.
Such technique has been proposed in an SDN setting, allowing the maximization of aggregated network utilization with minimal scheduling overhead~\cite{al-fares2010}. 

SDN can also be used to provide a fully automated system for controlling the configuration of routers.
This can be particularly useful in scenarios that apply virtual aggregation~\cite{ballani2009}.
This technique allows network operators to reduce the data replicated on routing tables, which 
is one of the causes of routing tables' growth~\cite{meyer2007}. 
A specialized routing application~\cite{skoldstrom2013-1} can calculate, divide and configure the routing tables of the different routing devices through a southbound API such 
as OpenFlow.

Traffic optimization is another interesting application for large scale service providers, where dynamic scale-out 
is required. For instance, the dynamic and scalable provisioning of VPNs in cloud infrastructures, using protocolols
such as ALTO~\cite{alimi2013}, can be simplified through an SDN-based approach~\cite{scharf2013}.
\coloredtext{Recent work has also shown that optimizing rules placement can increase network efficiency~\cite{Nguyen2014_4}.
Solutions such as ProCel~\cite{Nagaraj2014_4}, designed for cellular core networks, are capable of reducing the signaling traffic up to 70\%, which represents a significant achievement.}

Other applications that perform routing and traffic engineering include application-aware 
networking for \coloredtext{video and data streaming~\cite{jarschel2013,Edwards2014_4}} and improved QoS by employing multiple packet 
schedulers~\cite{ishimori2013} and other techniques~\cite{kim2010,jeong2012,egilmez2012, kumar2013}.
As traffic engineering is a crucial issue in all kinds of networks, upcoming methods, techniques and innovations can be expected in the context of SDNs.

\vspace{2mm}
\noindent \textit{Mobility \& wireless}

The current distributed control plane of wireless networks is suboptimal for managing the limited 
spectrum, allocating radio resources, implementing handover mechanisms, managing interference, and performing efficient load-balancing between cells.
SDN-based approaches represent an opportunity for making it easier to deploy and manage different types of wireless networks, such as WLANs and cellular 
networks~\cite{schulz-zander2014-atc,yap2010-1,ali-ahmad2013,gudipati2013,li2012,jin2013}.
Traditionally hard-to-implement but desired features are indeed becoming a reality with the SDN-based wireless networks. 
These include seamless mobility through efficient hand-overs~\cite{schulz-zander2014-atc,dely2011,li2012}, load balancing~\cite{schulz-zander2014-atc,gudipati2013}, creation of on-demand virtual access points (VAPs)~\cite{schulz-zander2014-atc,vestin2013}, downlink scheduling (e.g., an OpenFlow switch can do a rate shaping or time division) ~\cite{vestin2013}, dynamic spectrum usage~\cite{vestin2013}, enhanced inter-cell 
interference coordination~\cite{vestin2013,li2012}, device to device offloading (i.e., decide when and how LTE transmissions should be offloaded to users adopting the D2D paradigm~\cite{yang2013d2d})~\cite{ali-ahmad2013}, per client and/or base station resource block allocations (i.e.,  time and frequency slots in LTE/OFDMA networks, which are known as resource blocks)~\cite{gudipati2013,ali-ahmad2013,jin2013}, control and assign 
transmission and power parameters in devices or in a group basis (e.g., algorithms to optimize the transmission and power parameters of WLAN devices, define and assign transmission power values to each resource block, at each base station, in LTE/OFDMA networks) ~\cite{ali-ahmad2013,gudipati2013}, simplified administration~\cite{schulz-zander2014-atc,yap2010-1,gudipati2013}, easy management of heterogenous network technologies~\cite{yap2010-1,gudipati2013,yap2010-2}, interoperability between different networks~\cite{yap2010-2,jin2013}, shared wireless infrastructures~\cite{yap2010-2}, seamless subscriber mobility and cellular networks~\cite{li2012}, QoS and access control policies made feasible and easier~\cite{li2012,jin2013}, and easy deployment of new applications~\cite{schulz-zander2014-atc,gudipati2013,yap2010-2}. 


One of the first steps towards realizing these features in wireless networks is to provide programmable and 
flexible stack layers for wireless networks~\cite{bansal2012,gudipati2013}.
One of the first examples is OpenRadio~\cite{bansal2012}, which proposes a software 
abstraction layer for decoupling the wireless protocol definition from the hardware, allowing shared MAC 
layers across different protocols using commodity multi-core platforms. OpenRadio can be seen as the 
``OpenFlow for wireless networks''. Similarly, SoftRAN~\cite{gudipati2013} proposes to rethink 
the radio access layer of current LTE infrastructures. Its main goal is to allow operators to improve and 
optimize algorithms for better hand-overs, fine-grained control of transmit powers, resource block allocation, 
among other management tasks.

Light virtual access points (LVAPs) is another interesting way of improving the management capabilities of 
wireless networks, as proposed by the Odin~\cite{schulz-zander2014-atc} framework. 
In contrast to OpenRadio, 
it works with existing wireless hardware and does not impose any change to IEEE 802.11 standards. An LVAP is implemented as a unique BSSID associated with a specific client, which means that there is a one-to-one mapping between LVAPs and clients. This per-client access point (AP) abstraction simplifies the handling of client associations, authentication, handovers, and unified slicing of both the wired and wireless portions of the network. Odin achieves control logic isolation between slices, since LVAPs are the primitive type upon which applications make control decisions, and applications do not have visibility of LVAPs from outside their slice.
 This empowers infrastructure operators to provide services through Odin applications, such as a mobility manager, client-based load balancer, channel selection algorithm, and wireless troubleshooting application within different network slices. For instance, when a user moves from one AP to another, the
 network mobility management application can automatically and proactively act and move the client LVAP from one
AP to the other. In this way, a wireless client will not even notice that it started to use a different AP
because there is no perceptive hand-off delay, as it would be the case in traditional wireless networks.

Very dense heterogeneous wireless networks have also been a target for SDN.
These DenseNets have limitations due to constraints 
such as radio access network bottlenecks, control overhead, and high operational costs~\cite{ali-ahmad2013}.
A dynamic two-tier SDN controller hierarchy can be adapted to address some of these constraints~\cite{ali-ahmad2013}. Local controllers can be used to take fast and fine-grained decisions, while regional 
(or ``global'') controllers can have a broader, coarser-grained scope, i.e., that take slower but more 
global decisions.
In such a way, designing a single integrated architecture that encompasses LTE (macro/pico/femto) and WiFi 
cells, while challenging, seems feasible. 

\vspace{2mm}
\noindent \textit{Measurement \& monitoring}

Measurement and monitoring solutions can be divided in two classes. First, applications that provide new 
functionality for other networking services. Second, proposals that target to improve features of OpenFlow-based SDNs, 
such as to reduce control plane overload due to the collection of statistics.

An example of the first class of applications is improving the visibility of broadband 
performance~\cite{sundaresan2011,kim2013}. An SDN-based broadband home connection can 
simplify the addition of new functions in measurement systems such as BISmark~\cite{sundaresan2011}, 
allowing the system to react to changing conditions in the home network~\cite{kim2013}. As an example, a home 
gateway can perform reactive traffic shaping considering the current measurement results of the home network.

The second class of solutions typically involve different kinds of sampling and estimation 
techniques to be applied, in order to reduce the burden of the control plane with respect to the collection of data plane statistics.
Different techniques have been applied to achieve this goal, such as stochastic and deterministic packet sampling techniques~\cite{mehdi2011}, traffic matrix estimation~\cite{tootoonchian2010-1}, fine-grained monitoring of wildcard rules~\cite{wette2013}, \coloredtext{two-stage Bloom filters~\cite{DBLP:journals/comsur/TarkomaRL12} 
to represent monitoring rules and provide high measurement accuracy without incurring in extra memory or control plane traffic overhead~\cite{Yu2014_4}, and special monitoring functions (extensions to OpenFlow) in forwarding devices to reduce traffic and processing load on the control plane~\cite{kempf2012}.}
Point-to-point traffic matrix estimation, in particular, can help in network design and operational tasks such as load balancing, anomaly detection, capacity planning and 
network provisioning.
With information on the set of active flows in the network, routing information (e.g., from the routing application), flow paths, and flow counters in the switches it is possible to 
construct a traffic matrix using diverse aggregation levels for sources and destinations~\cite{tootoonchian2010-1}.

Other initiatives of this second class propose a stronger decoupling between basic primitives (e.g., matching and counting) and 
heavier traffic analysis functions such as the detection of anomaly conditions attacks~\cite{bianchi2013}.
A stronger separation favors portability and flexibility.
For instance, a functionality to detect abnormal flows should not be constrained by the basic primitives or 
the specific hardware implementation.
Put another way, developers should be empowered with streaming 
abstractions and higher level programming capabilities.

In that vein, some data and control plane abstractions have been specifically designed for measurement purposes.
OpenSketch~\cite{yu2013-1} is a special-purpose southbound API 
designed to provide flexibility for network measurements.
For instance, by allowing multiple measurement tasks to execute concurrently without impairing accuracy.
The internal design of an OpenSketch switch can be thought of as a pipeline with three stages (hashing, classification, and counting). 
Input packets first pass through a hashing function. 
Then, they are classified according to a matching rule.
Finally, the match rule identifies a counting index, which is used to calculate the counter location in the counting stage. While a TCAM with few 
entries is enough for the classification stage, the flexible counters are stored in SRAM.
This makes the OpenSketch's operation efficient (fast matching) and cost-effective (cheaper SRAMs to store counters).

\coloredtext{Other monitoring frameworks, such as OpenSample~\cite{suh2014os} and PayLess~\cite{chowdhury2014payless}, propose different mechanisms for delivering real-time, low-latency and flexible monitoring capabilities to SDN without impairing the load and performance of the control plane. 
The proposed solutions take advantage of sampling technologies like sFlow~\cite{sflow.orgforum2012} to monitor high-speed networks, and flexible collections of loosely coupled (plug-and-play) components to provide abstract network views
yielding high-performance and efficient network monitoring approaches~\cite{suh2014os,chowdhury2014payless,wette2013}.
}

\vspace{2mm}
\noindent \textit{Security \& Dependability}

An already diverse set of security and dependability proposals is emerging in the context of SDNs.
Most take advantage 
of SDN for improving services required to secure systems and networks, such as policy enforcement (e.g., access 
control, \coloredtext{firewalling, middleboxes as middlepipes~\cite{Jamjoom2014_4})~\cite{casado2006,wang2012-1,yao2011,stabler2012,Jamjoom2014_4}, DoS attacks detection and mitigation~\cite{braga2010-1,Giotis2014}}, random host mutation~\cite{stabler2012} (i.e., randomly and frequently mutate the IP addresses of end-hosts to break the attackers' assumption about static IPs, which is the common case)~\cite{jafarian2012}, monitoring of cloud infrastructures for fine-grained security inspections (i.e., automatically analyze and detour suspected traffic to be  further inspected by specialized network security appliances, such as deep packet inspection systems) ~\cite{shin2012}, \coloredtext{traffic anomaly detection~\cite{mehdi2011,braga2010-1,Giotis2014}}, \coloredtext{fine-grained flow-based network access control~\cite{Matias2014}, fine-grained policy enforcement for personal mobile applications~\cite{Sapio2014}} and so forth~\cite{casado2006,wang2012-1,braga2010-1,jafarian2012,shin2012,stabler2012,yao2011,mehdi2011}.
Others address OpenFlow-based networks issues, \coloredtext{such as flow rule prioritization, security 
services composition, protection against traffic overload, and protection against malicious administrators~\cite{porras2012,shin2013-1,shin2013-3,wang2013,Matsumoto2014_4}}.

There are essentially two approaches, one involves using SDNs to improve network security, and another for improving the 
security of the SDN \emph{itself}. 
The focus has been, thus far, in the latter.

%

\noindent \textit{Using SDN to improve the security of current networks}.
Probably the first instance of SDN was an application for security policies enforcement~\cite{casado2006}. 
An SDN allows the enforcement to be done on the first entry point to the network (e.g., the Ethernet switch to which the user is connected to). 
Alternatively, in a hybrid environment, security policy enforcement can be made on a wider network perimeter through programmable devices (without the need to migrate the entire infrastructure to OpenFlow)~\cite{wang2012-1}.
With either application, malicious actions are blocked before entering the critical regions of the network.

SDN has been successfully applied for other purposes, namely for the detection (and reaction) against DDoS flooding attacks~\cite{braga2010-1}, and active security~\cite{hand2013}.
OpenFlow forwarding devices make it easier to collect a variety of information from the network, in a timely 
manner, which is very handy for algorithms specialized in detecting DDoS flooding attacks.

The capabilities offered by software-defined networks in increasing the ability to collect statistics data from the network and of allowing applications to actively program the forwarding devices, are powerful for proactive and smart 
security policy enforcement techniques such as Active security~\cite{hand2013}.
This novel security methodology proposes a novel feedback loop to improve the control of defense mechanisms 
of a networked infrastructure, and is centered around five core capabilities: protect, sense, 
adjust, collect, counter.
In this perspective, active security provides a centralized programming interface that simplifies the integration of mechanisms for detecting attacks, by
a) collecting data from different sources (to identify attacks), 
b) converging to a consistent configuration for the security appliances, and 
c) enforcing countermeasures to block or minimize the effect of attacks.

\noindent \textit{Improving the security of SDN itself}.
There are already some research efforts on identifying the critical security threats of SDNs and in augmenting its security and dependability~\cite{porras2012,shin2013-1,kreutz2013}.
Early approaches try to apply simple techniques, such as classifying applications and using rule prioritization, to ensure that rules generated by security applications will not be overwritten by 
lower priority applications~\cite{porras2012}. 
Other proposals try to go a step further by providing a framework for developing security-related applications in SDNs~\cite{shin2013-1}.
However, there is still a long way to go in the development of secure and dependable SDN infrastructures~\cite{kreutz2013}.
An in-deep overview of SDN security issues and challenges can be found in Section~\ref{secSecurity}.

\vspace{2mm}
\noindent \textit{Data Center Networking}

From small enterprises to large scale cloud providers, most of the existing IT systems and services are strongly dependent on highly scalable and efficient data centers.
Yet, these infrastructures still pose significant challenges regarding computing, storage and networking.
Concerning the latter, data centers should be designed and deployed in such a way as to offer
high and flexible cross-section bandwidth and low-latency, 
QoS based on the application requirements,
high levels of resilience,
intelligent resource utilization to reduce energy consumption and improve overall efficiency,
agility in provisioning network resources, for example by means of network virtualization and orchestration with computing and storage,
and so forth~\cite{Kant20092939,greenberg2008cost,bari2013}.
Not surprisingly, many of these issues remain open due to the complexity and inflexibility of traditional network architectures.

The emergence of SDN is expected to change the current state of affairs.
Early research efforts have indeed showed that data center networking can significantly benefit from SDN in solving different problems such as live network migration~\cite{keller2012}, improved network management~\cite{keller2012,arefin2013}, eminent failure avoidance~\cite{keller2012,arefin2013}, rapid deployment from development to production networks~\cite{keller2012}, troubleshooting~\cite{keller2012,raghavendra2012},  optimization of network utilization~\cite{raghavendra2012,wang2012,das2013,arefin2013}, \coloredtext{dynamic and elastic provisioning of middleboxes-as-a-service~\cite{Jamjoom2014_4}, minimization of flow setup latency and reduction of controller operating costs~\cite{Krishnamurthy2014_4}}. 
SDN can also offer networking primitives for cloud applications, solutions to predict network transfers of applications~\cite{wang2012,das2013}, mechanisms for fast reaction to operation problems, network-aware VM placement~\cite{raghavendra2012,benson2011},  QoS support~\cite{raghavendra2012,benson2011}, realtime network monitoring and problem detection~\cite{raghavendra2012,das2013,arefin2013}, security policy enforcement services and mechanisms~\cite{raghavendra2012,benson2011}, and enable programmatic adaptation of transport protocols~\cite{wang2012,ghobadi2013}.

SDN can help infrastructure providers to expose more networking primitives to their customers, by allowing virtual network isolation, custom addressing, and the placement of middleboxes and virtual desktop cloud applications~\cite{benson2011,calyam2013}. 
To fully explore the potential of virtual networks in clouds, an essential feature is virtual network migration. 
Similarly to traditional virtual machine migration, a virtual network may need to be migrated when its virtual machines move from one place to another. Integrating live migration of virtual 
machines and virtual networks is one of the forefront challenges~\cite{keller2012}.
To achieve this goal it is necessary to dynamically reconfigure all affected networking devices (physical or 
virtual).
This was shown to be possible with SDN platforms, such as NVP~\cite{koponen}.

Another potential application of SDN in data centers is in detecting abnormal behaviors in network 
operation~\cite{arefin2013}. By using different behavioral models and collecting the necessary 
information from elements involved in the operation of a data center (infrastructure, operators, applications), 
it is possible to continuously build signatures for applications by passively capturing control traffic.
Then, the signature history can be used to identify differences in behavior.
Every time a difference is detected, operators can reactively or proactively take corrective measures.
This can help to isolate abnormal components and avoid further damage to the
infrastructure.

\vspace{2mm}
\noindent \textit{Towards SDN App Stores}

As can be observed in Tables~\ref{tab:managementapplications1} and~\ref{tab:managementapplications2}, most SDN applications rely on NOX and OpenFlow.
NOX was the first controller available for general use, making it a natural choice for most use-cases so far.
As indicated by the sheer number of security-related applications, security is probably one of the killer 
applications for SDNs. Curiously, while most use cases rely on OpenFlow, new solutions such as SoftRAN are 
considering different APIs, as is the case of the Femto API~\cite{smallcellforum2013,Chandrasekhar2008}. This diversity of 
applications and APIs will most probably keep growing in SDN.

There are other kinds of \manapps that do not easily fit in our taxonomy, such as Avior~\cite{parraga2013}, OESS~\cite{globalnoc2013}, and SDN App Store~\cite{duckett2013,hp2013-2}.
Avior and OESS are graphical interfaces and sets of software tools that make it easier to configure and manage controllers (e.g., Floodlight) and OpenFlow-enabled switches, respectively.
By leveraging their graphical functions it is possible to program OpenFlow enabled devices without coding in a particular programming language.

The SDN App Store~\cite{duckett2013,hp2013-2}, owned by HP, is probably the first SDN 
application market store. Customers using HP's OpenFlow controller have access to the online SDN App Store 
and are able to select applications to be dynamically downloaded and installed in the controller. 
The idea is similar to the Android Market or the Apple Store, making it easier for developers to provide 
new applications and for customers to obtain them.

\subsection{Cross-layer issues}
\label{sec:debuggingandtroubleshooting}

In this section we look at cross-layer problems such as debugging and troubleshooting, testing, verification, simulation and emulation.

\vspace{2mm}
\noindent \textit{Debugging and troubleshooting}

Debugging and troubleshooting have been important subjects in computing infrastructures, 
parallel and distributed systems, embedded systems, and desktop 
applications~\cite{sigelman2010,layman2013,erlingsson2012,tomaselli2013,tan2010,fonseca2007,trivedi2014}. 
The two predominant strategies applied to debug and troubleshoot are runtime debugging 
(e.g., \texttt{gdb}-like tools) and post-mortem analysis (e.g., tracing, replay and visualization).
Despite the constant evolution and the emergence of new techniques to improve debugging and troubleshooting, 
there are still several open avenues and research questions~\cite{layman2013}.


Debugging and troubleshooting in networking is at a very primitive stage.
In traditional networks, engineers and developers have 
to use tools such as \texttt{ping}, \texttt{traceroute}, \texttt{tcpdump}, \texttt{nmap}, 
\texttt{netflow}, and SNMP statistics for debugging and troubleshooting.
Debugging a complex 
network with such primitive tools is very hard.
Even when one considers frameworks such as
XTrace~\cite{fonseca2007}, Netreplay~\cite{anand2010} 
and NetCheck~\cite{zhuang2014}, which improve debugging capabilities in networks, it is still difficult 
to troubleshoot networking infrastructures. For instance, these frameworks require a huge effort in terms of 
network instrumentation. The additional complexity introduced 
by different types of devices, technologies and vendor specific components and features make matters worse.
As a consequence, 
these solutions may find it hard to be widely implemented and deployed in current networks.

\coloredtext{
SDN offers some hope in this respect. The hardware-agnostic software-based control capabilities} and the use of open standards for control communication can potentially make debug and troubleshoot easier.
The flexibility and 
programmability introduced by SDN is indeed opening new avenues for developing better tools to debug, troubleshoot, 
verify and test networks~\cite{handigol2012-1,wundsam2011,canini2012-1,rotsos2012-1,al-shaer2010,khurshid2012,kuzniar2012,altekar2010,kuzniar2012}.

Early debugging tools for OpenFlow-enabled networks, such as \texttt{ndb}~\cite{handigol2012-1}, OFRewind~\cite{wundsam2011} and NetSight~\cite{handigol2014}, make it easier to discover the source of network problems such as faulty device firmware~\cite{handigol2012-1}, inconsistent or 
non-existing flow rules~\cite{handigol2012-1,wundsam2011}, lack of reachability~\cite{handigol2012-1,wundsam2011}, and faulty routing~\cite{handigol2012-1,wundsam2011}.
Similarly to the well-known \texttt{gdb} software debugger, \texttt{ndb} provides basic debugging 
actions such as \textit{breakpoint}, \textit{watch}, \textit{backtrace}, \textit{single-step}, and 
\textit{continue}. These primitives help application developers to debug networks in a similar way 
to traditional software.
By using \texttt{ndb}'s postcards (i.e., a unique packet identifier composed of a truncated copy of the packet's header, the matching flow entry, the switch, and the output port), for instance, a programmer is able to quickly identify 
and isolate a buggy OpenFlow switch with hardware or software problems. 
If the switch is 
presenting abnormal behavior such as corrupting parts of the packet header, by 
analyzing the problematic flow sequences with a debugging tool one can find (in a matter of few seconds) where 
the packets of a flow are being corrupted, and take the necessary actions to solve the problem.

The OFRewind~\cite{wundsam2011} tool works differently. The idea is to record and 
replay network events, in particular control messages.
These usually account for less than 1\% of the data 
traffic and are responsible for 95\%-99\% of the bugs ~\cite{altekar2010}. 
This tool allows operators to 
perform fine-grained tracing of network behavior, being able to decide which subsets of the network will 
be recorded and, afterwards, select specific parts of the traces to be replayed. These replays provide 
valuable information to find the root cause of the network misbehavior.
\coloredtext{Likewise, NetRevert~\cite{Zhang2014_42} also records the state of OpenFlow networks. However, the primary goal is not to reproduce network behavior, but rather to provide rollback recovery in case of failures, which is a common approach used in distributed systems for eliminating transient errors in nodes~\cite{Plank1995,Plank1999}.}

Despite the availability of these debugging and verification tools, it is still difficult to answer questions such as: What is happening to my packets that are flowing from point A to point B? What path do they follow? What header modifications do they undergo on the way? 
To answer some of these questions one could recur to the \emph{history} of the packets. A packet's history corresponds 
to the paths it uses to traverse the network, and the header modifications in each hop of the path.
NetSight~\cite{handigol2014} is a platform whose primary goal is to allow applications that use the history of the packets to be built, in order to find out problems in a network.
This platform is composed of three essential elements: (1) NetSight, with its dedicated servers that receive and 
process the postcards for building the packet history, (2) the NetSigh-SwitchAssist, which can be used in 
switches to reduce the processing burden on the dedicated servers, and (3) the NetSight-HostAssist to generate 
and process postcards on end hosts (e.g., in the hypervisor on a virtualized infrastructure).

\texttt{netwatch}~\cite{handigol2014}, \texttt{netshark}~\cite{handigol2014} and \texttt{nprof}~\cite{handigol2014} are three examples of tools built over NetSight.
The first is a live network invariant monitor. 
For instance, an alarm can be trigged every time a packet violates any invariant (e.g., no loops).
The second, \texttt{netshark}, enables users to define and execute filters on the entire history of packets.
With this tool, a network operator can view a complete list of properties of packets at each hop, such as input port, output port, and packet header values.
Finally, \texttt{nprof} can be used to profile sets of network links to provide data for analyzing 
traffic patterns and routing decisions that might be contributing to link load.

\vspace{2mm}
\noindent \textit{Testing and verification}

Verification and testing tools can complement debugging and troubleshooting.
Recent research~\cite{khurshid2012,altekar2010,al-shaer2010,canini2012-1,kuzniar2012,ruchansky2013,zeng2014} 
has shown that verification techniques can be applied to detect and avoid problems in SDN, 
such as forwarding loops and black holes. Verification can be done at different layers (at the controllers, \manapps, or network devices).
\coloredtext{Additionally, there are different network properties  -- mostly topology-specific -- that can be formally verified, provided a network model is available.
Examples of such properties are connectivity, loop freedom, and access control~\cite{Casado2014_4}.
A number of tools have also been proposed to evaluate the performance of OpenFlow controllers by emulating the load of large-scale networks (e.g., Cbench~\cite{sherwood2011}, OFCBenchmark~\cite{jarschel2012}, PktBlaster~\cite{veryxtech}).
Similarly, benchmarking tools for OpenFlow switches are also available (e.g., OFLOPS~\cite{rotsos2012-1}, FLOPS-Turbo~\cite{Rotsos2014}).}



Tools such as NICE~\cite{canini2012-1} generate sets of diverse streams of packets 
to test as many as possible events, exposing corner cases such as race conditions.
Similarly, OFLOPS~\cite{rotsos2012-1} provides a set of features and functions that allow 
the development and execution of a rich set of tests on OpenFlow-enabled devices. 
Its ultimate 
goal is to measure the processing capacity and bottlenecks of control applications and forwarding devices.
With this tool, users are able to observe and evaluate forwarding table consistency, flow setup latency, 
flow space granularity, packet modification types, and traffic monitoring capabilities (e.g., counters).

FlowChecker~\cite{al-shaer2010}, OFTEN~\cite{kuzniar2012}, and VeriFlow~\cite{khurshid2012} are three examples of tools to verify correctness 
properties violations on the system. While the former two are based on offline analysis, the latter is capable of 
online checking of network invariants. Verification constraints include security and reachability issues, 
configuration updates on the network, loops, black holes, etc.

Other formal modeling techniques, such as Alloy, can be applied to SDNs to identify unexpected 
behavior~\cite{ruchansky2013}. For instance, a protocol specification can be weak 
when it under-specifies some aspects of the protocol or due to a very specific sequence of events.
In such situations, model checking techniques such as Alloy can help to find and correct unexpected 
behaviors.

\coloredtext{Tools such as FLOWGUARD~\cite{Hu2014_4} are specifically designed to detect and resolve security policy violations in OpenFlow-enabled networks.
FLOWGUARD is able to examine on-the-fly network policy updates, check indirect security violations (e.g., OpenFlow's \texttt{Set-Field} actions modification) and perform stateful monitoring.
The framework uses five resolution strategies for real-time security policy violation resolution,
flow rejecting, dependency breaking, update rejecting, flow removing, and packet blocking~\cite{Hu2014_4}.
These resolutions are applied over diverse update situations in OpenFlow-enabled networks.

More recently, tools such as VeriCon~\cite{ball2014vericon} have been designed to verify the correctness of SDN applications in a large range of network topologies and by analyzing a broad range of sequences of network events.
In particular, VeriCon confirms, or not, the correct execution of the SDN program.
}

One of the challenges in testing and verification is to verify forwarding tables in very large networks to find routing errors, which can cause traffic losses and security breaches, as quickly as possible.
In large scale networks, it is not possible to assume that the network snapshot, at any point, is consistent, 
due to the frequent changes in routing state.
Therefore, solutions such as HSA~\cite{kazemian2012}, Anteater~\cite{mai2011}, NetPlumber~\cite{kazemian2013}, VeriFlow~\cite{khurshid2012}, \coloredtext{and assertion languages~\cite{Beckett2014_4}} are not suited for this kind of environment.
Another important issue is related on how fast the verification process is done, especially in modern data centers
that have very tight timing requirements. Libra~\cite{zeng2014} represents one of the first attempts to address 
these particular challenges of large scale networks. 
This tool provides the means for capturing stable and consistent 
snapshots of large scale network deployments, while also applying long prefix matching techniques 
to increase the scalability of the system.
By using MapReduce computations, Libra is capable of verifying the correctness of a 
network with up to 10k nodes within one minute.

\coloredtext{
Anteater~\cite{mai2011} is a tool that analyzes the data plane state of network devices by encoding switch configurations as boolean satisfiability problems (SAT) instances, allowing to use a SAT solver to analyze the network state.
The tool is capable of verifying violations of invariants such as loop-free forwarding, connectivity, and consistency.
These invariants usually indicate a bug in the network, i.e., their detection helps to increase the reliability of the network data plane.
}

\vspace{2mm}
\noindent \textit{Simulation and Emulation}

Simulation and emulation software is of particular importance for fast prototyping and testing without the need for expensive physical devices.
Mininet~\cite{lantz2010} is the first system that provides a quick and easy way to prototype and evaluate SDN protocols and applications.
One of the key properties of Mininet is its use of software-based OpenFlow switches in virtualized containers, providing the exact same semantics of hardware-based OpenFlow  switches.
This means that controllers or applications developed and tested in the emulated environment can be (in theory) deployed in an OpenFlow-enabled network without any modification. Users can easily emulate an OpenFlow network with hundreds of nodes and dozens of switches by using a single personal computer.
Mininet-HiFi~\cite{handigol2012} is an evolution of Mininet that enhances the container-based (lightweight) 
virtualization with mechanisms to enforce performance isolation, resource provisioning, and accurate monitoring for 
performance fidelity. 
One of the main goals of Mininet-HiFi is to improve the reproducibility of networking research.

Mininet CE~\cite{antonenko2013} and SDN Cloud DC~\cite{teixeira2013} are extensions 
to Mininet for enabling large scale simulations.
Mininet CE combines groups of Mininet instances into one cluster 
of simulator instances to model global scale networks.
SDN Cloud DC enhances Mininet and POX to emulate an SDN-based 
intra-DC network by implementing new software modules such as data center topology discovery and network traffic generation. 
\coloredtext{Recent emulation platform proposals that enable large scale experiments following a distributed approach include MaxiNet~\cite{wette2014}, DOT~\cite{roy2014}, and CityFlow~\cite{Carter2014}.
The latter is a project with the main goal of building an emulated control plane for a city of one million inhabitants. 
Such initiatives are a starting point to provide experimental insights for large-scale SDN deployments.
}

The capability of simulating OpenFlow devices has also been added to the popular ns-3 simulator~\cite{ns3project2013}.  
Another simulator is \textit{fs-sdn}, which extends the \textit{fs} simulation engine~\cite{sommers2011} by incorporating a controller and switching components with OpenFlow support.
Its main goal is to provide a more realistic and scalable simulation platform as compared to Mininet.
Finally, STS~\cite{ucbsts2013} is a simulator designed to allow developers to specify and apply a variety of test cases, while allowing them to interactively examine the state of the network.

{\renewcommand{\arraystretch}{1.4}
\begin{table*}[!htp]
\caption{Debugging, verification and simulation}
\label{tab:debuggingverification}
\begin{center}
\footnotesize
\begin{tabularx}{\linewidth}{p{1.2cm}p{2.5cm}p{3.5cm}X}
\hline
\textbf{Group} & \textbf{Solution} & \textbf{Main purpose} & \textbf{Short description} \\
\hline
\multirow{6}{*}{Debugging} 
& \texttt{ndb}~\cite{handigol2012-1} & \textit{gdb} alike SDN debugging  & Basic debugging primitives that help developers to debug their networks. \\
& NetSight~\cite{handigol2014} & multi purpose packet history & Allows to build flexible debugging, monitoring and profiling applications. \\ 
& OFRewind~\cite{wundsam2011} & tracing and replay & OFRewind allows operators to do a fine-grained tracing of the network behavior. Operators can decide which subsets of the network will be recorded. \\

& PathletTracer~\cite{Zhang2014_4} & inspect layer 2 paths & Allows to inspect low-level forwarding behavior through on-demand packet tracing capabilities.\\

& SDN traceroute~\cite{Agarwal2014_4} & query OpenFlow paths & Allows users to discover the forwarding behavior of any Ethernet packet and debug problems regarding both forwarding devices and applications.\\

\hline
\multirow{26}{*}{Verification} 
& Assertion language~\cite{Beckett2014_4} & debug SDN apps & Enables assertions about the data plane on the apps with support to dynamic changing verification conditions.\\

& Cbench~\cite{sherwood2011} & evaluate OpenFlow controllers & The Cbench framework can be used to emulate OpenFlow switches which are configured to generate workload to the controller.\\

& FLOVER~\cite{son2013} & model checking for security policies & FLOVER provides a provably correct and automatic method for verifying security properties with respect to a set of flow rules committed by an OF controller.\\
& FlowChecker~\cite{al-shaer2010} & flow table config verification & A tool used to verify generic properties of global behaviors based on flow tables.  \\

& FLOWGUARD~\cite{Hu2014_4} & verify security policy & Provides mechanisms for accurate detection and resolution of firewall policy violations in OpenFlow-based networks.\\

& FlowTest~\cite{Fayaz2014_4} & verify network policies & Provides the means for testing stateful and dynamic network policies by systematically exploring the state space of the network data plane. \\

& NetPlumber~\cite{kazemian2013} & real time policy checking & NetPlumber uses a set of policies and invariants to do real time checking. It leverages header space analysis and keeps a dependency graph between rules.\\
& NICE~\cite{canini2012-1} & remove bugs in controllers & Its main goal is to test controller programs without requiring any type of modification or extra work for application programmers. \\
& OFCBenchmark~\cite{jarschel2012} & evaluate OpenFlow controllers & creates independent virtual switches, making is possible to emulate different scenarios. Each switch has its how configuration and statistics.\\
& OFTEN~\cite{kuzniar2012} & catch correctness property violations & A framework designed to check SDN systems, analyzing controller and switch interaction, looking for correctness condition violation. \\
 & OFLOPS~\cite{rotsos2012-1} & evaluate OpenFlow switches & A framework with a rich set of tests for OpenFlow protocol, enabling to measure capabilities of both switch and applications. \\

& OFLOPS-Turbo~\cite{Rotsos2014} & evaluate OpenFlow switches & A framework that integrates OFLOPS with OSNT~\cite{Shahbaz2013}, a 10GbE traffic generation and monitoring system based on NetFPGA. \\

& PktBlaster~\cite{veryxtech} & emulation / benchmarking & Integrated test and benchmarking solution that emulates large scale software-defined networks.\\

& SDLoad~\cite{Laurent2014_4} & evaluate OpenFlow controllers & A traffic generation framework with customizable workloads to realistically represent different types of applications. \\

& VeriCon~\cite{ball2014vericon} & verify SDN apps & Is a tool for verifying the correctness of SDN applications on large range of topologies and sequences of network events. \\

& VeriFlow~\cite{khurshid2012} & online invariant verification & It provides real time verification capabilities, while the network state is still evolving. \\

\hline
\multirow{16}{*}{\begin{minipage}{1.2cm}Simulation \& \\Emulation\end{minipage}} 

& DOT~\cite{roy2014} & network emulation & Leverages VMs for large scale OpenFlow-based network emulations with resource allocation guarantees.\\

& \textit{fs-sdn}~\cite{gupta2013} & fast simulation & Like Mininet, it provides a simulation environment, but with speed and scalability advantages.\\

& MaxiNet~\cite{wette2014} & network simulation & Similarly to Mininet CE, it is a combination of Mininet tools  for large scale simulation of network topologies and architectures.\\

& Mininet~\cite{lantz2010} & fast prototyping & It emulates and OpenFlow network using Open vSwitches to provide the exact same semantics of hardware devices. \\
& Mininet CE~\cite{antonenko2013} & global network modeling & It is a combination of tools to create a Mininet cluster for large scale simulation of network topologies and architectures.\\
& Mininet-HiFi~\cite{handigol2012} & reproducible experimentation & Evolution of Mininet to enable repeatable and high fidelity  experiments. \\
& ns-3~\cite{ns3project2013} & network simulation & The latest version of ns-3 simulator provides support to OpenFlow, enabling to create programmable network devices. \\
& SDN Cloud DC~\cite{teixeira2013} & cloud data center emulation & The SDN Cloud DC solution allows users to evaluate the performance of their controllers at scale.\\
& STS~\cite{ucbsts2013} & troubleshooting & It simulates a network, allows to generate tricky test cases, and allows interactively examine the state of the network.\\

& VND-SDN~\cite{fontes2014auth} & simulation and analysis & Makes it easier to define experiments via 
GUI authoring of SDN scenarios and automatic generation of NSDL.\\

\hline
\end{tabularx}
\end{center}
\end{table*}
}

%% file: text/5_challenges.tex
\section{Ongoing Research Efforts and Challenges}
\label{sec:challenges}

The research \coloredtext{developments} we have surveyed so far seek to overcome the challenges of realizing the vision and fulfilling the \coloredtext{promises} of SDN.
While Section~\ref{sec:layeredapproach} provided a perspective structured across the layers of the ``SDN stack'', this
section highlights research \coloredtext{efforts} we consider of particular importance for unleashing the full potential of SDN, and that therefore deserves a specific coverage in this survey.

\subsection{Switch Designs}

Currently available OpenFlow switches are very diverse and exhibit notable differences in terms of 
feature set (e.g., flow table size,  optional actions), performance (e.g., fast vs. slow 
path, control channel latency/throughput), interpretation and adherence to the protocol 
specification (e.g., \texttt{BARRIER} command), and architecture (e.g., hardware vs. software designs).

\vspace{2mm}
\noindent \textit{Heterogenous Implementations}

Implementation choices have a fundamental impact on the behavior, accuracy, and performance of switches, 
ranging from differences in flow counter behavior~\cite{curtis2011} to a number of 
other performance metrics~\cite{rotsos2012-1}. One approach to accommodate such heterogeneity 
is through NOSIX, a portable API that separates the application expectations from the switch heterogeneity~\cite{wundsam2012}. To do so, NOSIX provides a pipeline of multiple virtual flow 
tables and switch drivers. Virtual flow tables are intended to meet the expectations of applications and 
are ultimately translated by the drivers into actual switch flow tables. 
Towards taming the  complexity of multiple OpenFlow protocol versions with different sets of required and optional capabilities, a roadblock for SDN practitioners, tinyNBI~\cite{casey2014} has been proposed as a  simple API providing a unifying set of core abstractions of five OpenFlow protocol versions (from 1.0 to 1.4).
Ongoing efforts to introduce a new Hardware Abstraction Layer  (HAL) for non-OpenFlow capable devices~\cite{alienfp7}  include the development of open source artifacts like ROFL (Revised OpenFlow Library) and the xDPd (eXtensible DataPath daemon), a framework for creating new OpenFlow datapath implementations based on a diverse set of hardware and software platforms.
A related open source effort to develop a common library to implement OpenFlow 1.0 and 1.3 protocol endpoints (switch agents and controllers) is libfluid~\cite{libfluid}, winner of the OpenFlow driver competition organized by the ONF.

Within the ONF, the Forwarding Abstraction Working Group (FAWG) is pursuing another solution to the heterogeneity problem, through Table Type Patterns 
(TTPs)~\cite{onf2013}. A TTP is a standards-based and negotiated switch-level behavioral abstraction. 
It consists of the relationships between tables forming a graph structure, the types of tables in the graph, 
a set of the parameterized table properties for each table in the graph, the legal \texttt{flow-mod} and \texttt{table-mod} commands for 
each flow table, and the metadata mask that can be passed between each table pair in the graph.

\vspace{2mm}
\noindent \textit{Flow Table Capacity}

Flow matching rules are stored in flow tables inside network devices.
One practical challenge is to provide switches with large and efficient flow tables to store the 
rules~\cite{appelman2012}. TCAMs are a common choice to hold flow tables. While flexible 
and efficient in terms of matching capabilities, TCAMs are costly and usually small (from 4K to 32K 
entries). Some TCAM chips today integrate 18 M-bit (configured as 500k entries $*$ 36 bit per entry) 
into a single chip working at 133 Mhz~\cite{kannan2013}, i.e., capable of 133M lookups 
per second. However, these chips are expensive and have a high-power consumption~\cite{liao2012}, 
representing a major power drain in a switching device~\cite{agrawal2006}. These are some of the reasons 
why currently available OpenFlow devices have TCAMs with roughly 8K entries, where the actual capacity 
in terms of OpenFlow table size has a non-trivial relationship to the type of flow entries being used~\cite{owens2013,salisbury2012}. OpenFlow version 1.1 introduced multiple tables, 
thereby adding extra flexibility and scalability. Indeed, OpenFlow 1.0 implied state explosion due to 
its flat table model~\cite{onf2013}. However, supporting multiple tables in hardware is challenging 
and limited -- yet another motivation for the ongoing ONF FAWG work on TTPs~\cite{onf2013}.

\coloredtext{Some efforts focus on compression techniques to reduce the number of flow entries in TCAMs~\cite{Braun2014,Agarwal2014_42,Rudell1987espresso}.
The Espresso heuristic~\cite{Rudell1987espresso} can be used to compress wildcards of OpenFlow-based inter-domain routing tables, reducing the forwarding information base (FIB) by 17\% and, consequently, saving up to 40,000 flow table entries~\cite{Braun2014}. 
Shadow MACs~\cite{Agarwal2014_42} propose label switching for solving two problems, consistent updates and rule space exhaustion, by using opaque values (similar to MPLS labels) to encode fine-grained paths as labels. 
A major benefit of fixed-size labels is relying on exact-math lookups which can be easily and cost-effectively implemented by simple hardware tables instead of requiring rules to be encoded in expensive TCAM tables.

}

\vspace{2mm}
\noindent \textit{Performance}

\coloredtext{Today, the throughput of commercial OpenFlow  switches varies from 38 to 1000 \texttt{flow-mod} per second, with most devices achieving a throughput lower than 500 \texttt{flow-mod} per second~\cite{Bifulco2014, stephens2012-1}.}
This is clearly a limiting 
factor that shall be addressed in the switch design process -- support of OpenFlow in existing product 
lines has been more a retrofitting activity than a clean feature planning and implementation activity. Deployment experiences~\cite{Kobayashi2014151} have pointed to a series of challenges stemming from the limited embedded CPU power of current commercial OpenFlow switches. 
One approach to handle the problem consists of adding more powerful CPUs into the switches, as proposed 
in~\cite{mogul2012}. Others have proposed to rethink the distribution of control 
actions between external controllers and the OpenFlow agent inside the switch~\cite{curtis2011}. 
Our current understanding indicates that an effective way forward is a native design of SDN switches 
consistent with the evolution of the southbound API standardization activities~\cite{bosshart2013-1,onf2013}. 


\coloredtext{
\vspace{2mm}
\noindent \textit{Evolving Switch Designs \& Hardware Enhancements}

As in any soft\-wa\-re/hard\-wa\-re innovation cycle, a number of advancements are to be expected from the hardware 
perspective to improve SDN capabilities and performance. 
New SDN switch designs are appearing in a myriad of hardware 
combinations to efficiently work together with TCAMs, such as SRAM, RLDRAM, DRAM, GPU, FPGA, NPs, CPUs, 
among other specialized network processors~\cite{ferkouss2011,naous2008,memon2013,luo2009,rostami2012,pongracz2013}.} These early works suggest the need for additional efforts into new hardware architectures for future 
SDN switching devices. For instance, some proposals target technologies such as GPUs that have demonstrated 
20 Gbps with flow tables of up to 1M exact match entries and up to 1K wildcard entries~\cite{memon2013}. 
Alternatives to TCAM-based designs include new hardware architectures and components, as well as new and more 
scalable forwarding planes, such as the one proposed by the Rain Man firmware~\cite{stephens2012}.
Other design solutions, such as parallel lookup models~\cite{li2013}, can also be applied to SDN to reduce 
costs in switching and routing devices. Recent proposals on cache-like OpenFlow switch arrangements~\cite{katta2013} shed some light on overcoming the practical limitations of 
flow table sizes with clever switching designs. Additionally, counters represent another practical challenge 
in SDN hardware implementations. Many counters already exists, and they could lead to significant control 
plane monitoring overhead~\cite{curtis2011}. Software-defined counters 
(SDC)~\cite{mogul2012} have been proposed to provide both scalability and flexibility.

\coloredtext{
Application-aware SDN architectures are being proposed to generalize the standard OpenFlow forwarding abstractions by including stateful actions to allow processing information from layers 4 to 7~\cite{Mekky2014-4}.
To this end, application flow tables are proposed as data plane application modules that require only local state, i.e., do not depend on a global view of the network. Those tiny application modules run inside the forwarding devices (and can be installed on-demand), alleviating the overhead on the control plane and augmenting the efficiency of certain tasks, which can be kept in the data plane. 
Similarly, other initiatives propose solutions based on pre-installed state machines.
FAST (Flow-level State Transitions)~\cite{Moshref2014_4} allows controllers to proactively program state transitions in forwarding devices, allowing switches to run dynamic actions that require only local information.

Other approaches towards evolving switch designs include CAching in Buckets (CAB), a reactive wildcard caching proposal that uses a geometric representation of the rule set, which is divided into small logical structures (buckets)~\cite{Yan2014CAB}. 
Through this technique CAB is able to solve the rule dependency problem and achieve efficient usage of control plane resources, namely bandwidth, controller processing load, and flow setup latency.
 
New programmable Ethernet switch chips, such as XPliant Ethernet~\cite{McGillicuddy2014_4}, are emerging into this new market of programmable networks.
Its main aim is enabling new protocol support and the addition of new features through software updates, increasing flexibility.
One example of such flexibility is the support of GENEVE~\cite{Gross2014_4}, a recent effort towards generic network virtualization encapsulation protocols, and OpenFlow.
The throughput of the first family of XPliant Ethernet chip varies from 880 Gbps to 3.2 Tbps, supporting up to 64 ports of 40 GbE or 50 GbE, for instance. 
}


Microchip companies like Intel are already shipping processors with flexible SDN capabilities to the 
market~\cite{intelprocessors2012}. Recent advances in general-purpose CPU technology include a
data-plane development kit~(DPDK)~\cite{intelcorporation2014} that allows high-level programming of how data 
packets shall be processed directly within network interface cards. Prototype implementations of Intel 
DPDK accelerated switch shows the potential to deliver high-performance SDN software switches~\cite{pongracz2013}. This trend is likely to 
continue since high-speed and specialized hardware is needed to boost SDN performance and scalability for 
large, real-world networks. Hardware-programmable technologies such as FPGA are widely used to reduce time 
and costs of hardware-based feature implementations. NetFPGA, for instance, has been a pioneering technology 
used to implement OpenFlow 1.0 switches~\cite{naous2008}, providing a commodity 
cost-effective prototyping solution. Another line of work on SDN data planes proposes to augment switches 
with FPGA to (remotely) define the queue management and scheduling behaviour of packet 	
switches~\cite{sivaraman2013}.
\coloredtext{
Finally, recent developments have shown that state-of-the-art System-on-chip (SoC) platforms, such as the Xilinx Zynq ZC706 board, can be used to implement OpenFlow devices yielding 88 Gpbs throughput for 1K flow supporting dynamic updates~\cite{Zhou2014-4}. 
}

\vspace{2mm}
\noindent \textit{Native SDN Switch Designs}

Most of the SDN switch (re)design efforts so far follow an evolutionary approach to retrofit 
OpenFlow-specific programmable features into existing hardware layouts, following common wisdom 
on switch/router designs and consolidated technologies (e.g., SRAM, TCAM, FPGA). One departure from 
this approach is the ongoing work on \textit{forwarding metamorphosis}~\cite{bosshart2013-1}, 
a reconfigurable match table model inspired from RISC-like pipeline architecture applied to switching 
chips. This work illustrates the feasibility of realizing a minimal set of action primitives for flexible 
header processing in hardware, at almost no additional cost or power. Also in line with the core SDN goals 
of highly flexible and programmable (hardware-based) data planes, Protocol-Oblivious Forwarding (POF)~\cite{song2013-1} aims at overcoming some of the limitations of OpenFlow (e.g., expressiveness, 
support of user-defined protocols, memory efficiency), through generic flow instruction sets.
Open-source prototypes are available~\cite{song2013} as well as evaluation 
results showing the line-speed capabilities using a network processing unit (NPU)-based~\cite{Hauger2009} proof of concept implementation. 
\coloredtext{In this line, we already mentioned OpenState~\cite{bianchi2014}, another initiative that aims to augment the capability and flexibility of forwarding devices. 
By taking advantage of eXtended Finite State Machines (XFSM)~\cite{bianchi2012,tinnirello2012}, OpenState proposes an abstraction -- as a super set of OpenFlow primitives -- to enable stateful handling of OpenFlow rules inside forwarding devices. 
}

In the same way as TTPs allow controllers to compile the right set of low-lever instructions known to be supported by the switches, a new breed of switch referred to as P4 (programmable, protocol-independent packet
processor)~\cite{bosshart2013} suggests an evolution path for OpenFlow, based on a high-level compiler. 
This proposal would allow the functionality of programmable switches (i.e., pipeline, header parsing, field matching) to be not only specified by the controller but also changed in the field.
In this model, programmers are able to decide how the forwarding plane processes packets without caring about implementation details.
It is then the compiler that transforms the imperative program into a control flow graph that can be mapped to different target switches.

\subsection{Controller Platforms}

In the SDN model, the controller platform is a critical pillar of the architecture, and, as such, 
efforts are being devoted to turn SDN controllers into high-performance, scalable, distributed, 
modular, and highly-available programmer-friendly software.
\coloredtext{Distributed controller platforms, in particular, have to address a variety of challenges. 
Deserving special consideration are the latency between forwarding devices and controller instances,  fault-tolerance, load-balancing, consistency and synchronization,  
among other issues~\cite{koponen-1,schmid2013,Berde2014ONOS}.
Operators should also be able to observe and understand how the combination of different functions and modules can impact their network~\cite{Volpano2014_4}.
}



As the SDN community learns from the development and operational experiences with OpenFlow controllers (e.g., Beacon~\cite{erickson2013}), further advancements are expected in terms of raw performance of controller implementations, including the exploitation of hierarchical designs and optimized buffer sizing~\cite{azodolmolky2013-2}.  
\coloredtext{One approach to increase the performance of controllers is the IRIS IO engine~\cite{Park2014}, enabling  significant increases 
in the flow-setup rate of SDN controllers. 
Another way of reducing the control plane overhead is by keeping a compressed copy of the flow tables in the controller's memory~\cite{zhang2014-4}.
}


\vspace{2mm}\coloredtext{
\noindent \textit{Modularity \& Flexibility}}

\coloredtext{A series of ongoing research efforts target the modular and flexible composition of controllers. 
RAON~\cite{Park2014-1} proposes a recursive abstraction of OpenFlow controllers where each controller sees the controllers below as OpenFlow switches. 
Open research issues include the definition of suitable interfaces between the different layers in such a hierarchy of controllers. 

Other open issues to be further investigated in this context are the East/westbound APIs, and their use in enabling suitable hierarchical designs to achieve scalability, modularity, and security~\cite{ONF2014SDNarch}. 
For instance, each level of a hierarchy of controllers can offer different abstractions and scopes for either intra- and inter-data center routing, thus increasing scalability and modularity.
Similarly, from a security perspective, each hierarchical level may be part of a different trust domain.
Therefore, east/westbound interfaces between the different layers of controllers should be capable of enforcing both intra- and inter-domain security policies.

Another important observation is that, currently, the lack of modularity in most SDN controllers forces developers to re-implement basic network services from scratch in each new application~\cite{Casado2014_4}. 
} 
As in software engineering in general, lack of modularity results in controller 
implementations that are hard to build, maintain, and extend -- and ultimately become resistant to further 
innovations, resembling traditional ``hardware-defined'' networks. As surveyed in Section~\ref{sec:programminglanguages}, 
SDN programming abstractions (e.g.,~Pyretic~\cite{monsanto2013}) introduce modularity in SDN 
applications and simplify their development altogether. Further research efforts (e.g., 
Corybantic~\cite{auyoung2013}) try to achieve modularity in SDN control programs. Other 
contributions towards achieving modular controllers can be expected from other areas of computer science 
(e.g., principles from Operating System~\cite{monaco2013}) and best practices of modern 
cloud-scale software applications.

%
\vspace{2mm}
\noindent \textit{Interoperability and application portability}

Similarly to forwarding device vendor agnosticism that stems from standard southbound interfaces, it is important to foster interoperability between controllers.
Early initiatives towards more interoperable control platforms include portable programming languages such as Pyretic~\cite{monsanto2013} and east/westbound interfaces among controllers, such as SDNi~\cite{yin2012}, ForCES CE-CE interface~\cite{doria2010,wang2011-1}, 
and ForCES Intra-NE mechanisms~\cite{ogawa2013}.
However, these efforts are yet far from fully realizing controller interoperability and application portability.

\coloredtext{
In contrast to Pyretic~\cite{reich2013}, PANE~\cite{ferguson2013},
Maple~\cite{voellmy2013}, and Corybantic~\cite{auyoung2013}, which are
restricted to traffic engineering applications and/or impose
network state conflict resolution at the application level (making
application design and testing more complicated),
Statesman~\cite{Sun2014nms} proposes a framework to enable a
variety of loosely-coupled network applications to co-exist on the
same control plane without compromising network safety and
performance.
This framework makes application development simpler by automatically and transparently resolving conflicts.
In other words, Statesman allows a safe composition of uncoordinated
or conflicting application's actions. 
}

\coloredtext{
Another recent approach to simplify network management is the idea of compositional SDN hypervisors~\cite{Jin2014_4}.
Its main feature is allowing applications written in different languages, or on different platforms, to work together in processing the same traffic.
The key integration component is a set of simple prioritized lists of OpenFlow rules, which can be generated by different programming languages or applications. 
}

\vspace{2mm}
\noindent \textit{High-Availability}

In production, SDN controllers need to sustain healthy operation 
under the pressure of different objectives from the applications they
host. Many advances are called for in order
to deal with potential risk vectors of controller-based solutions~\cite{kreutz2013}. Certainly, 
many solutions will leverage on results from the distributed systems and security communities made over the last decade. 
\coloredtext{For instance, recent efforts propose consistent, fault-tolerant data stores for building reliable distributed controllers~\cite{botelho2013,Botelho2014,Berde2014ONOS}.
}

\coloredtext{Another possible approach towards building low latency, highly available SDN controllers is to exploit controller locality~\cite{schmid2013,levin2012}.
Classical models of distributed systems, such as LOCAL and CONGEST~\cite{Peleg2000}, can be explored to solve this problem.
Those models can be used to develop coordination protocols that enable each controller to take independent actions over events that take place in its local neighborhood~\cite{schmid2013}. 
}

\coloredtext{Another core challenge relates to the fundamental trade-offs between the consistency model of state distribution in distributed SDN controllers, the consistency requirements of control applications, and performance~\cite{levin2012}.
To ease development, the application should ideally not be aware of the vagaries of distributed state.
This implies a strong consistency model, which can be achieved with distributed data stores as proposed recently~\cite{botelho2013}.
However, keeping all control data in a consistent distributed data store is unfeasible due to the inherent performance penalties.
Therefore, hybrid solutions are likely to co-exist requiring application developers to be aware of the trade-offs and penalties of using, or not, a strong consistency model, a tenet of the distributed Onix controller~\cite{koponen-1}. 
}

\coloredtext{High availability can also be achieved through improved southbound APIs and controller placement heuristics and formal models~\cite{Ros2014_4,philip2014cross, Borokhovich2014_4}. These aim to maximize resilience and scalability by allowing forwarding devices to connect to multiple controllers in a cost-effective and efficient way~\cite{philip2014cross}.
Early efforts in this direction have already shown that forwarding devices connecting to two or three controllers can typically achieve high availability (up to five nines) and robustness in terms of control plane connectivity~\cite{Ros2014_4,Borokhovich2014_4}.
It has also been shown that the number of required controllers is more dependent on the topology than on network size~\cite{Ros2014_4}.
Another finding worth mentioning is the fact that for most common topologies and network sizes fewer than ten controllers seem to be enough~\cite{Ros2014_4}. 
}


%

\coloredtext{
\vspace{2mm}
\noindent \textit{Delegation of control}

To increase operational efficiency, SDN controllers can delegate control functions to report state and attribute value changes, threshold crossing alerts, hardware failures, and so forth.
These notifications typically follow a publish/subscribe model, i.e., controllers and applications subscribe (on-demand) to the particular class of notifications they are interested in.
In addition, these subsystems may provide resilience and trustworthiness properties~\cite{Kreutz2012_fit}. 

Some reasons for delegating control to the data plane include~\cite{ONF2014SDNarch}:
\begin{itemize}
\item Low latency response to a variety of network events;
\item The amount of traffic that must be processed in the data plane, in particular in large-scale networks such as data centers;
\item Low level functions such as those (byte- or bit-oriented) required by repetitive SDH (Synchronous Digital Hierarchy)~\cite{tesink2003sdh} multiplex section overhead;
\item Functions well-understood and standardized, such as encryption, BIP~\cite{Prasanna2002bip}, AIS~\cite{swallow2011rfc} insertion, MAC learning, and CCM (Codec Control Messages)~\cite{whiting2003ccm} exchanges;
\item Controller failure tolerance, i.e., essential network functions should be able to keep a basic network operation even when controllers are down;
\item Basic low-level functions usually available in data plane silicon, such as protection switching state machines, CCM counters and timers;
\item All those functions that do not add any value when moved from the data to the control plane.
\end{itemize}

Strong candidates for execution in the forwarding devices instead of being implemented in the control platforms thus include OAM, ICMP processing, MAC learning, neighbor discovery, defect recognition and integration~\cite{ONF2014SDNarch}.
This would not only reduce the overhead (traffic and computing) of the control plane, but also improve network efficiency by keeping basic networking functions in the data plane.

}

\subsection{Resilience}
\label{sec:resiliency}

Achieving resilient communication is a top purpose of networking.
As such, SDNs are expected to yield the same levels of availability as legacy and any new alternative technology. 
Split control architectures as SDN are commonly questioned~\cite{desai2010} about their
actual capability of being resilient to faults that may compromise
the control-to-data plane communications and thus result in
``brainless'' networks.
Indeed, the malfunctioning of particular SDN elements should not result in the loss
of availability. The relocation of SDN control plane functionality,
from inside the boxes to remote, logically centralized loci, becomes a
challenge when considering critical control plane functions such as
those related to link failure detection or fast reaction decisions.
The resilience of an OpenFlow network depends on fault-tolerance in the data plane (as in traditional networks) but also on the high availability of the (logically) centralized control plane functions. Hence, the resilience of SDN is challenging due to the multiple possible failures of the different pieces of the architecture. 

As noted in~\cite{kim2012}, there is a lack of sufficient research and experience in building and operating fault-tolerant SDNs.
Google B4~\cite{jain2013-1} may be one of the few examples that have proven that SDN can be resilient at scale. 
A number of related \coloredtext{efforts~\cite{kempf2012,sharma2013-1,panda2013,reitblatt2013,ku'zniar2013,dixit2013,ramos2013,araujosoftware,Krishnamurthy2014_4}} have started to tackle the concerns around control plane split architectures. 
The distributed controller architectures surveyed in Section~\ref{sec:controllers} are examples of approaches towards resilient SDN controller platforms with different tradeoffs in terms of consistency, durability and scalability. 

On a detailed discussion on whether the CAP theorem~\cite{Brewer2012CAP} applies to networks, Panda et al.~\cite{panda2013} argue that the trade-offs in building consistent, available and
partition-tolerant distributed databases (i.e., CAP theorem) are applicable
to SDN. 
The CAP theorem demonstrates that it is impossible for
datastore systems to simultaneously achieve strong consistency,
availability and partition tolerance.
While availability and partition tolerance problems are similar in both distributed databases and networks, the problem of consistency in SDN relates to the consistent application of policies.

Considering an OpenFlow network, when a switch detects a link failure (\texttt{port-down} event), a notification is sent to the controller, which then takes the required actions (re-route computation, pre-computed back-up path lookup) and installs updated flow entries in the required switches to redirect the affected traffic. Such reactive strategies imply (1) high restoration time due to the necessary interaction with the controller; and (2) additional
load on the control channel. 
One experimental work on OpenFlow for carrier-grade networks investigated the restoration process and measured a restoration times in the order of 100 ms~\cite{sharma2013-1}. The delay introduced by the controller may, in some cases, be prohibitive. 

\coloredtext{In order to meet carrier grade requirements (e.g., 50 ms of recovery time),  protection schemes are required to mitigate the effects of a separate control plane.
Suitable protection mechanisms (e.g., installation of pre-established  backup paths in the forwarding devices) can be implemented by means of OpenFlow group table entries using ``fast-failover'' actions. 
An OpenFlow fault management approach~\cite{kempf2012} similar to MPLS global path protection could also be a viable solution, provided that OpenFlow switches are extended with end-to-end path monitoring capabilities similarly to those specified by Bidirectional Forwarding Detection (BFD)~\cite{katz2010bfd}. 
Such protection schemes are a critical design choice for larger scale networks and may also require considerable additional flow space. 
By using primary and secondary path pairs programmed as OpenFlow fast failover group table entries, a path restoration time of 3.3 ms has been reported~\cite{Adrichem2014} using BFD sessions to quickly detect link failures.
}

On a related line of data plane resilience, SlickFlow~\cite{ramos2013} leverages the idea of using packet header space to carry alternative path information to implement resilient source routing in OpenFlow networks. Under the presence of failures along a primary path, packets can be rerouted to alternative paths by the switches themselves without involving the controller.  
Another recent proposal that uses in-packet information is INFLEX~\cite{araujosoftware}, an SDN-based architecture
for cross-layer network resilience which provides on-demand path fail-over by having end-points tag packets with virtual routing plane information that can be used by egress routers to re-route by changing tags upon failure detection.

\coloredtext{Similarly to SlickFlow, OSP~\cite{sgambelluri2013opt} proposes a protection approach for data plane resilience. It is based on protecting individual segments of a path avoiding the intervention of the controller upon
failure. The recovery time depends on the failure detection time, i.e., a few tens of milliseconds in the proposed scenarios.
In the same direction, other proposals are starting to appear for enabling fast failover mechanisms for link 
protection and restoration in OpenFlow-based networks~\cite{Sahri2014fast}.}

Language-based solutions to the data plane fault-tolerance problem have also been proposed~\cite{reitblatt2013}. In this work the authors propose a language that compiles regular expressions into OpenFlow rules to express what network paths packets may take and what degree of (link level) fault tolerance is required. Such abstractions around fault tolerance allow developers to build fault recovery capabilities into applications without huge coding efforts.

\subsection{Scalability}
\label{sec:scaling}

Scalability has been one of the major concerns of SDNs from the outset.
This is a problem that needs to be addressed in any system -- e.g., in traditional networks -- and is obviously also a matter of much discussion in the context of SDN~\cite{yeganeh2013}.

Most of the scalability concerns in SDNs are related to the decoupling of the control and data planes. 
Of particular relevance are reactive network configurations where the first packet of a new flow is sent by the first forwarding element to the controller.
The additional control plane traffic increases network load and makes the control plane a potential bottleneck. 
Additionally, as the flow tables of switches are configured in real-time by an outside 
entity, there is also the extra latency introduced by the flow setup process.
In large-scale networks controllers will need to be able to process millions of flows per second~\cite{Benson2010DC} without compromising the quality of its service.
Therefore, these overheads on the control plane and on flow setup latency are (arguably) two of the major scaling concerns in SDN.

As a result, several efforts have been devoted to tackle the SDN scaling concerns, including
DevoFlow~\cite{curtis2011}, 
Software-Defined Counters (SDCs)~\cite{mogul2012},
DIFANE~\cite{yu2010-1},
Onix~\cite{koponen-1}, 
HyperFlow~\cite{tootoonchian2010},
Kandoo~\cite{yeganeh2012},
Maestro~\cite{cai2011},
NOX-MT~\cite{tootoonchian2012}, and 
Maple~\cite{voellmy2013}. 
\coloredtext{Still related to scalability, the notion of elasticity in SDN controllers is also being pursued~\cite{dixit2013,Krishnamurthy2014_4}}.
Elastic approaches include dynamically changing 
the number of controllers and their locations under different conditions~\cite{bari2013-1}.

Most of the research efforts addressing scaling limitations of SDN can be classified in three categories: data 
plane, control plane, and hybrid.
While targeting the data plane, proposals such as 
DevoFlow~\cite{curtis2011} and Software-Defined Counters 
(SDC)~\cite{mogul2012} actually reduce the overhead of the control plane by delegating some work to the forwarding devices.
For instance, instead of requesting a decision from the
controller for every flow, switches can selectively identify the flows (e.g., elephant flows) that may need higher-level decisions from the control plane applications.
Another example is to introduce 
more powerful general purpose CPUs in the forwarding devices to enable SDCs. 
A general purpose CPU and software-defined counters offer new possibilities for reducing the control plane overhead by allowing 
software-based implementations of functions for data aggregation and compression, for instance.

Maestro~\cite{cai2011},
NOX-MT~\cite{tootoonchian2012},
Kandoo~\cite{yeganeh2012},
Beacon~\cite{erickson2013}, and
Maple~\cite{voellmy2013}
are examples of the effort on designing and deploying high performance controllers, i.e., trying to increase the performance of the control plane.
These controllers mainly explore well-known techniques from networking, computer architectures and high performance computing, such as buffering, pipelining and parallelism, to increase the throughput of the control platform.

The hybrid category is comprised of solutions that try to split the control logic functions between specialized data plane devices and controllers. 
In this category, DIFANE~\cite{yu2010-1} proposes authoritative 
(intermediate) switches to keep all traffic in the data plane, targeting a more scalable and efficient control plane. 
Authoritative switches are responsible for installing rules on the remaining switches, while the controller 
is still responsible for generating all the rules required by the logic of applications.
By dividing the controller work with these special switches, the overall system scales better.
 
Table~\ref{tab:scalabilitysolutions} provides a non-exhaustive list of proposals addressing scalability issues of 
SDN. 
We characterize these issues by application domain (control or data plane), their purpose, the throughput in terms of number of flows per second (when the results of the experiments are reported), and the strategies used.
As can be observed, the vast majority are control plane solutions that try to increase scalability by using distributed and multi-core architectures.

{\renewcommand{\arraystretch}{1.4}
\begin{table*}[!htp]
\caption{Summary and characterization of scalability proposals for SDNs.}
\label{tab:scalabilitysolutions}
\begin{center}
\footnotesize
\begin{tabularx}{\linewidth}{p{1.9cm}p{1.6cm}p{2.8cm}p{2.5cm}p{1.0cm}X}
\hline
\textbf{Solution} & \textbf{Domain} & \textbf{Proposes} & \textbf{Main purpose} & \textbf{Flows/s} & \textbf{Resorts to} \\\hline
Beacon~\cite{erickson2013} & control plane & a multi-threaded controller & improve controller performance & 12.8M & High performance flow processing capabilities using pipeline threads and shared queues. \\\hline
Beacon cluster~\cite{yazici2012} & control plane & coordination framework & create clusters of controllers & 6.2M & A coordination framework to create high-performance clusters of controllers. \\\hline
DevoFlow~\cite{curtis2011} & data plane & thresholds for counters, type of flow detection & reduce the control plane overhead & --- & Reduce the control traffic generated by counters statistics monitoring. \\\hline
DIFANE~\cite{yu2010-1} & control and data plane & authoritative specialized switches & improve data plane performance & 500K & Maintain flows in the data plane reducing controller work. \\\hline
Floodlight~\cite{erickson2013} & control plane & a multi-threaded controller & Improve controller performance & 1.2M & High performance flow processing capabilities. \\\hline
HyperFlow~\cite{tootoonchian2010} & control plane & a distributed controller & distribute the control plane & --- & Application on top of NOX to provide control message distribution among controllers. \\\hline
Kandoo~\cite{yeganeh2012} & control plane & a hierarchical controller & distribute the control plane hierarchically & --- & Use two levels of controller (local and root) to reduce control traffic. \\\hline
Maestro~\cite{cai2011} & control plane & a multi-threaded controller & improve controller performance & 4.8M & High performance flow processing capabilities. \\\hline
Maestro cluster~\cite{yazici2012} & control plane & coordination framework & create clusters of controllers & 1.8M & A coordination framework to create high-performance clusters of controllers. \\\hline
Maple~\cite{voellmy2013} & control plane & programming language & scaling algorithmic policies & 20M & Algorithmic policies and user- and OS-level threads on multicore systems (e.g., 40+ cores). \\\hline
NOX~\cite{erickson2013} & control plane & a multi-threaded controller & improve controller performance & 5.3M & High performance flow processing capabilities. \\\hline
NOX-MT~\cite{tootoonchian2012} & control plane & a multi-threaded controller & improve controller performance & 1.8M & High performance flow processing capabilities. \\\hline
NOX cluster~\cite{yazici2012} & control plane & coordination framework & create clusters of controllers & 3.2M & A coordination framework to create high-performance clusters of controllers. \\\hline
Onix~\cite{koponen-1} & control plane & a distributed control platform & robust and scalable control platform & --- & Provide a programmable and flexible distributed NIB for application programmers.\\\hline
SDCs~\cite{mogul2012} & data plane & Software-Defined Counters & reduce the control plane overhead & --- & Remove counters from the ASIC to a general purpose CPU, improving programmability. \\\hline
\end{tabularx}
\end{center}
\end{table*}
}

Some figures are relatively impressive, with some solutions achieving up to 20M flows/s. 
However, we should caution the reader that current evaluations consider only simple applications and count basically the number of \texttt{packet-in} and \texttt{packet-out} messages to measure throughput. 
The actual performance of controllers will be affected by other factors, such as the number and complexity of the applications running on the controller and security mechanisms implemented.
For example, a routing algorithm consumes more computing resources and needs more time to execute than a simple learning switch application.
Also, current evaluations are done using plain TCP connections.
The performance is very likely to change when basic security mechanisms are put in place, such as TLS, or more advanced mechanisms to avoid eavesdropping, man-in-the-middle and DoS attacks on the control plane.

Another important issue concerning scalability is data distribution among controller replicas in distributed architectures.
Distributed control platforms rely on data distribution mechanisms to achieve their goals.
For instance, controllers such as Onix, HyperFlow, and ONOS need mechanisms to keep a consistent state in the distributed control platform.
Recently, experimental evaluations have shown that high performance distributed and fault-tolerant data stores can be used to tackle such challenges~\cite{botelho2013}. 
Nevertheless, further work is necessary to properly understand state distribution trade-offs~\cite{levin2012}.

\subsection{Performance evaluation}
\label{sec:performance-eval}

As introduced in Section~\ref{sec:infrastructure}, there are already several OpenFlow implementations from 
hardware and software vendors being deployed in different types of networks, from small enterprise 
to large-scale data centers. Therefore, a growing number of experiments over SDN-enabled networks is expected 
in the near future. This will naturally create new challenges, as questions regarding SDN performance and 
scalability have not yet been properly investigated. Understanding the performance and limitation of the SDN 
concept is a requirement for its implementation in production networks. There are very few performance 
evaluation studies of OpenFlow and SDN architecture. Although simulation studies and experimentation are 
among the most widely used performance evaluation techniques, analytical modeling has its own benefits too. 
A closed-form description of a networking architecture paves the way for network designers to have a quick 
(and approximate) estimate of the performance of their design, without the need to spend considerable time 
for simulation studies or expensive experimental setup~\cite{Kobayashi2014151}.

Some work has investigated ways to improve the performance of switching capabilities in SDN.
These mainly consist of observing the performance of OpenFlow-enabled networks regarding different aspects, such as lookup performance~\cite{jarschel2011}, hardware acceleration~\cite{luo2009}, the influence of types of rules and packet sizes~\cite{bianco2010}, performance bottlenecks of current OpenFlow implementations~\cite{curtis2011}, how reactive settings impact the performance on data center networks~\cite{pries2012}, and the impact of configuration on OpenFlow switches~\cite{sherwood2011}.

Design choices can have a significant impact on the lookup performance of OpenFlow switching in Linux 
operating system using standard commodity network interface cards~\cite{jarschel2011}. 
Just by using commodity network hardware the packet switching throughput can be improved by up to 25\% when compared to one based on soft OpenFlow switching~\cite{jarschel2011}. Similarly, hardware acceleration based on network processors can also be applied to perform OpenFlow switching. 
In such cases, early reports indicate that performance, in terms of packet delay, can be improved by 20\% when compared to conventional designs~\cite{luo2009}.

By utilizing Intel's DPDK library~\cite{intelcorporation2014}, it has been shown that is possible to provide 
flexible traffic steering capability at the hypervisor level (e.g., KVM) without the performance limitations 
imposed by traditional hardware switching techniques~\cite{hwang2014}, such as SR-IOV~\cite{dong2008}.
This is particularly relevant since most of the current enterprise deployments of SDN are in virtualized data 
center infrastructures, as in VMware's NVP solution~\cite{koponen}.

Current OpenFlow switch implementations can lead to performance bottlenecks with respect to the CPU 
load~\cite{curtis2011}. Yet, modifications on the protocol specification can help reduce the occurrence 
of these bottlenecks. Further investigations provide measurements regarding the performance of the OpenFlow switch for different types of rules and packet sizes~\cite{bianco2010}.

In data centers, a reactive setting of flow rules can lead to an unacceptable performance when only eight 
switches are handled by one OpenFlow controller~\cite{pries2012}. This means that large-scale SDN deployments should probably not rely on a purely reactive ``modus operandi'', but rather on a combination of proactive and reactive flow setup.

To foster the evaluation of different performance aspects of OpenFlow devices, frameworks such as 
\coloredtext{OFLOPS~\cite{rotsos2012-1},  OFLOPS-Turbo~\cite{Rotsos2014},} Cbench~\cite{tootoonchian2012}, and OFCBenchmark~\cite{jarschel2012} have been proposed. They provide a set of tools to analyze the performance of OpenFlow switches and controllers.
Cbench~\cite{tootoonchian2012,sherwood2011} is a benchmark tool developed to evaluate the performance of OpenFlow controllers. By taking advantage of the Cbench, it is possible to identify performance improvements for OpenFlow controllers based on different environment and system configurations, such as the number of forwarding devices, network topology, overall network workload, type of equipments, forwarding complexity, and overhead of the applications being executed on top of controllers~\cite{tootoonchian2012}.
Therefore, such tools can help system designers make better decisions regarding the performance of devices and the network, while also allowing end-users to measure the device performance and better decide which one is best suited for the target network infrastructure.

Surprisingly, despite being designed to evaluate the performance of controllers, Cbench is currently a single-threaded tool. Therefore, multiple instances have to be started to utilize multiple CPUs. It also only establishes one controller connection for all emulated switches. Unfortunately, this means little can be derived from the results in terms 
of controller performance and behavior or estimation of different bounds at the moment. For instance, aggregated statistics are gathered for all switches but not for each individual switch. As a result, it is not possible to identify whether all responses of the controller are for a single switch, or whether the capacity of the controller is actually shared among the switches. Flexible OpenFlow controller benchmarks are available though. 
OFCBenchmark~\cite{jarschel2012} is one of the recent developments. It creates a set of message-generating virtual switches, which can be configured independently from each other to emulate a specific scenario and to maintain their own statistics.


Another interesting question to pose when evaluating the performance of SDN architectures is what is the required number of controllers for a given network topology and where to place the controllers~\cite{heller2012,philip2014cross}. 
By analyzing the performance of controllers in different network topologies, it is possible to conclude that one controller is often enough to keep the latency at a reasonable rate~\cite{heller2012}. 
Moreover, as observed in the same experiments, in the general case adding $k$ controllers to the network can reduce the latency by a factor of $k$. However, there are cases, such as large scale networks and WANs, where more controllers should be deployed to achieve high reliability and low control plane latency. 

Recent studies also show that the SDN control plane cannot be fully physically centralized due to responsiveness, reliability and scalability metrics~\cite{levin2012,philip2014cross}. 
Therefore, distributed controllers are the natural choice for creating a logically centralized control plane, while being capable of coping with the demands of large scale networks. 
However, distributed controllers bring additional challenges, such as the consistency of the global network view, which can significantly affect the performance of the network if not carefully engineered. 
Taking two applications as examples, one that ignores inconsistencies and another that takes inconsistency into consideration, it is possible to observe that optimality is significantly affected when inconsistencies are not considered and that the robustness of an application is increased when the controller is aware of the network state distribution~\cite{levin2012}.

Most of these initiatives towards identifying the limitations and bottlenecks of SDN architectures can take a lot of time and effort to produce consistent outputs due to the practical development and experimentation requirements.
As mentioned before, analytic models can quickly provide performance indicators and potential scalability bottlenecks for an OpenFlow switch-controller system before detailed data is available. 
While simulation can provide detailed insight into a certain configuration, the analytical model greatly simplifies a conceptual deployment decision. 
For instance, a Network calculus-based model can be used to evaluate the performance of an SDN switch and the interaction of SDN switches and controllers~\cite{azodolmolky2013-3}.  
The proposed SDN switch model captured the closed form of the packet delay and buffer length inside the SDN switch according to the parameters of a cumulative arrival process. 
Using recent measurements, the authors have reproduced the packet processing delay of two variants of OpenFlow switches and computed the buffer requirements of an OpenFlow controller.
Analytic models based on queuing theory for the forwarding speed and blocking probability of current OpenFlow 
switches can also be used to estimate the performance of the network~\cite{jarschel2011}.



\subsection{Security and Dependability}
\label{secSecurity}

Cyber-attacks against financial institutions, energy facilities, government units and research institutions 
are becoming one of the top concerns of governments and agencies around the globe~\cite{marchetti2012,amin2012,nicholson2012,choo2011,kushner2013,perez-pena2013}.
Different incidents, such as Stuxnet~\cite{kushner2013}, have already shown the persistence of threat
vectors~\cite{tankard2011}. Put another way, these attacks are capable of damaging a nation's wide 
infrastructure, which represent a significant and concerning issue. As expected, one of the most common means 
of executing those attacks is through the network, either the Internet or the local area network. It can be used 
as a simple transport infrastructure for the attack or as a potentialized weapon to amplify the impact of the 
attack. For instance, high capacity networks can be used to launch large-scale attacks, even though the attacker 
has only a low capacity network connection at his premises.

Due to the danger of cyber-attacks and the current landscape of digital threats, security and dependability are 
top priorities in SDN. While research and experimentation on software-defined networks is being conducted by 
some commercial players (e.g., Google, Yahoo!, Rackspace, Microsoft), commercial adoption is still in its early 
stage. Industry experts believe that security and dependability are issues that need to be addressed and further 
investigated in SDN~\cite{kreutz2013,sorensen2012,kerner2013}.

Additionally, from the dependability perspective, availability of Internet routers is nowadays a major concern 
with the widespread of clouds and their strong expectations about the network~\cite{agapi2011}. It is 
therefore crucial to achieve high levels of availability on \coloredtext{SDN control platforms if they are to become the main pillars of networked applications~\cite{Ros2014_4}}. 

Different threat vectors have already been identified in SDN architectures~\cite{kreutz2013}, as well 
as several security issues and weaknesses in OpenFlow-based networks\coloredtext{~\cite{kloti2013,wasserman2013,shin2013,porras2012,benton2013,shin2014,scotthayward2013sec,sezer2013are}}.
While some threat vectors are common to existing networks, others are more specific to SDN, such as attacks on 
control plane communication and logically-centralized controllers. It is worth mentioning that most threats vectors 
are independent of the technology or the protocol (e.g., OpenFlow, POF, ForCES), because they represent threats on 
conceptual and architectural layers of SDN itself.

%

\begin{figure}[t!]
\centering
\includegraphics[width=0.85\columnwidth]{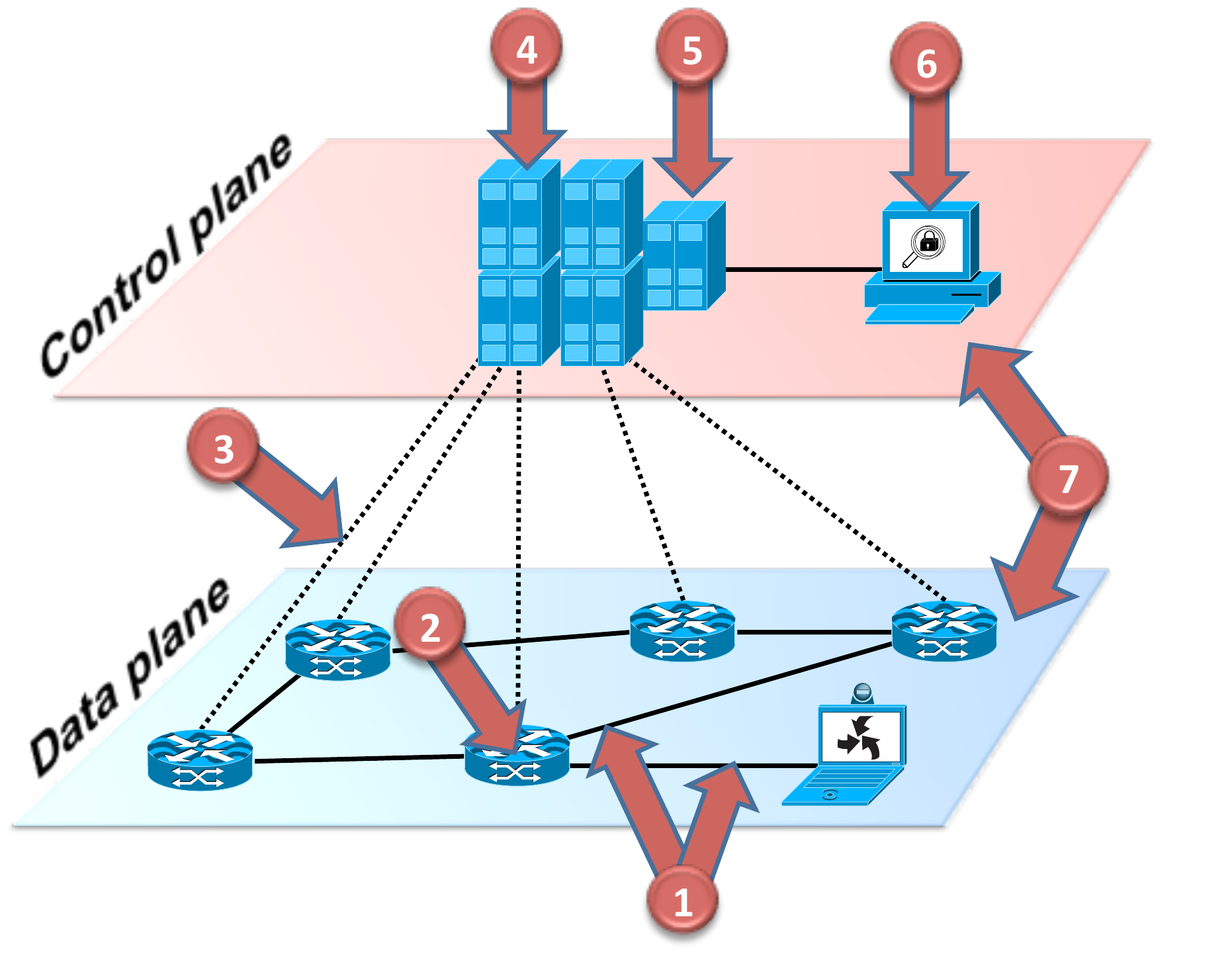}
\caption{Main threat vectors of SDN architectures}
\label{fig:threatvectorsmap}
\end{figure}

As shown in Figure~\ref{fig:threatvectorsmap} and Table~\ref{tab:newandoldproblems}, there are at least seven 
identified threats vector in SDN architectures. The first threat vector consists of forged or faked traffic flows 
in the data plane, which can be used to attack forwarding devices and controllers. The second allows 
an attacker to exploit vulnerabilities of forwarding devices and consequently wreak havoc with the network. Threat 
vectors three, four and five are the most dangerous ones, since they can compromise the network operation. Attacks 
on the control plane, controllers and applications can easily grant an attacker the control of the network. For 
instance, a faulty or malicious controller or application could be used to reprogram the entire network for data 
theft purposes, e.g., in a data center. The sixth threat vector is linked to attacks on and vulnerabilities in 
administrative stations. A compromised critical computer, directly connected to the control network, will empower 
the attacker with resources to launch more easily an attack to the controller, for instance. Last, threat vector 
number seven represents the lack of trusted resources for forensics and remediation, which can compromise 
investigations (e.g., forensics analysis) and preclude fast and secure recovery modes for bringing the network 
back into a safe operation condition.

As can be observed in Table~\ref{tab:newandoldproblems}, threat vectors 3 to 5 are specific to SDN as they 
stem from the separation of the control and data planes and the consequent introduction of a new entity in 
these networks -- the logically centralized controller. The other vectors were already present 
in traditional networks. However, the impact of these threats could be larger than today -- or at least it 
may be expressed differently -- and as a consequence it may need to be dealt with differently.

{\renewcommand{\arraystretch}{1.4}
\begin{table}[!ht]
\caption{SDN specific vs. non-specific threats}
\label{tab:newandoldproblems}
\begin{center}
\footnotesize
\begin{tabularx}{\linewidth}{p{1.2cm}p{1.2cm}X}
\hline
\textbf{Threat vectors}  & \textbf{Specific to SDN?}  & \textbf{Consequences in software-defined networks} \\\hline
Vector 1      & no      & Open door for DDoS attacks.\\\hline
Vector 2      & no      & Potential attack inflation.\\\hline
Vector 3      & yes     & Exploiting logically centralized controllers.\\\hline
Vector 4      & yes     & Compromised controller may compromise the entire network.\\\hline
Vector 5      & yes     & Development and deployment of malicious applications on controllers. \\\hline
Vector 6      & no      & Potential attack inflation.\\\hline
Vector 7      & no      & Negative impact on fast recovery and fault diagnosis.\\
\hline
\end{tabularx}
\end{center}
\end{table}
}

OpenFlow networks are subject to a variety of security and dependability problems such as spoofing~\cite{kloti2013}, tampering~\cite{kloti2013}, 
repudiation~\cite{kloti2013}, information disclosure~\cite{kloti2013}, denial of service~\cite{kloti2013,shin2013,benton2013}, elevation of privileges~\cite{kloti2013}, \coloredtext{and the assumption that all applications are benign and will not affect SDN operation~\cite{shin2014}}.
The lack of isolation, protection, access control and stronger security recommendations\coloredtext{~\cite{wasserman2013,shin2013,porras2012,benton2013,shin2014}} are some of the reasons for these vulnerabilities.
We will explore these next.


\vspace{2mm}
\noindent \textit{OpenFlow security assessment}

\coloredtext{
There is already a number of identified security issues in OpenFlow-enabled networks.
Starting from a STRIDE methodology~\cite{hernan2006}, it is possible to identify different attacks to OpenFlow-enabled networks.}
Table~\ref{tab:securitythreatsopenflow} summarizes these attacks (based on ~\cite{kloti2013}). 
For instance, information disclosure can be achieved through side channel attacks targeting the flow rule setup process. 
When reactive flow setup is in place, obtaining information about network operation is relatively easy.
An attacker that measures the delay experienced by the first packet of a flow and the subsequent can easily infer that the target network is a reactive SDN, and proceed with a specialized attack. 
This attack -- known as fingerprinting~\cite{shin2013} -- may be the first step to launch a DoS attack intended to exhaust the resources of the network, for example. 
If the SDN is proactive, guessing its forwarding rule policies is harder, but still feasible~\cite{kloti2013}. 
Interestingly, all reported threats and attacks affect all versions (1.0 to 1.3.1) of the OpenFlow specification. 
It is also worth emphasizing that some attacks, such as spoofing, are not specific to SDN. 
However, these attacks can have a larger impact in SDNs. 
For instance, by spoofing the address of the network controller, the attacker (using a fake controller) could  take over the control of the entire network. 
A smart attack could persist for only a few seconds, i.e., just the time needed to install special rules on all forwarding devices for its malicious purposes (e.g., traffic cloning). 
Such attack could be very hard to detect.

{\renewcommand{\arraystretch}{1.4}
\begin{table*}[!htp]
\caption{Attacks to OpenFlow networks.}
\label{tab:securitythreatsopenflow}
\begin{center}
\footnotesize
\begin{tabularx}{.85\textwidth}{ccX}
\hline
\textbf{Attack}  & \textbf{Security Property} & \textbf{Examples} \\\hline
\textbf{S}poofing & Authentication & MAC and IP address spoofing, forged ARP and IPv6 router advertisement\\\hline
\textbf{T}ampering & Integrity & Counter falsification, rule installation, modification affecting data plane.\\\hline
\textbf{R}epudiation & Non-repudiation & Rule installation, modification for source address forgery.\\\hline
\textbf{I}nformation disclosure & Confidentiality & Side channel attacks to figure out flow rule setup. \\\hline
\textbf{D}enial of service & Availability &  Flow requests overload of the controller. \\\hline
\textbf{E}levation of privilege & Authorization & Controller take-over exploiting implementation flaws. \\\hline
\end{tabularx}
\end{center}
\end{table*}
}

Taking counters falsification as another example, an attacker can try to guess installed flow 
rules and, subsequently, forge packets to artificially increase the counter.
Such attack would be specially critical for billing and load balancing systems, for instance.
A customer could be charged for more traffic than she in fact used, while a load balancing algorithm may take non-optimal decisions due to forged counters. 

Other conceptual and technical security concerns in OpenFlow networks include the lack of strong 
security recommendations for developers, the lack of TLS and access control support on most switch and controller 
implementations~\cite{wasserman2013}, the belief that TCP is enough because links are ``physically secure''~\cite{benton2013,wasserman2013}, the fact that many switches have listener 
mode activated by default (allowing the establishment of malicious TCP connections, for instance)~\cite{benton2013} or that flow table verification capabilities are harder to implement when TLS is not 
in use~\cite{wasserman2013,son2013}.
In addition, it is worth mentioning the high denial of service risk posed to centralized controllers~\cite{shin2013,son2013}, the vulnerabilities in the controllers themselves~\cite{son2013,kreutz2013}, \coloredtext{bugs and vulnerabilities in applications~\cite{Chandrasekaran2014_4}, targeted flooding attacks~\cite{jarraya2014}, insecure northbound interfaces that can lead to security breaches~\cite{jarraya2014}}, and the risk of resource depletion attacks~\cite{shin2013,benton2013}.
For instance, it has been shown that an attacker can easily compromise control plane communications through 
DoS attacks and launch a resource depletion attack on control platforms by exploiting a single  application such as a learning switch~\cite{benton2013,shin2013}.

\coloredtext{Another point of concern is the fact that current controllers, such as Floodlight, OpenDaylight, POX, and Beacon, have several security and resiliency issues~\cite{shin2014}.
Common application development problems (bugs), such as the sudden exit of an application or the continuous allocation of memory space, are enough to crash existing controllers.
On the security perspective, a simple malicious action such as changing the value of a data structure in memory can also directly affect the operation and reliability of current controllers.
These examples are illustrative that from a security and dependability perspective, there is still a long way to go.
}

\vspace{2mm}
\noindent \textit{Countermeasures for OpenFlow based SDNs}

Several countermeasures can be put in place to mitigate the security threats in SDNs.
Table~\ref{tab:countermeasuresthreats} summarizes a number of countermeasures that can be applied to different 
elements of an SDN/OpenFlow-enabled network. Some of these measures, namely rate limiting, event filtering, 
packet dropping, shorter timeouts, and flow aggregation, are already recommended in the most recent 
versions of the OpenFlow specification (version 1.3.1 and later). 
However, most of them are not yet supported or implemented in SDN deployments.

{\renewcommand{\arraystretch}{1.4}
\begin{table}[!htp]
\caption{Countermeasures for security threats in OpenFlow networks.}
\label{tab:countermeasuresthreats}
\begin{center}
\footnotesize
\begin{tabularx}{\columnwidth}{lX}
\hline
\multicolumn{1}{c}{\textbf{Measure}} & \multicolumn{1}{c}{\textbf{Short description}} \\\hline
Access control       & Provide strong authentication and authorization mechanisms on devices.  \\\hline
Attack detection     & Implement techniques for detecting different types of attacks. \\\hline
Event filtering         & Allow (or block) certain types of events to be handled by special devices. \\\hline
Firewall and IPS     & Tools for filtering traffic, which can help to prevent different types of attacks. \\\hline
Flow aggregation   & Coarse-grained rules to match multiple flows to prevent information disclosure and DoS attacks. \\\hline
Forensics support  & Allow reliable storage of traces of network activities to find the root causes of different problems.  \\\hline
Intrusion tolerance &  Enable control platforms to maintain correct operation despite intrusions. \\\hline
Packet dropping     & Allow devices to drop packets based on security policy rules or current system load. \\\hline
Rate limiting           & Support rate limit control to avoid DoS attacks on the control plane. \\\hline
Shorter timeouts  & Useful to reduce the impact of an attack that diverts traffic.  \\\hline
\end{tabularx}
\end{center}
\end{table}
}

Traditional techniques such as 
access control, attack detection mechanisms, event filtering (e.g.,  controller decides which asynchronous messages he is not going to accept), firewalls, and intrusion detection systems, can be used to mitigate the impact of or to avoid attacks.
They can be implemented in different devices, such as controllers, forwarding devices, middleboxes, and so forth.
Middleboxes can be a good option for enforcing security policies in an enterprise because they are (in general) more robust and special purpose (high performance) devices.
Such a strategy also reduces the potential overhead cause by implementing these countermeasures directly on controllers or forwarding devices.
However, middleboxes can add extra complexity to the network management, i.e., increase the OPEX at the cost of better performance.

Rate limiting, packet dropping, shorter timeouts and flow aggregations are techniques that can be applied on controllers and forwarding devices to mitigate different types of attacks, such as denial-of-service and information disclosure.
For instance, reduced timeouts can be used to mitigate the effect of an attack exploring the reactive operation mode of the network to make the controller install rules that divert traffic to a malicious machine. 
With reduced timeouts, the attacker would be forced to constantly generate a number of forged packets to avoid timeout expiration, making the attack more likely to be detected.
Rate limiting and packet dropping can be applied to avoid DoS attacks on the control plane or stop on-going attacks directly on the data plane by installing specific rules on the devices where the attacks is being originated.

Forensics and remediation encompass mechanisms such as secure logging, event correlation and consistent reporting. 
If anything wrong happens with the network, operators should be able to safely figure out the root cause of the problem and put the network to work on a secure operation mode as fast as possible.
Additionally, techniques to tolerate faults and intrusions, such as state machine replication~\cite{bolosky2011}, proactive-reactive recovery~\cite{sousa2010}, and diversity~\cite{garcia2013}, can be added to control platforms for increasing the robustness and security properties by automatically masking and removing faults.
Put differently, SDN controllers should be able to resist against different types of events (e.g., power outages, network disruption, communication failures, network partitioning) and attacks (e.g., DDoS, resource exhaustion)~\cite{kreutz2013,botelho2013}. 
One of the most traditional ways of achieving high availability is through replication. 
Yet, proactive-reactive recovery and diversity are two examples of crucial techniques that add value to the system for resisting against different kinds of attacks and failures (e.g., those exploring common vulnerabilities or caused by software aging problems).

Other countermeasures to address different threats and issues of SDN include enhancing the security and 
dependability of controllers, protection and isolation of applications\coloredtext{~\cite{sorensen2012,kreutz2013,porras2012,shin2014}}, trust management between controllers 
and forwarding devices~\cite{kreutz2013}, integrity checks of controllers and applications~\cite{kreutz2013}, forensics and remediation~\cite{sorensen2012,kreutz2013}, 
verification frameworks~\cite{chua2013,porras2012,korniak2011}, and resilient control planes~\cite{fonseca2012,korniak2011,kreutz2013,sorensen2012}.
Protection and isolation mechanisms should be part of any controller. Applications should be isolated from each other and from the controller. 
Different techniques such as security domains (e.g., kernel, security, and user level) and data access protection mechanisms should be put in place in order to avoid security threats from \manapps. 

Implementing trust between controllers and forwarding is another requirement for insuring that malicious elements 
cannot harm the network without being detected. 
An attacker can try to spoof the IP address of the controller and make switches connect to its own controller. 
This is currently the case since most controllers and switches only establish insecure TCP connections. 
Complementarly, integrity checks on controller and application software can help to ensure that safe code is being bootstrapped, which eliminates harmful software from being started once the system restarts. 
Besides integrity checks, other things such as highly specialized malware detection systems should be developed for SDN. 
Third-party \manapps should always be scanned for bad code and vulnerabilities because a malicious application represents a significant security threat to the network. 

It is worth mentioning that there are also other approaches for mitigating security threats in SDN, such as declarative languages to eliminate network 
protocol vulnerabilities~\cite{casey2013}. 
This kind of descriptive languages can specify semantic constraints, structural constraints and safe access properties of OpenFlow messages. 
Then, a compiler can use these inputs to find programmers' implementation mistakes on message operations. 
In other words, such languages can help find and eliminate implementation vulnerabilities of southbound specifications.

\coloredtext{Proposals providing basic security properties such as authentication~\cite{Toseef2014} and access control~\cite{Klaedtke2014-4} are starting to appear. 
C-BAS~\cite{Toseef2014} is a certificate-based AAA (Authentication, Authorization and Accounting) architecture for improving the security control on SDN experimental facilities.
Solutions in the spirit of C-BAS can be made highly secure and dependable through hybrid system architectures, which combine different technologies and techniques from distributed systems, security, and fault and intrusion tolerance~\cite{Kreutz2014PRDC,Kreutz2014sodis,Verissimo2003ADS}.
}

\subsection{Migration and Hybrid deployments}
\label{sec:hybrid}

The promises by SDN to deliver easier design, operation and management of computer networks are endangered
by challenges regarding incremental deployability, robustness, and scalability. 
A prime SDN adoption challenge relates to organizational barriers that may arise due to the first (and second) order 
effects of SDN automation capabilities and ``layer/domain blurring''. 
Some level of human resistance is to be expected and may affect the decision and deployment processes of SDN, especially by those that may regard the 
control refactorization of SDN as a risk to the current chain of control and command, or even to their job security. 
This complex social challenge is similar (and potentially larger) to known issues between the transport and IP 
network divisions of service providers, or the system administrator, storage, networking, and security teams 
of enterprise organizations. Such a challenge is observable on today's virtualized data centers, through the shift 
in role and decision power between the networking and server people. Similarly, the development and operations (DevOps) movement has caused a shift in the locus of influence, not only 
on the network architecture but also on purchasing, and this is an effect that SDN may exacerbate. These changes in role and power causes a second order effect on the sales division of vendors that are required to adapt accordingly.

Pioneering SDN operational deployments have been mainly greenfield scenarios and/or tightly controlled single 
administrative domains. Initial roll-out strategies are mainly based on virtual switch overlay models or 
OpenFlow-only network-wide controls. However, a broader adoption of SDN beyond data center silos -- and 
between themselves -- requires considering the interaction and integration with legacy control planes providing 
traditional switching; routing; and operation, administration, and management (OAM) functions. Certainly, 
rip-and-replace is not a viable strategy for the broad adoption of new networking technologies.

Hybrid networking in SDN should allow deploying OpenFlow for a subset of all flows only, enable OpenFlow on 
a subset of devices and/or ports only, and provide options to 
interact with existing OAM protocols, legacy devices, and neighboring domains. As in any technology transition 
period where fork-lift upgrades may not be a choice for many, migration paths are critical for 
adoption.

Hybrid networking in SDN spans several levels. The Migration Working Group of the ONF is tackling the scenario where hybrid switch architectures and hybrid (OpenFlow and non-OpenFlow) devices 
co-exist. Hybrid switches can be configured to behave as a legacy switch or as an OpenFlow switch and, in some 
cases, as both simultaneously. This can be achieved, for example, by partitioning the set of ports of a switch, 
where one subset is devoted to OpenFlow-controlled networks, and the other subset to legacy networks. For these 
subsets to be active at the same time, each one having its own data plane, multi-table support at the forwarding 
engine (e.g., via TCAM partitioning) is required. Besides port-based partitioning, it is also possible to rely 
on VLAN-based (prior to entering the OpenFlow pipeline) or flow-based partitioning using OpenFlow matching and 
the \texttt{LOCAL} and/or \texttt{NORMAL} actions to redirect packets to the legacy pipeline or the switch's 
local networking stack and its management stack. Flow-based partitioning is the most flexible option, as it 
allows each packet entering a switch to be classified by an OpenFlow flow description and treated by the 
appropriate data plane (OpenFlow or legacy).

There are diverse controllers, such as OpenDaylight~\cite{opendaylight2013}, HP VAN SDN~\cite{hp2013-1}, and OpenContrail~\cite{junipernetworks2013-1}, that have been designed to integrate current non-SDN technologies (e.g., SNMP, PCEP, BGP, NETCONF) with SDN interfaces such as OpenFlow and OVSDB.
Nonetheless, controllers such as ClosedFlow~\cite{Hand2014_4} have been recently proposed with the aim of introducing SDN-like programming capabilities in traditional network infrastructures, making the integration of legacy and SDN-enabled networks a reality without side effects in terms of programmability and global network control.
ClosedFlow is designed to control legacy Ethernet devices (e.g., Cisco 3550 switches with a minimum IOS of 12.2 SE) in a similar way an OpenFlow controller allows administrators to control OpenFlow-enabled devices.
More importantly, ClosedFlow does not impose any change on the forwarding devices. 
It only takes advantage of the existing hardware and firmware capabilities to mimic an SDN control over the network, i.e., allow dynamic and flexible programmability in the data plane.
The next step could be the integration of controllers like ClosedFlow and OpenFlow-based controllers, promoting interoperability among controllers and a smooth transition from legacy infrastructures to SDN-enabled infrastructure with nearly all the capabilities of a clean-slate SDN-enabled infrastructure.

Furthermore, controllers may have to be separated into distinct peer domains for different reasons, such as scalability, technology, controllers from different vendors, controllers with different service functionality, and diversity of administrative domains~\cite{ONF2014SDNarch}.
Controllers from different domains, or with distinct purposes, are also required to be backwards compatible either by retrofitting or extending existing multi-domain protocols (e.g. BGP) or by proposing new SDN-to-SDN protocols (aka east/westbound APIs).

Some efforts have been already devoted to the challenges of migration and hybrid SDNs. RouteFlow~\cite{rothenberg2012-1} implements an IP level control plane on top of an OpenFlow 
network, allowing the underlying devices to act as IP routers under different possible arrangements. \coloredtext{The Cardigan project~\cite{stringer2013,floss-meets-sdn} has deployed RouteFlow at a live Internet eXchange now for over a year.  
LegacyFlow~\cite{rothenberg2014} extends the OpenFlow-based controlled network to embrace non-OpenFlow nodes. 
There are also some other early use cases on integrating complex legacy system such as DOCSIS~\cite{Fuentes2014}, Gigabit Ethernet passive optical network and DWDM ROADM (Reconfigurable Optical Add/Drop Multiplexer)~\cite{Parniewicz2014,Belter2014}.}
The common 
grounds of these pieces of work are (1) considering hybrid as the coexistence of traditional environments of 
closed vendor's routers and switches with new OpenFlow-enabled devices; (2) targeting the interconnection of 
both control and data planes of legacy and new network elements; and (3) taking a controller-centric approach, 
drawing the hybrid line outside of any device itself, but into the controller application space.

Panopticon~\cite{levin2014} defines an architecture and methodology to consistently implement 
SDN inside enterprise legacy networks through network orchestration under strict budget constraints. The proposed 
architecture includes policy configurations, troubleshooting and maintenance tasks establishing transitional networks 
(SDN and legacy) in structures called Solitary Confinement Trees (SCTs), where VLAN IDs are efficiently used by 
orchestration algorithms to build paths in order to steer traffic through SDN switches. Defying the partial SDN 
implementation concept, they confirm that this could be a long-term operational strategy solution for enterprise 
networks.

HybNET~\cite{lu2013} presents a network management framework for
hybrid OpenFlow-legacy networks. It provides a common centralized
configuration interface to build virtual networks using VLANs. An
abstraction of the physical network topology is taken into account by
a centralized controller that applies a path finder mechanism, in
order to calculate network paths and program the OpenFlow switches via
REST interfaces and legacy devices using NETCONF~\cite{enns2011-1}.

\coloredtext{More recently, frameworks such as ESCAPE~\cite{Csoma2014}  and its extensions have been proposed to provide multi-layer service orchestration in multi-domains.
Such frameworks combine different tools and technologies such as Click~\cite{morris1999}, POX~\cite{mccauley2012}, OpenDaylight~\cite{opendaylight2013} and NETCONF~\cite{enns2011-1}.
In other words, those frameworks integrate different SDN solutions with traditional ones.
Therefore, they might be useful tools on the process of integrating or migrating legacy networking infrastructure to SDN.

Other hybrid solutions starting to emerge include Open Source Hybrid IP/SDN (OSHI)~\cite{salsano2014oshi}. 
OSHI combines Quagga for OSPF routing and SDN capable switching devices (e.g., Open vSwitch) on Linux to provide  backwards compatibility for supporting incremental SDN deployments, i.e., enabling interoperability with non-OF forwarding devices in carrier-grade networks.
}

While full SDN deployments are straightforward only in some green field deployments such as data center networks or by 
means of an overlay model approach, hybrid SDN approaches represent a very likely deployment model 
that can be pursued by different means, including~\cite{vissicchio2014}:

\begin{itemize}
\item Topology-based hybrid SDN: Based on a topological separation of the nodes controlled by traditional and SDN 
paradigms. The network is partitioned in different zones and each node belongs to only one zone. 
\item Service-based hybrid SDN: Conventional networks and SDN provide different services, where overlapping nodes, 
controlling a different portion of the FIB (or generalized flow table) of each node. Examples include network-wide 
services like forwarding  that can be based on legacy distributed control, while SDN provides edge-to-edge services 
such as enforcement of traffic engineering and access policies, or services requiring full traffic visibility 
(e.g., monitoring).
\item Class-based hybrid SDN: Based on the partition of traffic in classes, some controlled by SDN and the remaining 
by legacy protocols. While each paradigm controls a disjoint set of node forwarding entries,  each paradigm is 
responsible for all network services for the assigned traffic classes. 
\item Integrated hybrid SDN: A model where SDN is responsible for all the network services, and
uses traditional protocols (e.g., BGP) as an interface to node FIBs. For example, it can control forwarding
paths by injecting carefully selected routes into a routing system or adjusting protocol settings (e.g., IGP weights). 
Past efforts on RCPs~\cite{caesar2005} and the ongoing efforts within ODL~\cite{opendaylight2013} can be considered 
examples of this hybrid model.
\end{itemize}

In general, benefits of hybrid approaches include enabling flexibility (e.g., easy match on packet fields for 
middleboxing) and SDN-specific features (e.g., declarative management interface) while partially keeping the inherited 
characteristics of conventional networking such as robustness, scalability, technology maturity, and low deployment 
costs. On the negative side, the drawbacks of hybridization include the need for ensuring profitable interactions 
between the networking paradigms (SDN and traditional) while dealing with the heterogeneity that largely depends on 
the model.

Initial trade-off analyses~\cite{vissicchio2014} suggest that the combination of centralized and distributed paradigms may provide 
mutual benefits. However, future work is required to devise techniques and interaction mechanisms that maximize 
such benefits while limiting the added complexity of the paradigm coexistence.


\subsection{Meeting carrier-grade and cloud requirements}

A number of carrier-grade infrastructure providers (e.g., NTT, AT\&T, Verizon, Deutsche Telekom) are at the core of the SDN community with the ultimate goal of solving their long standing networking problems.
\coloredtext{In the telecom world, NTT can be considered one of the forefront runners in terms of investing in the adoption and deployment of SDN in all kinds of network infrastructures, from backbone, data center, to edge customers~\cite{Reinecke2014}. In 2013, NTT launched an SDN-based, on-demand elastic provisioning platform of network resources (e.g., bandwidth) for HD video broadcasters~\cite{bernier2013}.
}
Similarly, as a global cloud provider with data centers spread across the globe~\cite{nttdata2014}, the same company launched a similar service for its cloud customers, who are now capable of taking advantage of dynamic networking 
provisioning intra- and inter-data centers~\cite{wagner2014}.
AT\&T is another telecom company that is investing heavily in new services, such as user-defined network clouds, that 
take advantage of recent developments in NFV and SDN~\cite{atandtinc.2014}.
\coloredtext{As we mentioned before, SDN and NFV are complementary technologies that can be applicable to different types of networks, from local networks and data centers to transport networks~\cite{Haleplidis2014_4,onf2014-1,Cerrato2014-1,Xia2014_4,haleplidis2014sdnnfv,gemberjacobson2014}.
Recently, several research initiatives have worked towards combining SDN and NFV through Intel's DPDK, a set of libraries and drivers that facilitates the development of network-intensive applications and allows the implementation of fine-grained network functions~\cite{Cerrato2014}. 
Early work towards service chaining have been proposed by combining SDN and NFV technologies~\cite{Ruckert2014,Blendin2014,Skoldstrom2014,Jamjoom2014_4,john2013nsc}, and studies around the ForCES's~\cite{doria2010} applicability to SDN-enhanced NFV have also come to light~\cite{Haleplidis2014_4}.}
These are some of the early examples of the opportunities SDNs seem to bring to telecom and cloud providers. 

Carrier networks are using the SDN paradigm as the technology means for solving a number of long standing problems. 
Some of these efforts include 
new architectures for a smooth migration from the current mobile core infrastructure to SDN~\cite{pentikousis2013}, and techno-economic models for virtualization of these networks~\cite{naudts2012,onfsolutionbrief2013};  
carrier-grade OpenFlow virtualization schemes~\cite{skoldstrom2013,koponen}, including virtualized broadband access infrastructures~\cite{gharakheili2013}, techniques that are allowing the offer of network-as-a-service~\cite{pacnet2013}; 
\coloredtext{programmable GE-PON and DWDM ROADM~\cite{Parniewicz2014,Belter2014,Belter2014-1,Clegg2014}; large-scale inter-autonomous systems (ASs) SDN-enabled deployments~\cite{bennesby2012inter};}
flexible control of network resources~\cite{corporation2012}, including offering MPLS services using an SDN approach~\cite{das2011};
and the investigation of novel network architectures, from proposals to separate the network edge from the core~\cite{casado2012,6786608}, with the latter forming the fabric that transports packets as defined by an intelligent edge, to software-defined Internet exchange points~\cite{feamster2013,stringer2013}.

\coloredtext{Use-case analysis~\cite{Devlic2012} of management functions required by carrier networks have identified a set of requirements and existing limitations in the SDN protocol toolbox. 
For instance, it has been pinpointed that OF-Config~\cite{OFCONFIG} needs a few extensions in order to meet the carrier requirements, such as physical resource discovery, logical link configuration, logical switch instantiation, and device and link OAM configuration~\cite{Devlic2012}.
Similarly, OpenFlow extensions have also been proposed to realize packet-optical integration with SDN~\cite{shirazipour2012}.
In order to support SDN concepts in large-scale wide area networks, different extensions and mechanisms are required, both technology-specific (e.g., MPLS BFD) and technology agnostic, such as: resiliency mechanisms for surviving link failures~\cite{Adrichem2014}, failures of controller or forwarding elements; solutions for integrating residential customer services in different forms (i.e., support also current technologies); new energy-efficient networking approaches; QoS properties for packet classification, metering, coloring, policing, shaping and scheduling; and multi-layer aspects outlining different stages of packet-optical integration~\cite{John2014_4,John2014split,shirazipour2012}.
}
%

SDN technology also brings new possibilities for cloud providers.
By taking advantage of the logically centralized control of network resources~\cite{hong2013,jain2013-1} it is possible to simplify and optimize network management of data centers and achieve:
(i) efficient intra-datacenter networking, including fast recovery mechanisms for the  data and control planes~\cite{sharma2013-1,staessens2011,sharma2013}, adaptive traffic engineering with minimal modifications to DCs networks~\cite{benson2011mTE}, simplified fault-tolerant routing~\cite{mysore2009}, performance isolation~\cite{greenberg2009}, and easy and efficient resource migration (e.g., of VMs and virtual networks)~\cite{sharma2013-1};
(ii) improved inter-datacenter communication, including the ability to fully utilize the expensive high-bandwidth links without impairing quality of service~\cite{jain2013-1,sadasivarao2013};
(iii) higher levels of reliability (with novel fault management mechanisms, etc.)~\cite{mysore2009,staessens2011,sharma2013-1,Adrichem2014}; and 
(iv) cost reduction by replacing complex, expensive hardware by simple and cheaper forwarding devices~\cite{tanner2013,jain2013-1}.

Table~\ref{tab:carriergradeneeds} summarizes some of the carrier-grade network and cloud infrastructure providers' requirements.
In this table we show the current challenges and what is to be expected with SDN.
As we saw before, some of the expectations are already becoming a reality, but many are still open issues.
What seems to be clear is that SDN represents an opportunity for telecom and cloud providers, in providing flexibility, cost-effectiveness, and easier management of their networks.

{\renewcommand{\arraystretch}{1.4}
\begin{table*}[t!]
\caption{Carrier-grade and cloud provider expectations \& challenges}
\label{tab:carriergradeneeds}
\newcommand{\firstcolumnwidth}{2.5cm} 
\begin{center}
\footnotesize
\begin{tabularx}{0.99\textwidth}{p{\firstcolumnwidth}p{6cm}X}
\hline
\textbf{What} & \textbf{Currently} & \textbf{Expected with SDN} \\\hline
\multirow{8}{*}{\begin{minipage}{\firstcolumnwidth}Resource Provisioning\end{minipage}} 
                     & Complex load balancing configuration. & Automatic load balancing reconfiguration.~\cite{elby2012,jain2013-1} \\\cline{2-3}
		    & Low virtualization capabilities across hardware platforms & NFV for virtualizing network functionality across hardware appliances.~\cite{tanner2013,atandtinc.2014} \\\cline{2-3}
		    & Hard and costly to provide new services. & Create and deploy new network service quickly.~\cite{tanner2013,atandtinc.2014}\\\cline{2-3}
                     & No bandwidth on demand. & Automatic bandwidth on demand.~\cite{onfsolutionbrief2013} \\\cline{2-3}
                     & Per network element scaling. & Better incremental scaling.~\cite{elby2012,staessens2011} \\\cline{2-3}
		    & Resources statically pre-provisioned. & Dynamic resource provisioning in response to load.~\cite{elby2012,jain2013-1,tanner2013,naudts2012,hong2013} \\
\hline
\multirow{4}{*}{\begin{minipage}{\firstcolumnwidth}Traffic Steering\end{minipage}} 
                     & All traffic is filtered. & Only targeted traffic is filtered.~\cite{elby2012} \\\cline{2-3}
                     & Fixed only. & Fixed and mobile.~\cite{elby2012} \\\cline{2-3}
                     & Per network element scaling. & Better incremental scaling. ~\cite{naudts2012,staessens2011}\\\cline{2-3}
                     & Statically configured on a per-device basis. & Dynamically configurable.~\cite{jain2013-1,onfsolutionbrief2013,anwer2013} \\
\hline
\multirow{4}{*}{\begin{minipage}{\firstcolumnwidth}Ad Hoc Topologies\end{minipage}}   
                     & All traffic from all probes collected. & Only targeted traffic from targeted probes is collected. \\\cline{2-3}
                     & Massive bandwidth required. & Efficient use of bandwidth.~\cite{jain2013-1,onfsolutionbrief2013} \\\cline{2-3}
                     & Per network element scaling. & Better incremental scaling.~\cite{elby2012,onfsolutionbrief2013} \\\cline{2-3}
                     & Statically configured. & Dynamically configured.~\cite{elby2012,gerlach2013,bernier2013} \\
\hline
\multirow{7}{*}{\begin{minipage}{\firstcolumnwidth}Managed Router\\ Services\end{minipage}}   
                     & Complex configuration, management and upgrade. & Simplified management and upgrade.~\cite{jain2013-1,elby2012,tanner2013,onfsolutionbrief2013,staessens2011} \\\cline{2-3}
                     & Different kinds of routers, such as changeover (CO). & No need for CO routers, reducing aggregation costs.~\cite{elby2012,tanner2013,naudts2012} \\\cline{2-3}
                     & Manual provisioning. & Automated provisioning.~\cite{elby2012,onfsolutionbrief2013,anwer2013} \\\cline{2-3}
                     & On-premises router deployment. & Virtual routers (either on-site or not).~\cite{onfsolutionbrief2013,elby2012,naudts2012} \\\cline{2-3}
                     & Operational burden to support different equipments. & Reduced technology obsolescence.~\cite{naudts2012} \\\cline{2-3}
                     & Router change-out as technology or needs change. & Pay-as-you grow CAPEX model.~\cite{naudts2012} \\\cline{2-3}
                     & Systems complex and hard to integrate. & Facilitates simplified system integrations.~\cite{elby2012,tanner2013,gerlach2013}\\\hline
\multirow{2}{*}{\begin{minipage}{\firstcolumnwidth}Revenue Models\end{minipage}}   
                     & Fixed long term contracts. & More flexible and on-demand contracts.~\cite{onfsolutionbrief2013,corporation2012}\\\cline{2-3}
                     & Traffic consumption. & QoS metrics per-application.~\cite{onfsolutionbrief2013,staessens2011,staessens2011,velasco2013} \\
\hline
\multirow{4}{*}{\begin{minipage}{\firstcolumnwidth}Middleboxes \\Deployment \& \\Management\end{minipage}}   
                     & Composition of services is hard to implement. & Easily expand functionality to meet the infrastructure needs.~\cite{tanner2013}\\\cline{2-3}
                     & Determine where to place middleboxes a priori (e.g., large path inflation problems). & Dynamic placement using shortest or least congested path. ~\cite{qazi2013-1,velasco2013,gerlach2013} \\\cline{2-3}
                     & Excessive over-provisioning to anticipate demands. & Scale up to meet demands, and scale down to conserve resources (elastic middleboxes).~\cite{elby2012,naudts2012} \\
\hline
\multirow{3}{*}{\begin{minipage}{\firstcolumnwidth}Other Issues\end{minipage}}   
                     & Energy saving strategies are hard to implement. & Flexible and easy to deploy energy saving strategies.~\cite{staessens2011}\\\cline{2-3}
                     & Complex and static control and data plane restoration techniques. & Automated and flexible restoration techniques for both control and data plane.~\cite{staessens2011}\\\cline{2-3}
\hline
\end{tabularx}
\end{center}
\end{table*}
}


\subsection{SDN: the missing piece towards Software-Defined Environments}

The convergence of different technologies is enabling the emergence of fully programmable IT infrastructures.
It is already possible to dynamically and automatically configure or reconfigure the entire IT stack, from the network infrastructure up to the applications, to better respond to workload changes.
Recent advances makes on-demand provisioning of resources possible, at nearly all infrastructural layers.
The fully automated provisioning and orchestration of IT infrastructures as been recently named Software-Defined Environments (SDEs)~\cite{racherla2014,li2014}, by IBM.
This is a novel approach that is expected to have significant potential in simplifying IT management, optimizing the use of the infrastructure, reduce costs, 
and reduce the time to market of new ideas and products.
In an SDE, workloads can be easily and automatically assigned to the appropriate IT resources based on application characteristics, security and service level policies, and the best-available resources to deliver continuous, dynamic optimization and reconfiguration to address infrastructure issues in a rapid and responsive manner.
Table~\ref{tab:TraditionalAndSDE} summarizes the traditional approaches and some of the key features being enabled 
by SDEs~\cite{alba2014,arnold2014}.

{\renewcommand{\arraystretch}{1.4}
\begin{table*}[t!]
\caption{SDE pushing IT to the next frontier}
\label{tab:TraditionalAndSDE}
\newcommand{\firstcolumnwidth}{3.2cm} 
\begin{center}
\footnotesize
\begin{tabularx}{0.99\textwidth}{XX}
\hline
\textbf{Traditionally} & \textbf{Expected with SDEs} \\\hline
IT operations manually map the resources for apps for software deployment.  & Software maps resources to the workload and deploys the workload. \\\hline
Networks are mostly statically configured and hard to change.  & Networks are virtualized and dynamically configured on-demand. \\\hline
Optimization and reconfiguration to reactively address issues are manual.  & Analytics-based optimization and reconfiguration of infrastructure issues. \\\hline
Workloads are typically manually assigned to resources.  & Workloads are dynamically assigned. \\\hline
\end{tabularx}
\end{center}
\end{table*}
}

In an SDE the workloads are managed independently of the systems and underlying infrastructure, i.e., are not tied to a specific technology or vendor~\cite{li2014,racherla2014}.
Another characteristic of this new approach is to offer a programmatic access to the environment as a whole, selecting the best available resources based on the current status of the infrastructure, and enforcing the policies defined.
In this sense, it shares much of the philosophy of SDN.
Interestingly, one of the missing key pieces of an SDE was, until now, Software-Defined Networking.

The four essential building blocks of an SDE~\cite{li2014,racherla2014,arnold2014} are:

\begin{itemize}
\item Software-Defined Networks (SDN)~\cite{dixon2014,ibmsystemsandtechnologygroup2014},
\item Software-Defined Storage (SDS)~\cite{alba2014},
\item Software-Defined Compute (SDC)~\cite{racherla2014}, and 
\item Software-Defined Management (SDM)~\cite{ibmsystems2014}.
\end{itemize}

In the last decade the advances in virtualization of compute and storage, together with the availability of sophisticated cloud orchestration tools have enabled SDS, SDC and SDM.
These architectural components have been widely used by cloud providers and for building IT infrastructures in different enterprise environments.
However, the lack of programmable network control has so far hindered the realization of a complete Software-Defined Environment.
SDN is seen as the technology that may fill this gap, as attested by the emergence of cloud-scale network virtualization platforms based on this new paradigm~\cite{koponen}.


%
%

\begin{figure}[ht!]
\centering
\includegraphics[width=0.95\columnwidth]{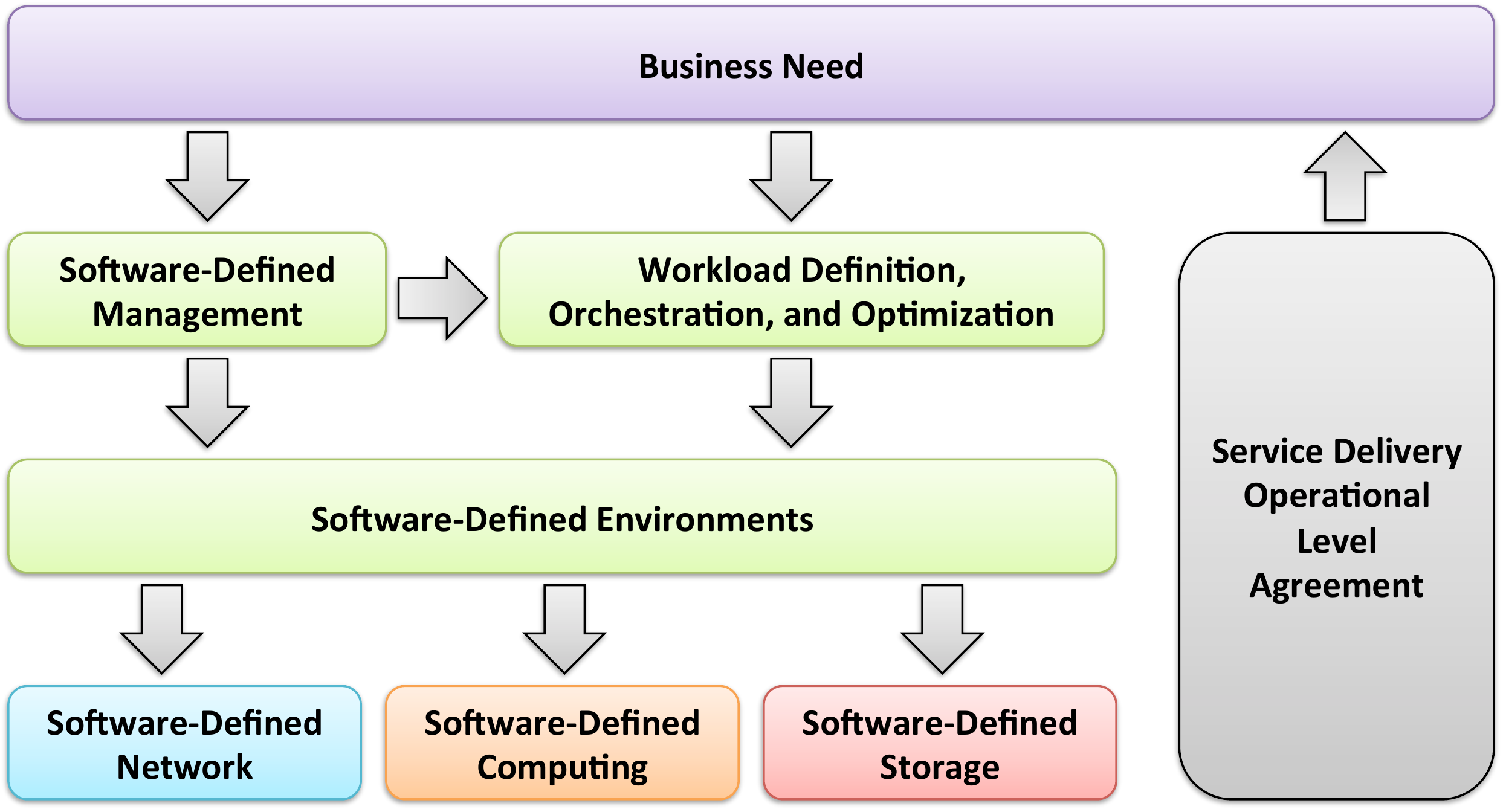}
\caption{Overview of an IT infrastructure based on a SDE.}
\label{fig:SDEenabledIT}
\end{figure}

The IBM SmartCloud Orchestrator is one of the first examples of an SDE~\cite{li2014,racherla2014}.
It integrates compute, storage, management and networking in a structured way.
Figure~\ref{fig:SDEenabledIT} gives a simplified overview of an SDE, by taking the approach developed by IBM as its basis.
The main idea of an SDE-based infrastructure is that the business needs that define the workloads trigger the reconfiguration of the global IT infrastructure (compute, storage, network). 
This is an important step towards a more customizable IT infrastructure that focuses on the business requirements rather than on the limitations of the infrastructure itself.

%% file: text/8_conclusion.tex
\section{Conclusion}

Traditional networks are complex and hard to manage.  One
of the reasons is that the control and data planes are vertically
integrated and vendor specific. Another, concurring reason, is that
typical networking devices are also tightly tied to line products and
versions.  In other words, each line of product may have its own
particular configuration and management interfaces, implying long
cycles for producing product updates (e.g., new firmware) or upgrades
(e.g., new versions of the devices). All this has given rise to vendor
lock-in problems for network infrastructure owners, as well as posing
severe restrictions to change and innovation.

Software-Defined Networking (SDN) created an opportunity for solving
these long-standing problems.  Some of the key ideas of SDN are the
introduction of dynamic programmability in forwarding devices through
open southbound interfaces, the decoupling of the control and data
plane, and the global view of the network by logical centralization of
the ``network brain''.  While data plane elements became dumb, but
highly efficient and programmable packet forwarding devices, the
control plane elements are now represented by a single entity, the
controller or network operating system.  Applications implementing the
network logic run on top of the controller and are much easier to
develop and deploy when compared to traditional networks. Given the
global view, consistency of policies is straightforward to enforce.
SDN represents a major paradigm shift in the development and
evolution of networks, introducing a new pace of innovation in networking infrastructure.

In spite of recent and interesting attempts to survey this new chapter
in the history of networks~\cite{lara2014,jarraya2014,nunes2014}, the
literature was still lacking, to the best of our knowledge, a single
extensive and comprehensive overview of the building blocks, concepts,
and challenges of SDNs.
%
%
Trying to address this gap, the present paper used a layered approach
to methodically dissect the state of the art in terms of concepts,
ideas and components of software-defined networking, covering a broad
range of existing solutions, as well as future directions.

We started by comparing this new paradigm with traditional networks
and discussing how academy and industry helped shape software-defined
networking.  Following a bottom-up approach, we provided an in-depth
overview of what we consider the eight fundamental facets of the SDN
problem: 1) hardware infrastructure, 2) southbound interfaces, 3)
network virtualization (hypervisor layer between the forwarding
devices and the network operating systems), 4) network operating
systems (SDN controllers and control platforms), 5) northbound
interfaces (common programming abstractions offered to network
applications), 6) virtualization using slicing techniques provided by
special purpose libraries and/or programming languages and compilers,
7) network programming languages, and finally, 8) network
applications.


SDN has successfully managed to pave the way towards a next generation networking, spawning an innovative research and development
environment, promoting advances in several areas: switch and
controller platform design, evolution of scalability and performance
of devices and architectures, promotion of security and dependability.

We will continue to witness extensive activity around SDN in the near
future. Emerging topics requiring further research are, for example:
the migration path to SDN, extending SDN towards carrier transport networks, realization of the
network-as-a-service cloud computing paradigm, or software-defined
environments (SDE).
\coloredtext{As such, we would like to receive feedback from the networking/SDN community as this novel paradigm evolves, to make this a ``live document'' that gets updated and improved based on the community feedback.
We have set up a github page\footnote{\coloredtext{https://github.com/SDN-Survey/latex/wiki}} for this purpose, and we invite our readers to join us in this communal effort.}